\documentclass[iop,numberedappendix,appendixfloats]{emulateapj}

\shorttitle{Herschel-detected SDSS Quasars}
\shortauthors{Ma \& Yan}

\usepackage{natbib}
\usepackage{enumitem}
\usepackage{amsmath}
\usepackage{color}
\usepackage{soul}

\usepackage[load-configurations=astronomy]{siunitx}
\DeclareSIUnit\micron{\ensuremath{\mathrm{\micro\meter}}}
\DeclareSIUnit\sig{\ensuremath{\mathrm{\sigma}}}
\DeclareSIUnit\mag{mag}
\DeclareSIUnit\parsec{pc}
\DeclareSIUnit\jansky{Jy}
\DeclareSIUnit\deg{deg}
\DeclareSIUnit\yr{yr}
\DeclareSIUnit\dex{dex}
\DeclareSIUnit\Lsun{\ensuremath{\mathrm{L_\sun}}}
\DeclareSIUnit\Msun{\ensuremath{\mathrm{M_\sun}}}

\newcommand{\Herschel}{\textit{Herschel}}
\newcommand{\Spitzer}{\textit{Spitzer}}

\newcommand{\tophline}{\hline\vspace{-2.0ex}\\\hline\vspace{-1.5ex}\\}
\newcommand{\midhline}{\vspace{-2.0ex}\\\hline\vspace{-1.5ex}\\}
\newcommand{\bottomhline}{\vspace{-2.0ex}\\\hline\vspace{-1.5ex}\\}
\newcommand{\grouphline}{\vspace{-2.0ex}\\\vspace{-1.5ex}\\}

\definecolor{revcolor}{rgb}{1.0, 1.0, 0.45}

\makeatletter
\renewcommand\hyper@natlinkbreak[2]{#1}  
\makeatother

\begin{document}
\title{Co-evolution of Extreme Star Formation and Quasar: hints from
    {\it Herschel} and the Sloan Digital Sky Survey}

\author{Zhiyuan Ma\altaffilmark{1, \dag}
    and Haojing Yan\altaffilmark{1,\dag\dag}}
\altaffiltext{1}{Department of Physics and Astronomy, University of
    Missouri-Columbia}
\altaffiltext{\dag}{{\email{zmzff@mail.missouri.edu}}}
\altaffiltext{\dag\dag}{{\email{yanha@missouri.edu}}}
\begin{abstract}

Using the public data from the \Herschel{} wide field surveys, we study the
far-infrared properties of optical-selected quasars from the Sloan Digital
Sky Survey. Within the common area of $\sim\SI{172}{\deg\squared}$, we
have identified the far-infrared counterparts for \num{354} quasars, among
which \num{134} are highly secure detections in the \Herschel{}
\SI{250}{\micron} band (signal-to-noise ratios $\geq5$). This sample is
the largest far-infrared quasar sample of its kind, and spans a wide
redshift range of $0.14{\leq}z\leq 4.7$. Their far-infrared spectral energy
distributions, which are due to the cold dust components within the host
galaxies, are consistent with being heated by active star formation. In
most cases ($\gtrsim80$\%), their total infrared luminosities as inferred
from only their far-infrared emissions ($L_{IR}^{(cd)}$) already exceed
\SI{d12}{\Lsun}, and thus these objects qualify as ultra-luminous infrared
galaxies. There is no correlation between $L_{IR}^{(cd)}$ and the absolute
magnitudes, the black hole masses or the X-ray luminosities of the quasars,
which further support that their far-infrared emissions are not due to
their active galactic nuclei. A large fraction of these objects
($\gtrsim50\text{--}60\%$) have star formation rates $\gtrsim
\SI{300}{\Msun\per\yr}$. Such extreme starbursts among optical quasars,
however, is only a few per cent. This fraction varies with redshift, and
peaks at around $z\approx2$. Among the entire sample, \num{136} objects
have secure estimates of their cold-dust temperatures ($T$), and we find
that there is a dramatic increasing trend of $T$ with increasing
$L_{IR}^{(cd)}$. We interpret this trend as the envelope of the general
distribution of infrared galaxies on the ($T$, $L_{IR}^{(cd)}$) plane.

\end{abstract}


\section{Introduction}

Ultra-Luminous InfraRed Galaxies (ULIRGs), discovered as a distinct
class in the early 1980's \citep{Houck1984a, Houck1985, Aaronson1984}
by the Infrared Astronomical Satellite (IRAS) survey of the sky in 12
to \SI{100}{\micron}, are believed to host extreme star formation
regions that are heavily enshrouded by dust. They are characterized by
their exceptionally high IR luminosities ($L_{IR}>\SI{d12}{\Lsun}$;
integrated over restframe 8 to \SI{1000}{\micron}), which are believed
to be predominantly due to the re-radiation of star light processed by
dust (see \citealt{Lonsdale2006} for a review), and thus imply very
high star formation rates (SFR) of $>
100\text{--}\SI{1000}{\Msun\per\yr}$ (using the conversion of
\citealt{Kennicutt1998}) completely hidden by dust.

However, it has been noticed ever since their discovery that a
significant fraction of ULIRGs, especially those very luminous ones,
have optical signatures indicative of classic AGN activities. For
examples, \citet{Carter1984} notes that ten among a sample of 13 IRAS
sources with \SI{60}{\micron} flux density
$f_{60}\geq\SI{1.2}{\jansky}$ are Seyfert galaxies.
\citet{Sanders1988} show that ten galaxies among the 324 sources with
$f_{60}\geq\SI{5.4}{\jansky}$ from the IRAS Bright Galaxy Survey are
ULIRGs and that they all have a mixture of starburst and AGN
signatures, which has led them to propose an evolutionary scenario
that ULIRGs are the prelude to quasars. In their redshift survey of
the IRAS $S_{60}>\SI{0.6}{\jansky}$ galaxy sample,
\citet{Lawrence1999} have found that $\sim 20\%$ of their 95 ULIRGs
are AGN\@. The consensus is that ULIRGs that have ``warm'' IR colors,
i.e., whose emissions tend to peak at restframe mid-IR (MIR) rather
than far-IR (FIR), generally host (optical) AGN, which could be the
main energy sources that power their strong IR emissions
\citep[e.g.,][]{deGrijp1985, Osterbrock1985, Kim1998}. On the other
hand, ``cold'' ULIRGs that have their IR emissions peak at FIR should
be mostly powered by starbursts \citep[e.g.,][]{Elston1985,
    Heckman1987}.

An interesting question then is whether any AGN ULIRGs, especially
quasar ULIRGs, have starbursts that dominate their IR emissions.
Quasars represent the most extreme process of supermassive black hole
accretion, while starbursts are the most extreme process of star
formation. It is expected that the interplay of these two extremes
will have important consequences. This is made particularly important
by the quasar evolutionary scenario of \citet{Sanders1988}, as such
objects could be the transitional type between non-quasar ULIRGs and
``fully exposed'' quasars. Sanders et al.\ themselves believe that AGN
heating is the main mechanism for the strong IR emission of ULIRG
quasars. In their discussion of PG quasar continuum distributions from
UV to millimeter (mm), \citet{Sanders1989} further propose that a
warped galactic disk (beyond the central $\sim\SI{10}{\parsec}$ to a
few \si{\kilo\parsec}) heated by the central AGN can explain the
entire range of IR emission from \SI{5}{\micron} to
\SI{1}{\milli\meter}. However, \citet{Rowan-Robinson1995} argues that
this scheme could only be viable when the total luminosity in IR is
comparable or less than that in UV-optical; instead, he has
successfully modeled the IRAS-detected PG quasars by attributing their
mid-IR emissions to AGN heating and the FIR emission to starburst
heating, respectively. More analysis using larger samples from the
Infrared Space Observatory (ISO) observations support this view. For
example, \citet{Haas2003} have studied 64 PG quasars and have
concluded that starburst heating is more likely the cause of the
observed cold dust ($\sim 30\text{--}\SI{50}{\kelvin}$) FIR emissions
among the ULIRG part of the sample. However, they have also pointed
out that AGN heating should be the main power source for those extreme
ones that qualify as ``Hyper-luminous Infrared Galaxies'' (HyLIRG;
usually defined by $L_{IR}>\SI{d13}{\Lsun}$).

There are also two pieces of important, albeit indirect, evidence
supporting that the FIR emissions of quasar ULIRGs are likely due to
starbursts. First, a significant fraction of such objects have a large
amount of molecular gas \citep[e.g.,][]{Sanders1988, Evans2001,
    Scoville2003, Xia2012}, which is a strong indicator of active star
formation. Second, most quasar ULIRGs have polycyclic aromatic
hydrocarbon (PAH) features, which are also strongly indicative of
on-going star formation \citep[e.g.,][]{Schweitzer2006, Shi2007,
    Hao2007, Netzer2007, Cao2008}. While none of these are sufficient
to assert that starbursts dominate the strong FIR continua of quasar
ULIRGs, it is clear that they at least contribute significantly.

The picture above is largely based on the ULIRGs in the nearby
universe where they can be studied in detail. It would not be
surprising if any of it changes at high redshifts, as both quasars and
ULIRGs evolve strongly. The number density of quasars rises rapidly
from $z=0$ to $z=1$, and reaches the peak at $z\approx 2\text{--}3$
\citep[e.g.,][]{Osmer2004}. Similarly, while ULIRGs are very rare
objects today, they are much more numerous in earlier epochs. The deep
ISO surveys have revealed a large number of IR-luminous galaxies,
among which $>10\%$ are ULIRGs and many are at $z>1$
\citep[e.g.,][]{Rowan-Robinson2004}. The discovery of the so-called
``submillimeter (submm) galaxies'' (SMGs) at 450 and \SI{850}{\micron}
(see \citealt{Blain2002} for a review) added a new population to the
ULIRG family, as most of them are at $z\approx 2\text{--}3$ and have
$L_{IR}>\SI{d12}{\Lsun}$ dominated by the emission from cold dust.
Furthermore, \Spitzer{} observations suggest that high stellar
mass ($>\SI{d11}{\Msun}$) and otherwise ``normal'' star-forming
galaxies at $z\sim 2$ are likely all ULIRGs
\citep[e.g.,][]{Daddi2005}, which increases the ULIRG number density
at high redshifts to a more dramatic level than expected. On the other
hand, molecular gas has also been detected in quasar ULIRGs from
$z\gtrsim 1$ to 6 (\citealt{Solomon2005}, and the references therein;
see also e.g., \citealt{Wang2010, Wang2011a, Wang2011} for the recent
results at $z\sim 6$), lending support to the starburst-powered
interpretation of the FIR emission of such objects at high redshifts
as well.

If quasars and dust-enshrouded starbursts do co-exist, it is important
to investigate their co-evolution, which would require a large sample
over a wide redshift range. \Herschel{} Space Observatory
\citep{Pilbratt2010} has offered an unprecedented opportunity to
investigate this problem at the FIR wavelengths that were largely
unexplored by the previous studies. \Herschel{} had two imaging
instruments, namely, the Photodetector Array Camera and Spectrometer
\citep[PACS,][]{Poglitsch2010} and the Spectral and Photometric
Imaging REceiver \citep[SPIRE,][]{Griffin2010}. The PACS bands are 100
(or 70) and \SI{160}{\micron}, and the SPIRE bands are 250, 350 and
\SI{500}{\micron}. Together they sample the peak of heated dust
emission from $z \approx 0$ to 6 and beyond. There already have been a
number of studies on the FIR emission of quasars using \Herschel{}
observations \citep[e.g.,][]{Serjeant2010a, Leipski2010a, Leipski2013,
    Dai2012, Netzer2014}, however the current collection of quasars
that have individual \Herschel{} detections are still very scarce in
number and few have spanned a sufficient redshift range (for examples,
\citet{Leipski2013} present 11 objects at $z>5$; \citet{Dai2012}
include 32 objects at $0.5\leq z\leq 3.6$; \citet{Netzer2014} report
ten within a narrow window at $z\approx 4.8$).

In this paper, we present a large sample of optical quasars that are
detected by the \Herschel{}, and provide our initial analysis of their
FIR properties. The quasars are from the Sloan Digital Sky Survey
\citep[SDSS;][]{York2000}, and the FIR data are from the public
releases of four major wide field surveys by \Herschel{}, namely, the
Herschel Astrophysical Terahertz Large Area Survey
\citep[H-ATLAS;][]{Eales2010}, the Herschel Multi-tiered Extragalactic
Survey \citep[HerMES;][]{Oliver2012}, the Herschel Stripe 82 Survey
\citep[HerS;]{Viero2014}, and the PACS Evolutionary Probe
\citep[PEP;][]{Lutz2011}. We describe the data and the sample
construction in \S\ref{sec:data}, and present our analysis of the FIR
dust emission in \S\ref{sec:modeling}. The implications of our results
are detailed in \S\ref{sec:result}, and we conclude with a summary in
\S\ref{sec:summary}. The catalog of our sample is available as online
data in its entirety. All quoted magnitudes in the paper are in the
AB system. We adopt the following cosmological parameters throughout:
$\Omega_M=0.27$, $\Omega_\Lambda=0.73$ and
$H_0=\SI{71}{\kilo\meter\per\second\per\mega\parsec}$.

\section{Data description and sample construction}\label{sec:data}

In brief, we built our sample by searching for the counterparts of the
SDSS quasars in the \Herschel{} wide field survey data. For the
sake of simplicity, hereafter we refer to these objects as ``IR
quasars''. We describe below the data used in our study and the
constructed IR quasar sample.

\subsection{Parent quasar samples}

\begin{figure*}[t]
\plotone{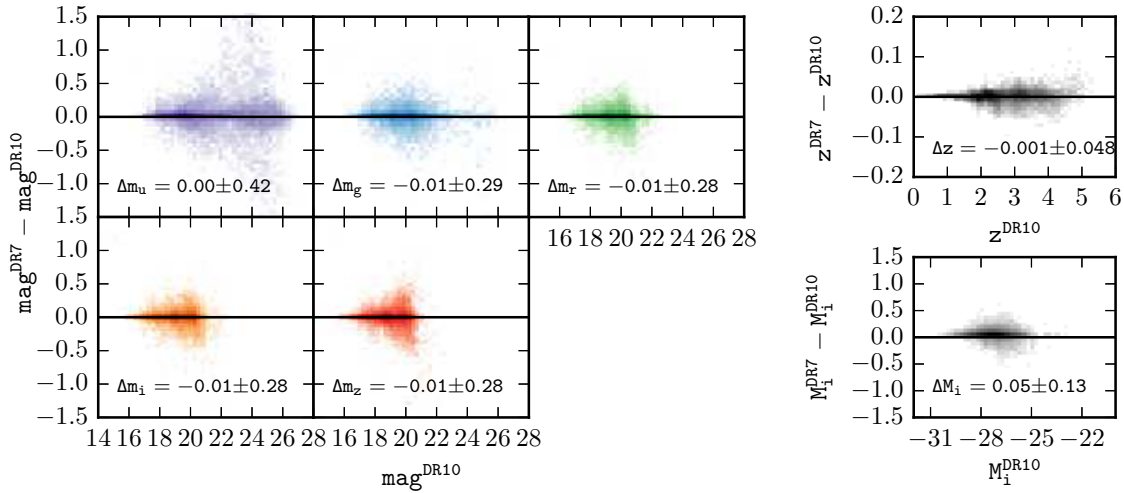}
\caption{Comparison of the photometry, the redshifts and the absolute
    magnitudes from the DR7Q and the DR10Q for the overlapped population in
    these two quasar catalogs (\num{16356} objects in total). For this
    particular subset, we adopt the DR10Q values in this paper unless noted
    otherwise. For the sake of consistency,  we use the absolute magnitudes
    $k$-corrected to $z=2$ for the DR7Q quasars as well, adapting the
    values from the catalog presented in \citet{Shen2011}.
}
\label{fig:dr7qdr10q_overlap}
\end{figure*}

\begin{figure}[t]
\plotone{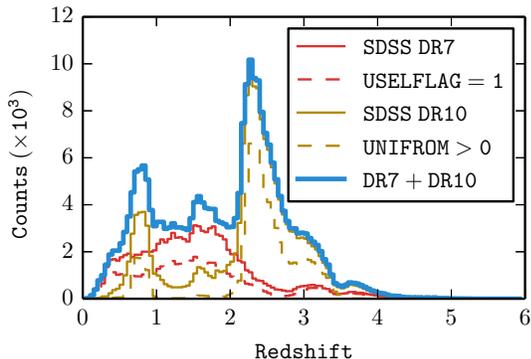}
\caption{Redshift distribution of the SDSS DR7Q (red curves) and DR10Q
    catalogs (yellow curves). The solid ones are for the entire
    catalogs, while the dashed ones are for the ``homogeneous'' subsamples
    as described in \cite{Schneider2010} and \citet{Paris2014}.}
\label{fig:hist_sdss}
\end{figure}

The parent quasar samples that we used are based on the SDSS Data
Release 7 and 10 quasar catalogs (hereafter DR7Q and DR10Q,
respectively), which are summarized as follows.

\begin{description}[leftmargin=*]

\item [DR7Q] As detailed in \citet{Schneider2010}, this quasar catalog
    is based on the SDSS DR7. It concludes the quasar survey in the
    SDSS-I and SDSS-II over \SI{9380}{\deg\squared}, and supersedes
    all previously released SDSS quasar catalogs. It includes
    \num{105783} quasars between $z=0.065$ and 5.46 (the median at
    $z=1.49$), all with absolute $i$-band magnitudes ($M_i$) brighter
    than \SI{-22}{\mag}.

\item [DR10Q] This quasar catalog is derived from the on-going Baryon
    Oscillation Spectroscopic Survey (BOSS) as part of the SDSS-III\@.
    While it was released in the SDSS DR10, its main target selection
    was based on the SDSS DR8. The detailed description of the catalog
    can be found in \citet{Paris2014}. It includes new quasars with
    $M_i[z=2] < \SI{-20.5}{\mag}$ from the SDSS-III, where $M_i[z=2]$
    is the absolute magnitude $k$-corrected to $z=2$ \citep[for
    details, see][]{Paris2014}. It also includes a large number of
    known quasars of similar characteristics (mostly from SDSS-I and
    II) that were re-observed by BOSS\@. In brief, the catalog
    contains \num{166583} quasars over \SI{6373}{\deg\squared}, with
    redshifts ranging from 0.05 to 5.86.

\end{description}

The quasars from these two catalogs are largely independent, however
there are still \num{16356} of them being duplicates, which we define
as the ones falling within a matching radius of $0.4\arcsec$.
Figure~\ref{fig:dr7qdr10q_overlap} shows the comparison of the
photometry, the redshift measurements and the absolute magnitudes from
these two catalogs for this overlapped population, which all agree
reasonably well.
For the sake of simplicity, we adopt the DR10Q values in this work
for these duplicates unless noted otherwise.

In the end, we produced a merged sample of \num{256010} unique
quasars, which represents the largest optical quasar sample selected
based on the most homogeneous data set over the widest area to date.
The redshift distributions of DR7Q, DR10Q and the merged catalog are
shown in Figure~\ref{fig:hist_sdss}. For simplicity, hereafter we
refer to the merged sample as the ``SDSS quasar sample'' and the
quasars therein as the ``SDSS quasars''. We note that DR7Q and DR10Q
are statistically different samples, and therefore any statistical
results from this merged catalog should be inferred with caution. This
is exacerbated by the fact that the quasars were not selected
uniformly within either DR7Q or DR10Q, as detailed in
\citet{Schneider2010} and \citet{Paris2014}, respectively. For
example, while most of the quasars in DR7Q were selected using the
algorithms as described in \citet{Richards2002} (with
\texttt{USELFLAG=1} in the catalog), a non-negligible fraction of them
were selected early in the SDSS campaign when such algorithms had not
yet been fully developed. DR10Q is even more non-uniform in this
sense, because only about half of its quasars (called the ``CORE''
sample) were selected uniformly (with \texttt{UNIFORM>0} in the
catalog) through the XDQSO method \citep{Bovy2011} and the other half
(called the ``BONUS'' sample) were selected using various different
methods \citep[see][]{Ross2012}. Nevertheless, the non-uniformity does not affect our current
work since we limit our study to the FIR properties of
optical-selected quasars, whose being selected did not use any FIR
information and thus should not favor or against any given FIR
property.

\subsection{Herschel Data}

\begin{table*}
\centering
\caption{Summary of \Herschel{} wide field data and IR QSO sample
    \label{tab:herschelcatalogues}}
\begin{tabular}{lrrrrrrrc}
\tophline
Survey & Field & Level & Coverage & \SI{5}{\sig} limit &
All\,\tablenotemark{a} & Match & SNR & $S_{250}>56.6$ \\
       &       &       &
\multicolumn{1}{c}{(\si{\deg\squared})} &
\multicolumn{1}{c}{(\si{\milli\jansky})} &
       &
\multicolumn{1}{c}{$<\SI{3}{\arcsec}$} &
\multicolumn{1}{c}{$>5$} &
\multicolumn{1}{c}{\si{\milli\jansky}} \\
\midhline
HerMES &          COSMOS\,\tablenotemark{b} &L1 &  5.00 & 31.4  &   216 &  47 &  7 &  3 \\
\grouphline
       &     GOODS-North\,\tablenotemark{b} &L2 &  0.63 & 27.1  &    17 &   4 &  1 &  1 \\
\grouphline
       &      Bootes-HerMES                 &L5 & 11.29 & 31.3  &  426 &  50 & 18 &  2  \\
\grouphline
       &      EGS-HerMES\,\tablenotemark{b} &L5 &  3.12 & 29.7  &   53 &   6 &  3 &  0  \\
       &   (Groth-Strip)                    &L3 &       & 30.9 &       &   3 &  2 &  0 \\
\grouphline
       & ELAIS-N1-HerMES                    &L5 &  3.74 & 30.4  &   22 &   3 &  3 &  0  \\
\grouphline
       &   Lockman-SWIRE\,\tablenotemark{b} &L5 & 19.73 & 37.0  &  276 &  35 & 12 &  5  \\
       & (Lockman-East-ROSAT)               &L3 &       & 31.5 &       &   1 &  0 &  0 \\
       &      (Lockman-North)               &L3 &       & 31.4 &       &   2 &  1 &  0 \\
\grouphline
       &             FLS                    &L6 &  7.30 & 35.4 &   89 &  17 &  5 &  3  \\
\grouphline
       &   XMM-LSS-SWIRE                    &L6 & 21.61 & 48.5 &  402 &  25 &  3 &  3  \\
       &              (UDS)                 &L4 &       & 33.8 &      &   3 &  1 &  0 \\
       &             (VVDS)                 &L4 &       & 33.7 &      &   8 &  2 &  0 \\
\grouphline
HerS   &        Stripe82              &$\sim$L7 & 80.76 & 51.8 & 4519 & 132 & 58 & 53  \\
\grouphline
HATLAS &             SDP                   &N/A & 19.28 & 34.1 &  690 &  18 & 18 & 12  \\
\midhline
Total  &                                   &    &172.46 &      & 6710 & 354 &134 & 82 \\
\bottomhline
\end{tabular}
\tablecomments{
    (1) The detection limits are based on the SPIRE \SI{250}{\micron}
    source count histograms where the count drops to 50\% of the
    peak value. The covered areas are calculated based on the pixel counts
    in the \SI{250}{\micron} maps.
    (2) The HerMES program has several ``nested fields'', which are
    the sub-fields of deeper observations embedded in the shallower but
    wider parent fields. In this table, the nested fields are labeled
    by parenthesis and are placed after their parent fields. The
    numbers of matched quasars in these nested fields and the wider
    regions outside are quoted separately.
}
\footnotetext{\,``All'' lists the number of SDSS quasars fall within the
    covered region.}
\footnotetext{\,The PEP data are available in these fields.}
\end{table*}

We utilized \emph{all} high-level, publicly available data from the
wide field \Herschel{} surveys. In all cases we adopted the latest
catalogs released by the survey teams to construct the FIR spectral
energy distributions (SEDs). The basic characteristics of these
surveys are summarized in Table~\ref{tab:herschelcatalogues}, and
their relevant details are briefly described below.

\subsubsection{HerMES}

HerMES was the largest \Herschel{} guaranteed time key program, and
surveyed \SI{100}{\deg\squared} in six levels of depth and spatial
coverage combinations (``L1'' to ``L6''), using both SPIRE and PACS\@.
Its latest data release, ``DR2'', contains the SPIRE maps and the
source catalogs in all three bands \citep{Wang2013o}. The PACS data
have not yet been released. In this work, we used the wide-field data
and excluded those galaxy cluster fields.

The HerMES DR2 includes two sets of source catalogs based on different
methods, one using the SUSSEXtractor (SXT) point source extractor and
the other using the iterative source detection tool StarFinder (SF)
combined with the De-blended SPIRE photometry algorithm (DESPHOT). We
adopted the band-merged version of the latter (denoted as ``xID250''in
DR2), which has the DESPHOT multi-band (250, 350 and
\SI{500}{\micron}) photometry at the positions of the SF
\SI{250}{\micron} sources. The \SI{5}{\sig} detection limits in
\SI{250}{\micron} range from $\sim 30$ to \SI{48}{\milli\jansky} in
our fields of interest.

\subsubsection{H-ATLAS}\label{sec:hatlas}

H-ATLAS was the largest \Herschel{} open-time key program. It surveyed
\SI{550}{\deg\squared} using both SPIRE and PACS\@. Currently, this
program has released the image maps and the source catalogs in the
field observed during the \Herschel{} Science Demonstration Phase
\citep[H-ATLAS SDP;][]{Ibar2010, Pascale2011, Rigby2011}, which covers
$\sim$\SI{19.3}{\deg\squared} and has reached the \SI{5}{\sig} limit
of \SI{33.4}{\milli\jansky} in \SI{250}{\micron}.

The catalog that we adopted is the band-merged one with both the SPIRE
and the PACS photometry. For the SPIRE bands, the source extraction
was done using the Multi-band Algorithm for source eXtraction
\citetext{MADX; Maddox et al.\ in prep.}, which employed a localized
background removal and PSF filtering procedures to the map and
extracted the sources using the \SI{250}{\micron} positions as the
priors. For the PACS bands, aperture photometry was performed on the
100 and \SI{160}{\micron} maps at the SPIRE \SI{250}{\micron}
positions.

\subsubsection{HerS}

HerS observed $\sim$ \SI{80}{\deg\squared} in the SDSS Stripe 82
region \citep{Abazajian2009, Annis2014} using SPIRE, reaching the
nominal \SI{5}{\sig} limit of \SI{50.4}{\milli\jansky} in
\SI{250}{\micron}. Both the image maps and the source catalogs have
been released \citep{Viero2014}. We adopted the band-merged catalog,
which was based on the SF \SI{250}{\micron} detections and the DESPHOT
photometry \citep{Roseboom2010}.

\subsubsection{PEP}\label{sec:pep}

PEP was also a \Herschel{} guaranteed time key program, which used
PACS to survey six well-studied extragalactic fields and also a number
of galaxy clusters. Both the image maps and the source catalogs have
been made public through Data Release 1 (DR1) of the team. These data
are not listed in Table~\ref{tab:herschelcatalogues}. All the PEP
fields are within the HerMES fields, however they only covered a small
fraction.

Whenever possible, we used their 100 and \SI{160}{\micron}
measurements in the COSMOS, GOODS-North, EGS, Lockman Hole fields to
supplement the HerMES SPIRE photometry to better constraint the FIR
SEDs of the detected quasars. These measurements were taken from the
SF ``blind extraction'' catalogs as described in \citet{Magnelli2009}.

\subsection{Herschel-detected quasars}

Our IR quasar sample was derived by matching the positions of the SDSS
quasars to those of the \Herschel{} sources in the band-merged,
\SI{250}{\micron}-based catalogs as described above.

\subsubsection{Matching radius}

\begin{figure}[t]
\plotone{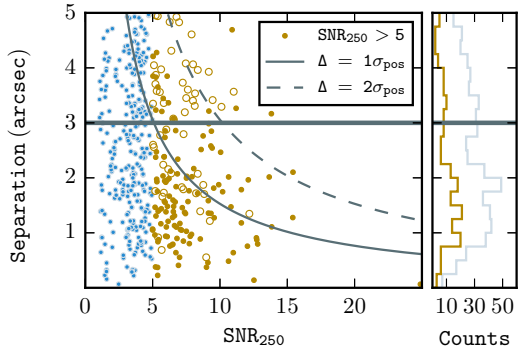}
\caption{Distribution of the separation between the SDSS positions and
    the matched SPIRE \SI{250}{\micron} positions when expanding the
    matching radius to \SI{5}{\arcsec}. The left panel shows the
    separation as a function of the SNR of the \SI{250}{\micron}
    detection. The solid and the dashed lines are the theoretical
    $\sim\SI{1}{\sig}$ and $\sim\SI{2}{\sig}$ positional uncertainties
    ($\sigma_{pos}$). The circles in yellow color are the sources
        with $\text{SNR}_{250}>5$, while the light blue ones represent
        the rest. Among the yellow circles, the open and the solid
        ones are those deemed by our visual inspection to be
        ``blended'' (i.e., affected by source blending) and ``clean''
        (free of source blending), respectively. The horizontal thick
    black line going through both panels indicates the adopted
    matching radius of \SI{3}{\arcsec}.}
\label{fig:hist_sep_match}
\end{figure}

The matching was performed using
\texttt{TOPCAT}/\texttt{STILTS}\footnote{\texttt{TOPCAT}
    \url{http://www.starlink.ac.uk/topcat};\\ \texttt{STILTS}
    \url{http://www.starlink.ac.uk/stilts} }.
We used a matching radius of \SI{3}{\arcsec}, which is justified
below.

While the \Herschel{} instruments have large beam sizes
\footnote{The FWHM beam sizes are \SI{18}{\arcsec}, \SI{25}{\arcsec}
    and \SI{36}{\arcsec} for the SPIRE 250, 350 and \SI{500}{\micron},
    respectively, and 6--\SI{7}{\arcsec} and 11--\SI{14}{\arcsec} for
    the PACS 100 and \SI{160}{\micron}, respectively.},
the source centroids can still be determined to high accuracy. The
positional uncertainty of a given source depends on its
signal-to-noise ratio (SNR) \citep[see, e.g.,][]{Ivison2007}, which
follows
\begin{equation}
    \Delta\alpha = \Delta\delta = \frac{0.6\theta}{\text{SNR}} \,,
\end{equation}
where $\Delta\alpha$ and $\Delta\delta$ are the nominal \SI{1}{\sig}
uncertainties of RA and Dec, respectively, and $\theta$ is the beam
size. The SPIRE \SI{250}{\micron} beam size is
$\theta=\SI{18}{\arcsec}$, which means that the \SI{1}{\sig}
positional uncertainty of a given \SI{250}{\micron} source is
\begin{equation}\label{eq:sepsnr}
    \sigma_{pos} = \sqrt{(\Delta\alpha)^2 + (\Delta\delta)^2} =
    \frac{15.27\arcsec}{\text{SNR}}\,.
\end{equation}

A matching radius of \SI{3}{\arcsec} thus corresponds to
$\sim\SI{1}{\sig}$ uncertainty for objects with $\text{SNR}>5$, or
$\sim\SI{0.59}{\sig}$ for those with $\text{SNR}>3$\footnote{We note
    that \citet{Smith2011} use a somewhat different relation between
    the astrometric uncertainty and the SNR, $\sigma_{pos} =
    0.655\theta/\text{SNR}$, which means that \SI{1}{\sig} uncertainty
    would be $2\farcs 4$ for $\text{SNR}=5$ in \SI{250}{\micron}. Our
    results shown in Figure~\ref{fig:hist_sep_match} indicate that
    such a matching radius would be slightly too stringent, and
    therefore we adhered to our choice.}. To further validate our
choice, we performed a test using an enlarged matching radius of
\SI{5}{\arcsec}, and Figure~\ref{fig:hist_sep_match} demonstrates the
results. The separations between the \SI{250}{\micron} and the SDSS
positions versus the \SI{250}{\micron} SNR are indeed consistent with
the expectation from Equation~\eqref{eq:sepsnr}, and the vast majority
of the matches within \SI{3}{\arcsec} fall within the $1\sigma_{pos}$
curve. Furthermore, the separation shows a double-peak feature roughly
divided at \SI{3}{\arcsec}, indicating that the matches beyond this
point are likely affected by other factors, such as the source
blending problem (see below).

\citet{Wang2013o} have addressed the source positional accuracy in the
HerMES DR2 catalogs through end-to-end simulations. For the SF
catalogs in \SI{250}{\micron}, they find that the real matches between
the input and the output have the positional offsets peak at around
\SI{5}{\arcsec}. We note that this is derived using the real matches
at all SNR levels. If a $\text{SNR}>5$ threshold is applied, this peak
is likely to shift to a smaller value. In any case, the matching
radius of \SI{3}{\arcsec} that we adopted is a rather conservative
choice.

In a number of fields the PACS data are also available. The H-ATLAS
band-merged catalog in the SDP field already includes the PACS
photometry (see \S\ref{sec:hatlas}) and thus no further action was
taken. The HerMES COSMOS, GOODS-N, EGS, and Lockman Hole fields have
PACS data from the PEP program (see \S\ref{sec:pep}), and we matched
the coordinates in the PEP 100 and \SI{160}{\micron} ``blind''
catalogs to both the SPIRE ``xID250'' catalogs and the SDSS quasars,
again using a matching radius of \SI{3}{\arcsec}.

\subsubsection{Source blending}

Due to the large beam sizes of the instruments, \Herschel{} images
still suffer from a severe source blending problem. To evaluate how
this problem could impact our sample, we visually inspected the
\Herschel{} and the SDSS images of all the matches in the test case
shown in Figure~\ref{fig:hist_sep_match}, i.e., using a larger
matching radius of \SI{5}{\arcsec}. We searched for the signatures of
possible source blending, such as the presence of close companions in
the SDSS images, and the offset of the source centers among the
\Herschel{} bands, etc. For this test, we only used the sources
    that have $\text{SNR}>5$ in \SI{250}{\micron}. The result is also
    shown in Figure~\ref{fig:hist_sep_match}.
As it turns out, enlarging the matching radius from \SI{3}{\arcsec} to
\SI{5}{\arcsec} will include 47 more objects, however only 16 of these 47
objects (34\%) are ``clean'' cases. For comparison, 83\% of the
sources within the matching radius of \SI{3}{\arcsec} are ``clean''.
This lends
further support to our choice of the matching radius. The drawback of
using this conservative value is that it could exclude some objects
that have genuine \Herschel{} detections but are blended with close
neighbor(s). While it is possible to include such
sources after using the methods described in \citet{Yan2014} to
de-blend, we defer this improvement to our future work.

In summary, our conservative choice of the matching radius has
resulted in a sample that is free of significant blending problem.
This has also simplified the photometry in the SPIRE 350 and
\SI{500}{\micron} bands. While the source catalogs that we adopted
from the survey teams were all derived using PSF-fitting based on the
\SI{250}{\micron} detections, the photometry in 350 and
\SI{500}{\micron} would still be prone to large errors introduced by
blending due to their much larger beam sizes (\SI{25}{\arcsec} and
\SI{36}{\arcsec}, respectively). Our relatively clean sample allows us
to assume that the fluxes measured in these two redder bands are
solely contributed by the source detected in \SI{250}{\micron}.

\subsubsection{IR quasar sample summary}\label{sec:samplesummary}

Our \Herschel{}-detected SDSS quasar sample, derived using a matching
radius of \SI{3}{\arcsec} as described above
(Table~\ref{tab:herschelcatalogues}), contains
\num{354}\footnote{We
        note that there are 18 more matched objects from HerMES DR2
        catalogs
        discarded due to $\text{SNR}_{250}<1$} objects in
total. Most of our studies are based on a subsample of those, called
the ``SNR5'' sample, which only includes \num{134} quasars that have
$\text{SNR}\geq 5$ in \SI{250}{\micron} and thus is the sample of the
most robust \Herschel{} detections. Based on this SNR5 sample, we
further imposed a cut in the \SI{250}{\micron} flux density,
$S_{250}\geq\SI{56.6}{\milli\jansky}$, and formed a ``bright''
subsample of \num{82} objects for various discussions below. This flux
density threshold was adopted based on the lowest flux density of the
SNR5 objects in the shallowest HerMES field L6-XMM-LSS-SWIRE\@.

\section{Dust emission modeling}\label{sec:modeling}

We inferred the dust emission properties of the IR quasars by fitting
their FIR SEDs. The thermal emission over the full IR regime can be
viewed as the collective result of all heated dust components of various
temperatures, and the FIR part is dominated by the coldest component. In
this paper, we focus on this FIR part in the IR quasars and hence our
conclusions are pertaining to their cold-dust components. We used two
distinct types of models to fit the FIR SEDs, one being a
single-temperature, modified blackbody (MBB) spectrum and the other
being three different sets of starburst templates. Fitting an MBB spectrum
to the FIR SED is always valid, regardless of the exact dust heating
sources (i.e., due to photons from either star formation or AGN activity).
However, it has the drawback that the SED must be properly sampled in
order to obtain well constrained results. Fitting starburst templates,
on the other hand, is only appropriate if the FIR emission is dominated
by heating from star formation, and the motivation of using these
models was to test if the FIR emissions are consistent with being caused
by star formation. These two types of SED fitting approach provide
independent check to each other, and we will show later
that they also lead to insights into the heating sources.

\subsection{SED fitting using MBB model}\label{sec:cmcirsed}

\begin{figure}[t]
\plotone{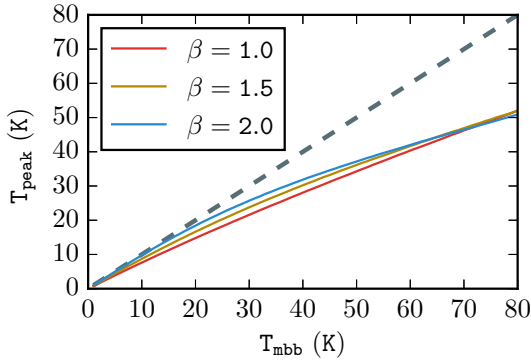}
\caption{Relation between the temperature of the modified blackbody
    spectrum $T_{mbb}$ and the temperature inferred from the Wein's
    displacement law $T_{peak}$, for the three cases when the
    emissivity $\beta$ is 1.0, 1.5 and 2.0, respectively. The dashed
    line represents the equality if these two quantities were the
    same.}
\label{fig:tpeak_vs_tmbb}
\end{figure}

\begin{figure*}[!t]
\plotone{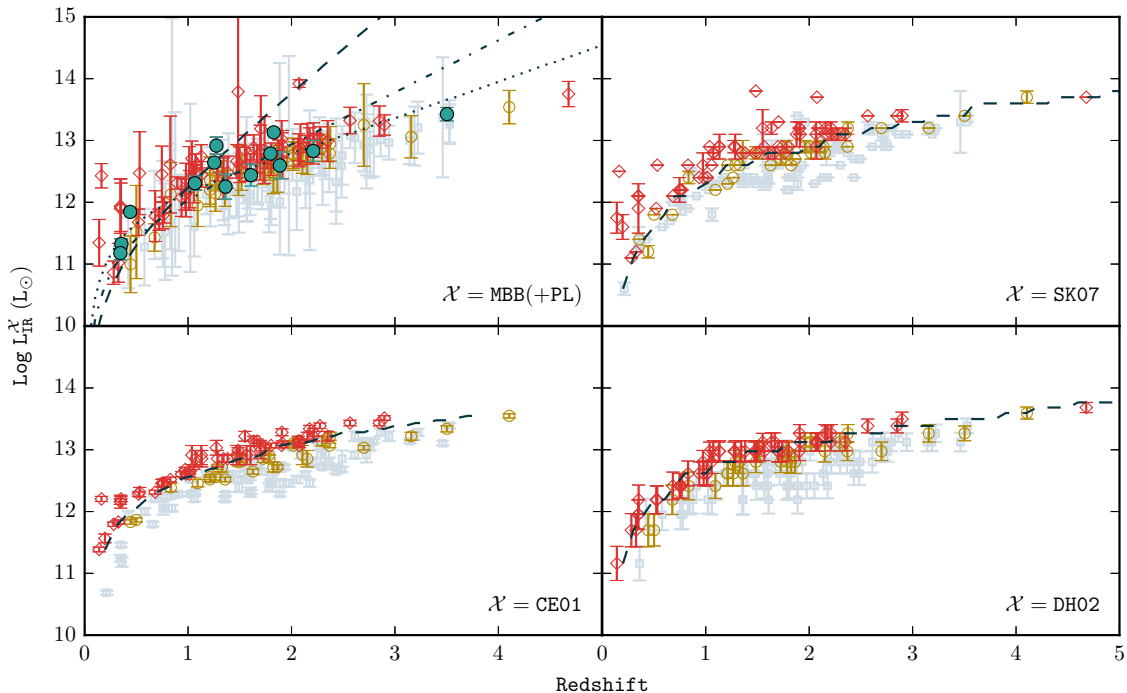}
\epsscale{1.1}
\caption{Derived IR luminosities ($L_{IR}^{\mathcal{\mathcal{X}}}$) of the IR quasars
    in our sample,
    where ``$\mathcal{\mathcal{X}}$'' denotes one of the four models in use, namely,
    MBB(+PL), SK07, CE01, and DH02\@.
    The errors between the MBB results and
    those from the starburst templates are not directly comparable
    because of the different approaches adopted in evaluating the
    errors. The colored symbols represent the SNR5 sample, while the
    grey squares represent the rest. Among the SNR5 objects, the red
    diamonds indicate those that are in the bright subsample
    ($S_{250}\geq\SI{56.6}{\milli\jansky}$), while the yellow circles
    indicate those that are not. The dark green solid circles in the MBB
    panel are the objects with PACS data available and hence a
    power-law component was added to the MBB spectrum in the fitting.
    To illustrate the impact of the survey
    limit, the limits of $L_{IR}^{\mathcal{X}}$ corresponding to the fiducial flux
    density limit of $S_{250}=\SI{56.6}{\milli\jansky}$ are shown as
    the dashed lines in the three panels for the starburst template
    fits (SK07, CE01 and DH02). In each of these cases, the limit is derived from the entire
    library by using the template with the lowest possible
    $L_{IR}^{\mathcal{X}}$ at
    a given redshift. In the panel for the MBB fit, the limits are
    given using three different $T_{mbb}$ of 15, 25, and
    \SI{35}{\kelvin}, shown as the dashed, the dot-dashed and the
    dotted lines, respectively.
}
\label{fig:lir_vs_z}
\end{figure*}

\begin{figure}[t!]
\plotone{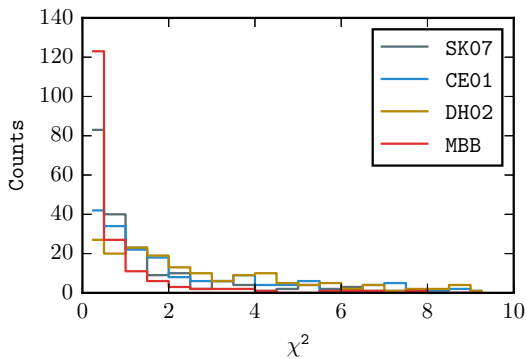}
\epsscale{1.0}
\caption{Distribution of $\chi^2$ of the best-fit models for all the IR
    quasars in our sample, using the four different methods as detailed in the
    text.}
\label{fig:hist_chi2}
\end{figure}

\begin{figure*}[t]
\plotone{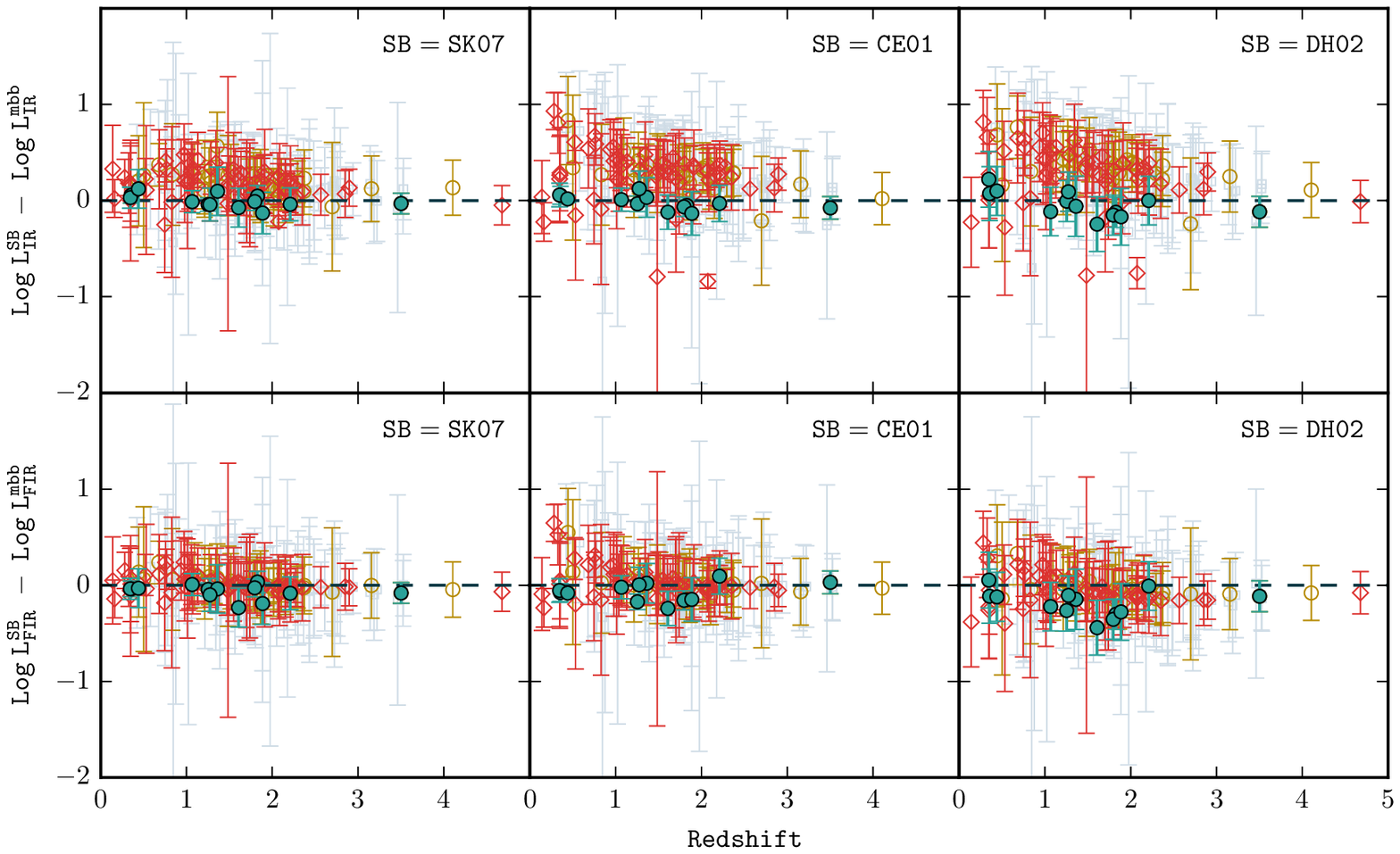}
\caption{Comparisons of IR luminosities derived using the MBB
        model and those based on the three sets of starburst
        templates. The symbols are the same as in
    Figure~\ref{fig:lir_vs_z}. The upper panels show the
        comparisons to $L_{IR}^{SB}$ (i.e., computed over the full IR
        range of $8\text{--}\SI{1000}{\micron}$), while the
        bottom panels show the comparison to $L_{FIR}^{SB}$ (computed
        over the FIR regime of $60\text{--}\SI{1000}{\micron}$),
    where ``SB'' is one of SK07, CE01 and DH02. See
    \S\ref{sec:result_lir} for details.}
\label{fig:lir_vs_z_ratio}
\end{figure*}

We used the \texttt{cmcirsed}
code\footnote{\url{http://herschel.uci.edu/cmcasey/sedfitting.html}}
of \citet{Casey2012} to perform the MBB fitting, which allowed us to
derive the IR luminosity, the dust temperature, and the dust mass.
This procedure was only carried out for the objects that have
photometry in all the three SPIRE bands (\num{187} objects in total, among
which \num{103} are in the SNR5 subsample), because the MBB fit would become
unconstrained with less bands.

The FIR emission due to MBB can be written as
\begin{equation}\label{eq:mbb}
     S(\lambda) = N_{mbb}\frac{(1-\mathrm{e}^
         {-(\frac{\lambda_0}{\lambda})^{\beta}})(\frac{c}{\lambda})^3}
     {\mathrm{e}^{hc/(\lambda kT_{mbb})}-1} \,,
\end{equation}
where $T_{mbb}$ is the characteristic temperature of the MBB,
$N_{mbb}$ is the scaling factor that is related to the intrinsic
luminosity, $\beta$ is the emissivity, and $\lambda_0$ is the
reference wavelength where the opacity is unity. As most of our
quasars only have three SPIRE bands available,
we had to limit the degrees of freedom. We adopted the default
emissivity of $\beta=1.5$, which is the
value typically assumed for cold dust
\citep{Casey2012}. By default, \texttt{cmcirsed}
sets $\lambda_0=\SI{200}{\micron}$. We adopted
$\lambda_0=\SI{100}{\micron}$, following \citet{Draine2006}. While the
exact choice of $\lambda_0$ only marginally affects the estimates of
the total IR luminosity and the dust mass, it will significantly
impact the estimate of the dust temperature. We will further discuss
this effect in Appendix~\ref{sec:app_lambda0}.

We note that the above form is for general opacity. In the optical
thin case, at $\lambda \gg \lambda_0$, the term
$(1-\mathrm{e}^{-(\frac{\lambda_0}{\lambda})^{\beta}})$ reduces to
$(\frac{\lambda_0}{\lambda})^{\beta}$, which is often adopted in the
submm/mm regime. Throughout this work, we used the general opacity
form as in Equation~\eqref{eq:mbb}.

The \texttt{cmcirsed} code has the capability of superposing a
power-law component (PL) to the MBB spectrum to accommodate the
possible warm dust component whose effect could be present in the
mid-IR regime (typically at $<\SI{50}{\micron}$ in restframe).
Thirteen quasars (seven of them are in the SNR5 sample) have PACS data
in addition to the three band SPIRE data, and we utilized this
capability when fitting these objects. In this case, the MBB+PL model
then reads
\begin{equation}
    S(\lambda) = N_{bb}\frac{(1-\mathrm{e}^
        {-(\frac{\lambda_0}{\lambda})^{\beta}})(\frac{c}{\lambda})^3}
    {\mathrm{e}^{hc/(\lambda kT_{mbb})}-1}
    +
    N_{pl}\lambda^{\alpha}\mathrm{e}^{-(\frac{\lambda}{\lambda_c})^2}
    \,,
\end{equation}
where $\alpha$ is the PL slope, $N_{pl}$ is the
normalization of the PL part, and $\lambda_c$ is the turnover wavelength
\citep[for detail, see ][]{Casey2012}.
Again, to limit the degrees of freedom, $\alpha$ had to be fixed,
and we adopted the default value of $\alpha=2.0$ \citep{Casey2012}.

Recalling that the FIR emission is dominated by the cold-dust component,
we can obtain the total IR luminosity of this component as
\begin{equation}\label{eq:lir}
    L_{IR}^{(cd)}\equiv L_{IR}^{mbb}\equiv
    \int_{8\micron}^{1000\micron}\mathcal{L}^{mbb}(\lambda)\mathrm{d}\lambda
\end{equation}
by integrating the best-fit model $\mathcal{L}^{mbb}(\lambda)$ from 8 to
\SI{1000}{\micron}.
For the 13 objects that have PACS data, we also calculated their
    total IR luminosities as
\begin{equation}\label{eq:lir2}
    L_{IR}\equiv L_{IR}^{mbb+pl}\equiv
    \int_{8\micron}^{1000\micron}\mathcal{L}^{mbb+pl}(\lambda)\mathrm{d}\lambda\,.
\end{equation}

We emphasize that $L_{IR}^{(cd)}$ (here represented by the MBB fit) is the
luminosity of the cold-dust component over the entire IR range
($8\text{--}\SI{1000}{\micron}$). While the bulk of its emission is in the FIR regime, this
cold-dust component emits beyond FIR and thus it is necessary to integrate
over $8\text{--}\SI{1000}{\micron}$ to capture its total IR luminosity. Note that this is
\emph{not} the total IR luminosity ($L_{IR}$) of the galaxy that includes
the contributions of all dust components over $8\text{--}\SI{1000}{\micron}$.
In other words, we have $L_{IR}^{(cd)}<L_{IR}$.

The best-fit temperature from \texttt{cmcirsed} was taken as the
temperature of the cold dust, i.e., $T_{dust}\equiv T_{mbb}$. We also calculated the ``peak''
temperature inferred from Wien's displacement law, which is given by
\begin{equation}\label{eq:tpeak}
    T_{peak} = b/\lambda_{peak} \,,
\end{equation}
where the coefficient $b=\SI{2.898d3}{\micro\meter\kelvin}$. As
compared to $T_{mbb}$, $T_{peak}$ is less sensitive to the specifics
of the dust emission models in use and therefore could be a better
proxy to the dust temperature when comparing results derived based on
different types of templates. For this reason, while we mostly use
$T_{dust}\equiv T_{mbb}$ in this paper, we also use $T_{peak}$ in some
occasions. The relation between $T_{peak}$ and $T_{mbb}$ is shown in
Figure~\ref{fig:tpeak_vs_tmbb}. In case of $\beta=1.5$, the relation
can be roughly described by the following broken linear function:
\begin{equation}\label{eq:tpeak_tmbb}
    T_{peak} \simeq
    \begin{cases}
        0.7653T_{mbb} + 0.9529\,, & T_{mbb} < \SI{35}{\kelvin} \\
        0.5485T_{mbb} + 8.5153\,, & T_{mbb} > \SI{35}{\kelvin}\,.
    \end{cases}
\end{equation}

\subsection{SED fitting using starburst templates}\label{sec:sbtemp}

We also fitted the FIR SEDs using three different libraries of
starburst galaxies separately, namely, the theoretical model of
\citet[hereafter SK07; 7220 templates]{Siebenmorgen2007}, and the empirical templates
of \citet[hereafter CE01; 105 templates]{Chary2001} and \citet[hereafter
DH02; 64 templates]{Dale2002}. For a given quasar, the restframe templates were
redshifted according to the quasar's redshift and convolved with the
\Herschel{} passband response curves, and then were compared to the
observed IR quasars SEDs. The best-fit template was chosen by
minimizing the $\chi$-square:
\begin{equation}
    \chi^2 =
    \sum\limits_{i=1}^{N}\left(\frac{f_{obs}-f_{th}}{\sigma_{obs}}\right)^2\,.
\end{equation}
We note again that the starburst model fitting was also done for the
objects that have photometry in only two or even one SPIRE band.
During the fitting, we did not re-scale the templates.
By construction, these templates all have associated $L_{IR}$ values
(over 8-\SI{1000}{\micron}), and
the value of the best-fit template was adopted as the total IR
luminosity of the object in question.
For clarity, we will refer to it as $L_{IR}^{SB}$, where ``SB'' can be
    replaced by ``SK07'', ``CE01'' or ``DH02'' when appropriate.
The error of $L_{IR}^{SB}$ was
estimated by taking the difference between the $L_{IR}$ value of the
best-fit template and that of the one of the second smallest $\chi^2$.
As compared to the formal likelihood method, this simple approach has
the advantage that it works consistently when the parameter space is
discrete, and that it includes the possible systematic errors
intrinsic to the template set.

We should emphasize that $L_{IR}^{SB}$ thus derived is the total IR
luminosity of the galaxy. If using the starburst models is appropriate, we
should have $L_{IR}^{SB}>L_{IR}^{mbb}$, where $L_{IR}^{mbb}\equiv
L_{IR}^{(cd)}$. While $L_{IR}^{(cd)}$ cannot be separated from
$L_{IR}^{SB}$ in any of these starburst models, we will show that
$L_{IR}^{SB}$ can help interpret the origin of $L_{IR}^{(cd)}$.

\subsection{Dust mass and gas mass}

The MBB fit by \texttt{cmcirsed} also resulted in estimate of dust
mass (hereafter $M_{dust}$). As this quantity is a strong dependent of
dust temperature ($M_d\propto L_{IR}T^{-5}$; see \citealt{Casey2012}),
it should be used with caution. For example, in our analysis in
\S\ref{sec:result}, we will only use those that have $T_{mbb}/\Delta
T_{mbb}\geq 3$.

Applying a nominal gas-to-dust ratio, the gas mass ($M_{gas}$) can
also be obtained.
We adopted the nominal Milky Way gas-to-dust-mass ratio of 140
\citep[e.g.][]{Draine2007} for this work.

\section{Results and discussions}\label{sec:result}

The major physical properties obtained in \S\ref{sec:modeling}
    are all given in the online table accompanying this paper. A
    summary of the information contained in this table is given in
    Appendix~\ref{sec:snr5_sed} (see Table~\ref{tab:onlinedata}). Some
    examples of the SED fitting are also provided in
    Appendix~\ref{sec:snr5_sed}. Here we discuss in detail these
results, some potential selection effects, and their implications.

\subsection{IR luminosity}\label{sec:result_lir}

\begin{figure*}[t]
\plotone{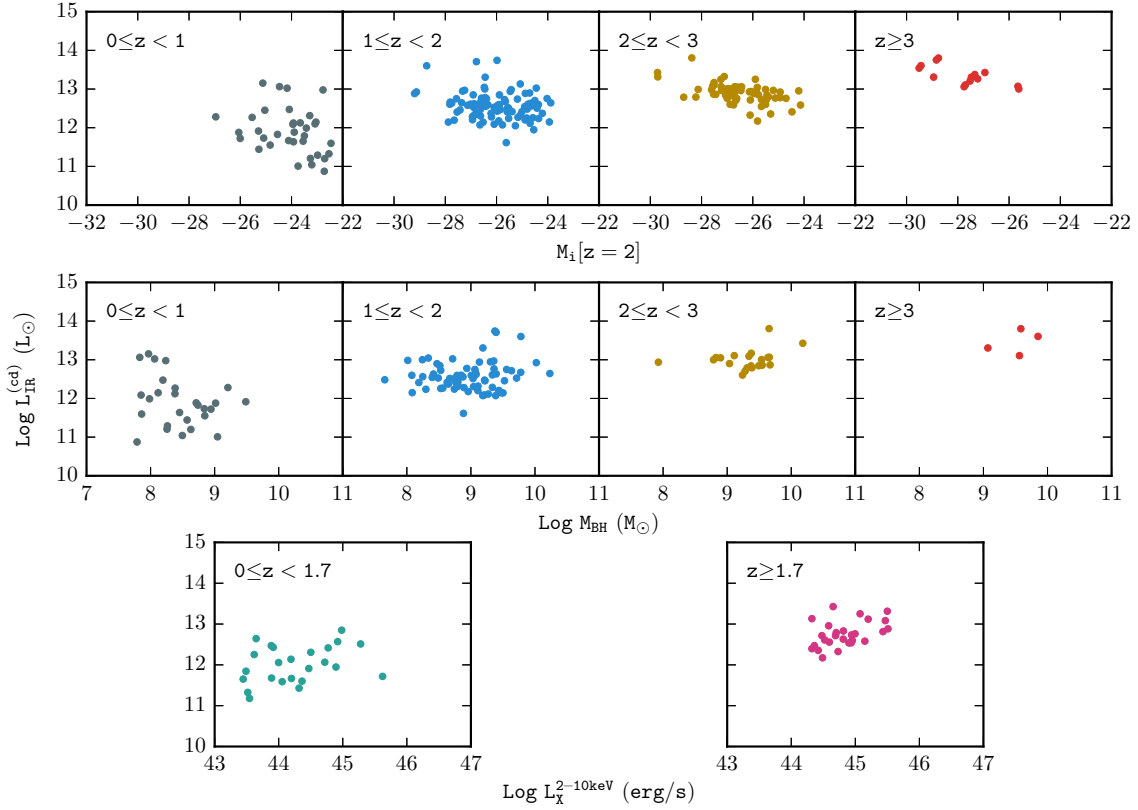}
\caption{No correlation between $L_{IR}^{(cd)}$ and the absolute magnitudes
    of quasars (upper panels), their black hole masses (middle panels), or
    the X-ray luminosities (lower panels). The results are shown in four
    redshift bins as labeled, and the error bars are omitted for clarity.
}
\label{fig:lir_vs_mibh}
\end{figure*}

For clarity, the various flavors of IR luminosities discussed in
\S\ref{sec:modeling} are summarized here:
\begin{itemize}
    \item $L_{IR}$: the general designation of the total IR luminosity over
        $8\text{--}\SI{1000}{\micron}$;
    \item $L_{IR}^{(cd)}$: the contribution of the cold-dust component to
        the total IR luminosity over $8\text{--}\SI{1000}{\micron}$, which
        is also referred to as the total IR luminosity of the cold-dust
        component;
    \item $L_{IR}^{mbb}$: the MBB best-fit to the FIR SED (as represented
        by the SPIRE data points) integrated over
        $8\text{--}\SI{1000}{\micron}$, and by definition
        $L_{IR}^{(cd)}\equiv L_{IR}^{mbb}$;
    \item $L_{IR}^{mbb+pl}$: the MBB+PL best-fit to the FIR SED (as
        represented by the SPIRE and the PACS data points) integrated over
        $8\text{--}\SI{1000}{\micron}$, which is the measurement of
        $L_{IR}$ using the MBB+PL model;
    \item $L_{IR}^{SB}$: the measurement of $L_{IR}$ using the starburst
        models (``SB'' is one of ``SK07'', ``CE01'' and ``DH02'', depending
        on the model set in use), and effectively is the best-fit SB
        template integrated over $8\text{--}\SI{1000}{\micron}$.
\end{itemize}

For one of our purposes later, we also define the corresponding quantities
in the FIR instead of over the entire IR range, i.e., by integrating over
$60\text{--}\SI{1000}{\micron}$ only. We designate these quantities with
the subscript of ``FIR'', e.g., $L_{FIR}^{(cd)}$, $L_{FIR}^{mbb}$,
$L_{FIR}^{SB}$, etc\@.

The derived $L_{IR}^{mbb}$ and $L_{IR}^{SB}$ (referred to as $L_{IR}^{\mathcal{X}}$)
values of our IR quasars are shown in Figure~\ref{fig:lir_vs_z}, and the
distributions of the best-fit $\chi^2$ are shown in
Figure~\ref{fig:hist_chi2}, respectively. We note that deriving
$L_{IR}^{mbb}$ (and hence $L_{IR}^{(cd)}$) was not always possible because
the MBB fit requires the FIR SED being well sampled by the three SPIRE
bands, while obtaining $L_{IR}^{SB}$ could always be done because of the
nature of the method (see \S\ref{sec:cmcirsed} \& \ref{sec:sbtemp}). We
also note that the SNR5 subsample, as expected, has the smallest errors in
$L_{IR}^{\mathcal{X}}$. In addition, the majority of the objects outside of the SNR5
sample still have $\chi^2\leq 10$ and thus are also deemed as having
reliable $L_{IR}^{\mathcal{X}}$ measurements.

Regardless of the exact model set in use, the majority of our IR quasars
have obtained good fits and the derived IR luminosities also agree to the
extend that we expect. This is more clearly shown in the upper panels of
Figure~\ref{fig:lir_vs_z_ratio}, where $L_{IR}^{SB}$ are compared against
$L_{IR}^{mbb}$. On average, $L_{IR}^{mbb}$ values are lower by 0.13, 0.23
and 0.25 dex as compared to $L_{IR}^{SB}$ from the SK07, the CE01 and the
DH02 models, respectively. As explained earlier, this is due to the fact
that the MBB model only includes the cold-dust component
($L_{IR}^{mbb}\equiv L_{IR}^{(cd)}$), whereas the starburst models also
include all other components of higher temperatures and thus give the total
IR luminosity ($L_{IR}$). This is also demonstrated by the 13 objects that
have PACS data (dark green points), for which we carried out the MBB+PL fit
and thus obtained the total IR luminosity in the form of $L_{IR}^{mbb+pl}$.
As one can see, there are no offsets between $L_{IR}^{mbb+pl}$ and
$L_{IR}^{SB}$.

To further demonstrate this point, the lower panel of
Figure~\ref{fig:lir_vs_z_ratio} shows similar comparisons as in the upper
panel, but between $L_{FIR}^{mbb}$ and $L_{FIR}^{SB}$. The agreements are
excellent and no systematic offsets are found. This can be understood as
follows. The emission from the hot-dust components should be minimal in the
FIR regime, and hence integrating the starburst models in the FIR should
only capture the contribution from the cold-dust component, i.e.,
$L_{FIR}^{SB}=L_{FIR}^{(cd)}=L_{FIR}^{mbb}$.

As we have been emphasizing, the MBB fit is independent of the heating
source, while the SB fit being valid hinges upon the heating source being
star formation. Therefore, the agreement between $L_{FIR}^{mbb}$ and
$L_{FIR}^{SB}$ strongly suggests that the SB fits are valid and that the
FIR emissions in these IR quasars are due to star formation. The corollary
then is that $L_{IR}^{mbb}$ ($\equiv L_{IR}^{(cd)}$) is due to star
formation, because it is the total luminosity of the same cold-dust
component that gives rise to the FIR emission. In the rest of this paper,
the quoted IR luminosity is $L_{IR}^{mbb}$ unless explicitly stated
otherwise (mostly in \S\ref{sec:result_frac}, \ref{sec:result_lird} and
Appendix~\ref{sec:app_stack}), and we use $L_{IR}^{mbb}$ and
$L_{IR}^{(cd)}$ interchangeably depending on the context. While
$L_{IR}^{mbb}$ could underestimate the true $L_{IR}$ (see the top panel of
Figure~\ref{fig:lir_vs_z_ratio}) by a factor of 1.35 (as compared to
$L_{IR}^{SK07}$) to 1.70-1.78 (as compared to $L_{IR}^{CE01}$ or
$L_{IR}^{DH02}$), we prefer to be conservative due to the lack of
observational constraints in the mid-IR for our entire sample. This,
however, is not necessarily a drawback because the mid-IR emission, unlike
the FIR one, could be seriously contaminated by the AGN contribution (see
Appendix~\ref{sec:app_decomp}).

Our IR quasars have $L_{IR}^{\mathcal{X}}$ values ranging from $\sim 10^{10.5}$ to
\SI{d13.8}{\Lsun} (after discarding two objects whose SEDs are barely
constrained). Most of them ($\gtrsim80\%$) are ULIRGs ($L_{IR} >
\SI{d12}{\Lsun}$), and some of them ($\gtrsim 15\text{--}23\%$) are even
HyLIRGs ($L_{IR} > \SI{d13}{\Lsun}$). As Figure~\ref{fig:lir_vs_z}
indicates, there is a trend of $L_{IR}^{\mathcal{X}}$ versus redshifts. Obviously,
the lack of IR quasars with low $L_{IR}^{\mathcal{X}}$ at high redshifts is caused by
the selection effect due to the survey limit. For illustration,
Figure~\ref{fig:lir_vs_z} shows the $L_{IR}^{\mathcal{X}}$ selection limit
corresponding to a \SI{250}{\micron} flux density limit of
$S_{250}=\SI{56.6}{\milli\jansky}$ (which is what we adopted to select the
bright subsample from the SNR5 sample). Interestingly, there seems to be a
deficit of very luminous IR quasars at $z<1$, which reflects a genuine IR
luminosity evolution that is broadly consistent with the evolution of
ULIRGs, i.e., there are more ULIRGs at $z>1$ than at lower redshifts.

Our conclusion that the FIR emission of IR quasars are powered by star
forming activity in dust-rich environments has also been suggested by
previous studies at high redshifts \citep[e.g.,][]{Wang2011}. If this is
indeed the case, using the standard $L_{IR}$ to SFR conversion of
\citet{Kennicutt1998}, i.e., $\text{SFR}_{IR}=\num{1.0d-10}L_{IR}/L_\odot$
for a Chabrier initial mass function (IMF)\footnote{The conversion would be
    a factor of $\sim 1.7$ higher if using a Salpeter IMF, which was
    adopted in \citet{Kennicutt1998}.}, the SFR of the HyLIRGs in our
sample would be $\sim 1.0\text{--}\SI{6.3d3}{\Msun\per\yr}$ \footnote{The
    most conservative SFR estimates would be using $L_{FIR}^{mbb}$ (over 60
    to \SI{1000}{\micron}) instead of $L_{IR}^{mbb}$ (over 8 to
    \SI{1000}{\micron}), which would reduce the SFR values by a factor of
    $\sim 1.5$.}.

There might be concerns whether such extreme SFRs are physical. The SFR
could indeed be overestimated in two ways. First, one could argue that AGN
heating is still an important contributor to $L_{IR}^{(cd)}$ and hence the
SFR cannot be calculated without subtracting this contribution. While there
is no viable model quantitatively showing that this could be the case (in
fact, all available models assume the opposite), we cannot yet assert that
this is impossible. Our argument of $L_{FIR}^{mbb}=L_{FIR}^{SB}$ presented
earlier is only a necessary condition that $L_{IR}^{(cd)}$ is due to the
heating from star formation but not a sufficient one, and therefore we
cannot rule out such a possibility based on this argument alone. However,
we can demonstrate that AGN heating is very unlikely dominant in
$L_{IR}^{(cd)}$. If it is dominant, it is expected that $L_{IR}^{(cd)}$
should be positively correlated with the AGN activity, i.e., the stronger
the AGN is, the larger $L_{IR}^{(cd)}$ should be.
Figure~\ref{fig:lir_vs_mibh} shows $L_{IR}^{(cd)}$ versus the quasar
absolute $i$-band magnitude (normalized to $z=2$, adapted from
\citet{Shen2011} and \citet{Paris2014} for DR7 and DR10, respectively, and
all are based on PSF magnitudes after the Galactic extinction correction)
in four redshift bins. Apparently, no such a correlation can be seen. In
the lowest redshift bin, the distribution of the objects is completely
chaotic. In the bins at higher and higher redshifts, we tend to see only
those objects that are more and more IR luminous, which is simply due to
the selection effect in a flux-limited survey. Even among these the most
luminous ones, no correlation among $L_{IR}^{(cd)}$ and $M_i$ can be
vouched for. Furthermore, Figure~\ref{fig:lir_vs_mibh} also shows
$L_{IR}^{(cd)}$ versus the back hole mass ($M_{BH}$) for the quasars that
have these estimates (taken from \citet{Shen2011} for the DR7Q quasars).
Similarly, no correlation exists.
Finally, the bottom panels of Figure~\ref{fig:lir_vs_mibh} show
$L_{IR}^{(cd)}$ versus the hard-band X-ray luminosity in the restframe
    $2\text{--}\SI{10}{\keV}$ ($L_X^{2\text{--}\SI{10}{\keV}}$) for a
    limited number of quasars that we can derive this quantity based on the
    data available in the literature
\footnote{$L_X^{2\text{--}\SI{10}{\keV}}$ were derived based on the
        data from the Chandra Source Catalog Release 1 \citep{Evans2010}
    and the 3XMM-DR5 catalog \citep{Rosen2015}. Briefly, a power-law
        in the form of $I_{\nu}\propto\nu^{-\alpha}$ was fit to the flux
        densities at different energy bands, and the total energy in
        restframe $2\text{--}\SI{10}{\keV}$ was calculated by integrating
        the best-fit power-law over this energy range. The best-fit
        $\alpha$ has a median of $\sim 0.7$, which correspond to the photon
        index $\Gamma \sim 1.7$.}. As there are only 40
    such quasars, we split them into two redshift bins, $0<z\leq 1.7$ and
    $z>1.7$, respectively, so that each bin receives approximately the same
    number of objects for statistics (19 and 21 objects, respectively).
    They all have $L_X^{2\text{--}\SI{10}{\keV}}>\SI{d43}{erg\per\second}$,
    which is well above the conventional X-ray AGN selection threshold of
    $\SI{d42}{erg\per\second}$, above which the X-ray luminosity is
    believed to be predominantly due to AGN. Therefore,
    $L_X^{2\text{--}\SI{10}{\keV}}$ is a strong indicator of the AGN
    activity. Again, no correlation between $L_{IR}^{(cd)}$ and
$L_X^{2\text{--}\SI{10}{\keV}}$ can be seen. This is also very
    consistent with the recent results of \citet{Symeonidis2014} and
\citet{Azadi2015} in the similar $L_X^{2\text{--}\SI{10}{\keV}}$
    regime.

Therefore, while we do not have direct evidence to assert that AGN has no
contribution to $L_{IR}^{(cd)}$, we do have evidence (albeit still
indirect) against that AGN contribution can be dominant. This further
strengths our conclusion of $L_{IR}^{(cd)}$ being due to star formation
based on the earlier argument of $L_{FIR}^{mbb}=L_{FIR}^{SB}$. As an
additional check of consistency in our conclusion, we have also tested the
AGN/starburst decomposition approach, using the method of
\citet{Mullaney2011} on the objects that have data in the PACS and/or the
\Spitzer{} MIPS bands. The results are detailed in
Appendix~\ref{sec:app_decomp}. Note that all the AGN/starburst
decomposition schemes available in the literature to date (including
\citet{Mullaney2011}) assume that the AGN contribution drops off in the
FIR, and hence the decomposition is not entirely appropriate in asserting
whether AGN contribute significantly to $L_{FIR}$. Nevertheless, our result
shows that the starburst-contributed IR luminosities as derived in the
decomposition scheme (designated as $L_{IR}^{DCSB}$) are also consistent
with our $L_{IR}^{mbb}$ and $L_{IR}^{SB}$ as derived using the SPIRE data
alone, and therefore we do not find any evidence against our conclusion.

The other possibility is that the most luminous objects are actually
gravitationally lensed, which means that their intrinsic luminosities
must be lower and so are their SFR estimates. Currently, we do not
have further data to address this issue. In \S\ref{sec:tlir}, however,
we will show that it is also unlikely that the most luminous objects
are predominantly the result of lensing.

\subsection{Dust temperature}

\begin{figure}[t]
\plotone{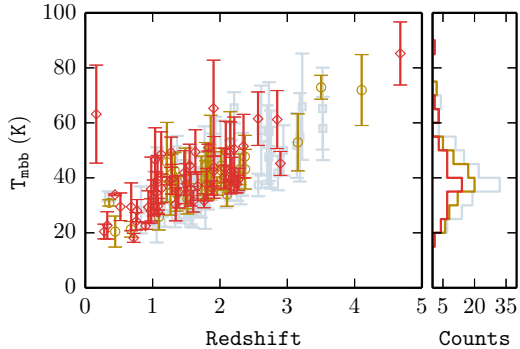}
\caption{Distribution of the derived dust temperatures ($T_{mbb}$).
    Only objects with $T_{mbb}\geq 3T_{mbb}^{err}$ are shown (134 in
    total). The
    colored symbols (86 objects) represent those that are within the
    SNR5 sample, while the grey squares (48 objects)
    represent the rest. Among the SNR5 objects, the red diamonds (55
    objects)
    indicate those that are in the bright subsample
    ($S_{250}>\SI{56.6}{\milli\jansky}$), while the yellow circles (31
    objects)
    indicate those that are not. The right panel shows the histograms
    of $T_{mbb}$ for the displayed objects (all: grey; SNR5: yellow;
    ``bright'' SNR5: red).}
\label{fig:tmbb_vs_z}
\end{figure}

\begin{figure*}[t]
\plotone{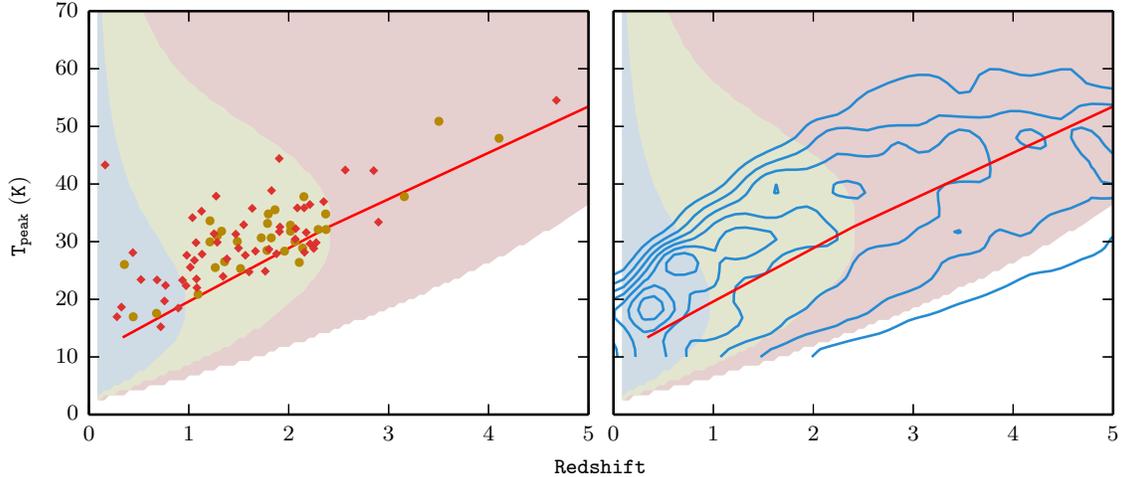}
\caption{Demonstration of the redshift bias in the dust temperature
    derived using the SPIRE bands. $T_{peak}$ is used for this
    purpose, and for simplicity a fiducial sensitivity threshold of
    $S_{250}^{th}=\SI{56.6}{\milli\jansky}$ is adopted in this figure.
    Using the MBB fit of the \texttt{cmcirsed} tool, $T_{peak}$ can be
    directly derived, and the results for the SNR5 sample are shown,
    with those from the bright subsample ($S_{250}\geq
    \SI{56.6}{\milli\jansky}$) being coded in red diamonds and the rest being
    coded in yellow dots.  The adopted $S_{250}^{th}$ imposes a selection
    boundary in the ($T_{peak}$, $z$) plane for a given $L_{IR}^{(cd)}$ limit,
    and thus forms a ``occupation region'' for the objects whose
    $L_{IR}^{(cd)}$ are less than this limit. The shaded color regions denote
    such regions for the $L_{IR}^{(cd)}$ limits of \num{d12} (light blue),
    \num{d13} (light green) and \SI{d14}{\Lsun} (pink), respectively.
    Such occupation regions have ``bumps'' whose tips are aligned on a
    straight ridge line (shown in red), and this ridge line indicates
    the temperatures preferred by the SPIRE bands at different
    redshifts.
    The right panel shows the results from the simulation as detailed
    in the text. The blue curves are the number density contours of
    the simulated $\text{SNR}=5$ (in \SI{250}{\micron}) objects that
    have reliable $T_{peak}$ estimates through the MBB fitting, and
    the most dense contours are all distributed close to the ridge
    line, which further confirms the redshift bias in the dust
    temperature derived based on the SPIRE bands.
    }
\label{fig:tpeak_vs_z}
\end{figure*}

Due to the limitation of our data, it was not always possible to
obtain a good temperature estimate using either $T_{mbb}$ or
$T_{peak}$. To constrain the temperature well, at least the three
SPIRE bands (250, 350 and \SI{500}{\micron}) should sample the peak
region of the FIR SED on both the ``rising'' and the ``falling'' sides
(i.e., short and long wavelength sides, respectively). However, this
is not always satisfied in our cases. To secure our discussion in this
section, here we only include the objects that have $T_{mbb}/\Delta
T_{mbb} \geq 3$ (i.e., equivalent to ``SNR'' in temperature estimate
$\geq 3$).

The left panel of Figure~\ref{fig:tmbb_vs_z} shows the distribution of
$T_{mbb}$ from the MBB fit. The majority of them are in the range of
$30\text{--}\SI{50}{\kelvin}$, which is generally consistent with the
picture that the FIR emission in these objects is dominated by star
formation activity heating the dust. There also seems to be an
``evolutionary trend'' in that $T_{mbb}$ increases with redshifts,
however this is due to the bias in how a secure estimate of
temperature could be derived in our case: the peak of the MBB FIR
emission at the same temperature would shift to longer wavelengths
with increasing redshifts, and therefore to keep the peak being well
sampled by the three SPIRE bands on both the rising and the falling
side of the SED, a higher temperature would be needed to shift the
peak to shorter wavelengths.

We carried out an extensive simulation using the MBB models to further
investigate this temperature bias, and the results are summarized in
Figure~\ref{fig:tpeak_vs_z}. To connect to the argument above,
$T_{peak}$ is used for this demonstration. For a given $L_{IR}^{(cd)}$, our
goal is to study the constraint on $T_{peak}$ when the detection limit
is imposed. We generated a large set of MBB models, with $L_{IR}^{(cd)}$
ranging from \num{d10} to \SI{d14}{\Lsun} (in a step-size of 0.2~dex).
At each $L_{IR}^{(cd)}$, we varied $T_{mbb}$ from $10$ to \SI{100}{\kelvin}
(step-size \SI{1}{\kelvin}). These models were redshifted to $z=0$ to 6
(step-size $0.1$), and the flux densities $S_{250}$ were
calculated from these simulated spectra. We then imposed a desired
detection limit, selected the simulated objects that have $S_{250}$
above this threshold, and, after calculating their $T_{peak}$, put
them on the $T_{peak}-z$ plane as shown in
Figure~\ref{fig:tpeak_vs_z}. For illustration purpose, here we use a
fiducial threshold of $S_{250}^{th}=\SI{56.6}{\milli\jansky}$, and we
only show three cases in $L_{IR}^{(cd)}$, namely, $L_{IR}^{(cd)}=\num{d12}$,
\num{d13} and \SI{d14}{\Lsun}, respectively.
The area filled in pink shows the region occupied by the
$L_{IR}^{(cd)}=\SI{d14}{\Lsun}$ objects thus selected. For the same detection
threshold, simulated objects of a smaller luminosity will occupy a
smaller region. For example, the green region within the pink region
is where the objects with $L_{IR}^{(cd)}=\SI{d13}{\Lsun}$ reside, and the
blue region within the green region is where the objects with
$L_{IR}^{(cd)}=\SI{d12}{\Lsun}$ reside.
We refer to such a color-coded region as the ``occupation region'' of
a given ($L_{IR}^{(cd)}$, $S_{250}^{th}$) pair. These regions have boundaries
at both high and low $T_{peak}$, or in other words, for the adopted
$S_{250}^{th}$, an object of the given $L_{IR}^{(cd)}$ can be detected at a
specified redshift only when its dust temperature is within the shown
boundaries. Lowering $S_{250}^{th}$, or equivalently, increasing
$L_{IR}^{(cd)}$, would expand the boundaries of an occupation region in the
way that the blue region would ``grow'' to the green region and to the
pink region.

Note that the occupation regions have distinct ``bumps'' whose tips
are aligned on a straight line, which is shown in red in
Figure~\ref{fig:tpeak_vs_z}. This ``ridge'' line indicates the
direction towards which an occupation region would ``grow'' fastest in
area when decreasing $S_{250}^{th}$ and/or increasing $L_{IR}^{(cd)}$.
Changing $L_{IR}^{(cd)}$ and/or $S_{250}^{th}$ will only shift the bumps
along the ridge line, and the ridge line itself does not change as
long as the bands involved in measuring the dust temperature stay the same.

To improve the above simulation further, we added photometric errors
(according to the HerMES survey limits) to the synthesized SEDs
derived from the simulated spectra, and fit them using the MBB models.
As it turns out, the simulated objects that have reliable temperature
estimates ($T_{mbb}/\Delta T_{mbb}\geq 3$) through the fit are all
distributed on the narrow ridge line that connects the ``bumps''. This
is shown in the right panel of Figure~\ref{fig:tmbb_vs_z} for the case
of $\text{SNR}=5$ in \SI{250}{\micron}. This again shows that the dust
temperature \emph{estimates} based on the SPIRE data prefer some
certain temperature at a given redshift, although the SPIRE
\emph{detections} can span a wide temperature range. Using a formalism
similar to Equation~\eqref{eq:tpeak}, this ``preferred'' temperature
as a function of redshift can be approximately described by the
following equation:
\begin{equation}
    T_{peak}^{pref} =
    \frac{\SI{2.898d3}{\micron\kelvin}}{\SI{303.7}{\micron}}(1+z)\,,
\end{equation}
where \SI{303.7}{\micron} is the ``preferred'' peak wavelength due to
the usage of the three SPIRE bands. In other words, this means that
the closer the peak of the FIR emission is to
$\SI{303.7}{\micron}/(1+z)$, the more reliable the temperature
estimation will be.

\subsection{Relation between dust temperature and IR
    luminosity}\label{sec:tlir}

\begin{figure*}[!t]
\epsscale{0.85}
\plotone{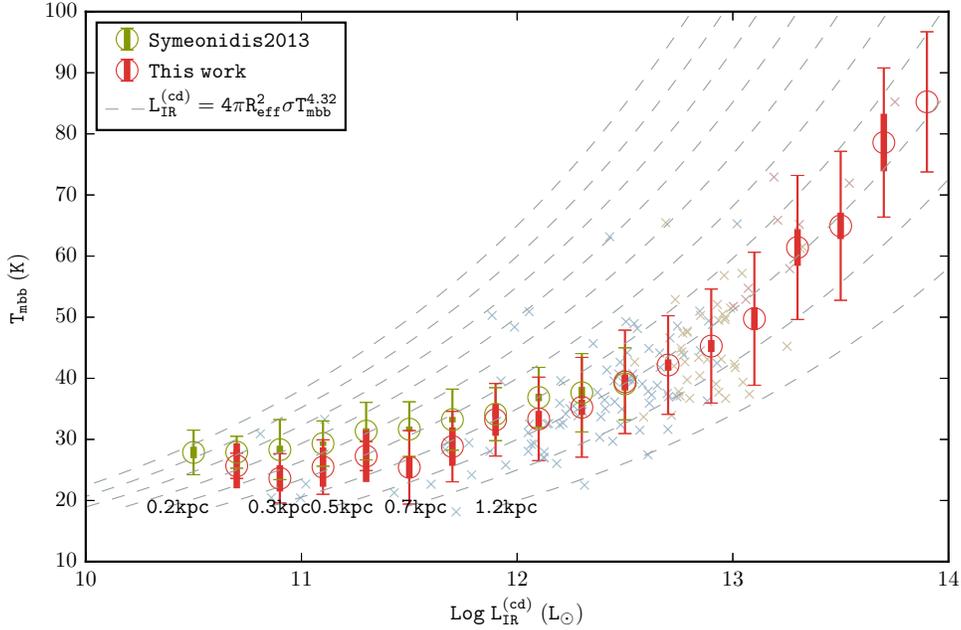}
\caption{
    Strong $T_{mbb}$-$L_{IR}^{(cd)}$ relation as inferred from our IR
    quasars that have reliable $T_{mbb}$ and $L_{IR}^{(cd)}$ measurements.
    The light grey crosses represent the individual quasars (134
        objects in total),
    while the big red circles show their mean $T_{mbb}$ values in
    $L_{IR}^{(cd)}$ bins of \SI{0.2}{dex}. The thick error bars on these red
    circles are the standard deviations around the mean values, while
    the thin error bars are the summed (in quadrature) errors of all
    objects in each bin. For comparison, the results from
    \citep{Symeonidis2013} are shown as the green symbols. The dashed
    lines represent the MBB equivalent of Stefan-Boltzmann law as
    derived based on Equation~\eqref{eq:mbb}, using a family of
    $R_{eff}$ values as labeled.}
\label{fig:lir_vs_tmbb}
\end{figure*}

\begin{figure*}[!t]
\epsscale{0.90}
\plotone{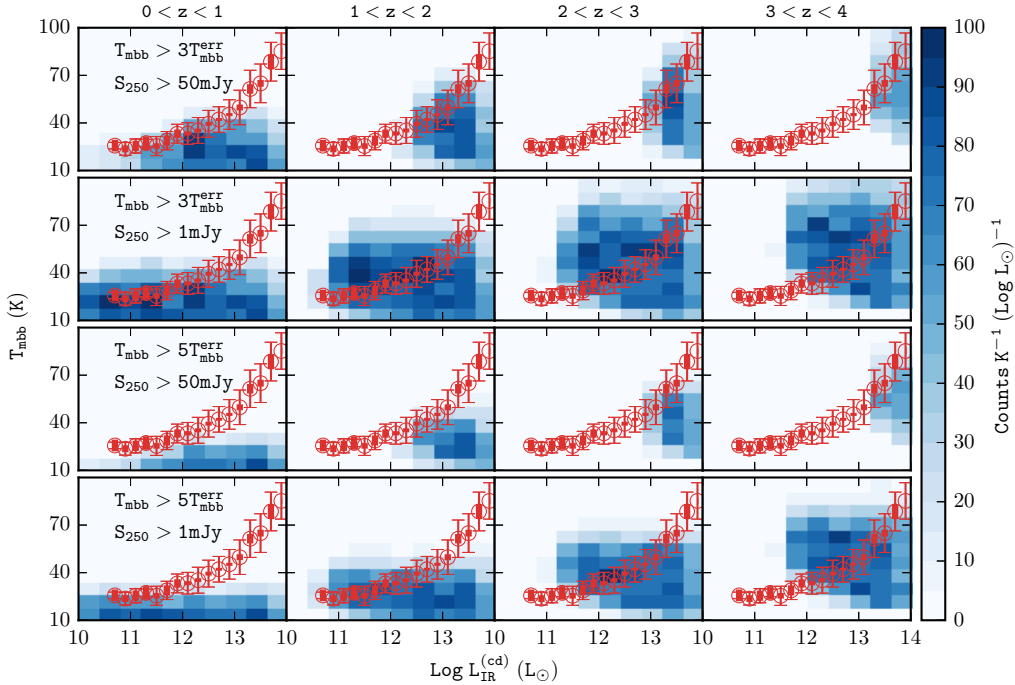}
\caption{Simulation results showing that the observed
    $T_{mbb}$-$L_{IR}^{(cd)}$ relation is not due to selection effect. As
    labeled in each panel, the results are given in four redshift bins
    (in column) for the combinations of two $T_{mbb}$ thresholds and
    two $S_{250}$ thresholds (in row). The red symbols are the same
    $T_{mbb}$-$L_{IR}^{(cd)}$ relation as in Figure~\ref{fig:lir_vs_tmbb},
    while the blue blocks in the background represent the densities of
    the simulated objects recovered by the labeled selection
    thresholds (the coding of the color depth is shown to the right).
    Under our current sample selection (the first row), the highest
    densities of the simulated objects all occur below the
    $T_{mbb}$-$L_{IR}^{(cd)}$ relation, i.e., in the regions of higher
    $L_{IR}^{(cd)}$ and lower $T_{mbb}$ than those defined by the
    $T_{mbb}$-$L_{IR}^{(cd)}$ relation. In other words, while the objects of
    higher $L_{IR}^{(cd)}$ and lower $T_{mbb}$ would have the highest
    probability of being included by our selection, they do not
    present in our sample, i.e., there is a genuine lack of such
    objects in the universe. On the other hand, our selection (mostly
    the $S_{250}$ threshold due to the survey limit) is against the
    objects of higher $T_{mbb}$. Therefore, it is most likely that the
    $T_{mbb}$-$L_{IR}^{(cd)}$ relation, discovered among IR quasars, is the
    envelope of the general distribution of IR galaxies.}
\label{fig:lir_vs_tmbb_simu}
\end{figure*}

\begin{figure*}[t]
\plotone{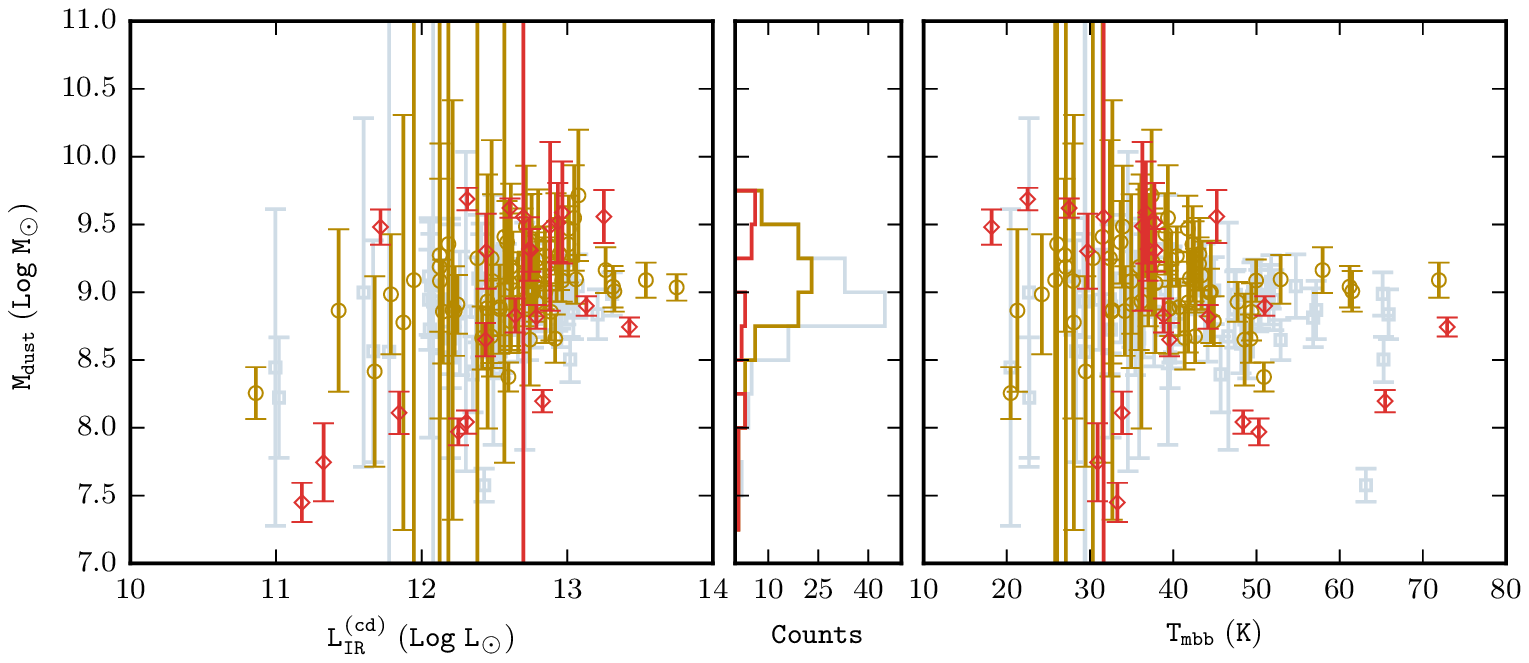}
\caption{Distribution of dust mass ($M_{dust}$) of the IR quasars with
    respect to $L_{IR}^{(cd)}$ (left panel) and $T_{mbb}$ (right panel). The
    middle panel shows the histogram. The gray squares, yellow
    circles, and red diamonds are for the objects with
    $5>T_{mbb}/T_{mbb}^{err} > 3$, $10>T_{mbb}/T_{mbb}^{err} > 5$, and
    $T_{mbb}/T_{mbb}^{err} > 10$, respectively. The gray, yellow, and
    red histograms in the middle panel are for the objects with
    $T_{mbb}/T_{mbb}^{err} > 3$, 5, and 10, respectively.}
\label{fig:mdust_vs_lir_tmbb}
\end{figure*}

Here we examine the $T_{dust}$-$L_{IR}^{(cd)}$ relation of IR quasars. For
simplicity, we use $T_{mbb}$ in this discussion, and only include the
\num{134} objects that have good estimates of both $L_{IR}^{(cd)}$
($\chi^2<10$) and $T_{mbb}$ (``SNR''$>3$). This is shown in
Figure~\ref{fig:lir_vs_tmbb}, where the crosses are the individual
objects and the red circles represent the average at a given $L_{IR}^{(cd)}$
(step-size of \SI{0.2}{\dex} in $\mathrm{Log}\,L_{IR}^{(cd)}$). We also plot the mean
result from \citet[light green circles]{Symeonidis2013}, who have
analyzed a sample of IR luminous ($L_{IR}>\SI{d10}{\Lsun}$) galaxies
at $0.1 < z < 2$ using the deep PACS and SPIRE data in the COSMOS, the
GOODS-N and the GOODS-S fields. \citet{Symeonidis2013} find that their
$L_{IR}$-$T$ relation\footnote{We note that \citet{Symeonidis2013} adopt
    $\lambda_0=\SI{100}{\micron}$ and $\beta=1.5$ as we do here.}
has only a modest increasing trend towards high luminosities, and
their interpretation is that the increase of $L_{IR}$ is caused by an
increase in the dust mass and/or the IR emitting radius rather than by
an increase in the intensity of the dust heating radiation field.

However, the picture is rather different for our sample because our IR
quasars span a much wider range in luminosity. Our $T_{mbb}$-$L_{IR}^{(cd)}$
relation agrees reasonably with that of \citet{Symeonidis2013} in the
overlapped, low luminosity range ($L_{IR}^{(cd)}\lesssim \SI{d12}{\Lsun}$),
and then dramatically rises to higher luminosities.
The existence of such a relation is against the possibility that our
sample could be significantly affected by gravitational lensing,
because the magnification cannot be correlated with the dust
temperature (see \S\ref{sec:result_lir}).

We argue that this $T_{mbb}$-$L_{IR}^{(cd)}$ relation cannot be attributed to
the selection effect of our sample. To demonstrate this point, we
simulated a large number of objects of different $T_{mbb}$ and
$L_{IR}^{(cd)}$ over the redshift range of our sample, and recovered
them using various selection criteria in
$T_{mbb}$ and $S_{250}$. Figure~\ref{fig:lir_vs_tmbb_simu} shows the
results in four redshift bins, for two different $T_{mbb}$ ``SNR''
thresholds of three and five, and two different $S_{250}$ thresholds
of 1 and \SI{50}{\milli\jansky}, respectively. The key points can be
summarized as follows. First, adopting a higher $T_{mbb}$ ``SNR''
(e.g., $T_{mbb}>5T_{mbb}^{err}$ instead of $T_{mbb}>3T_{mbb}^{err}$)
would be against objects with high $T_{mbb}$. Second, adopting a
higher $S_{250}$ threshold (e.g., $S_{250}>\SI{50}{\milli\jansky}$
instead of $S_{250}>\SI{1}{\milli\jansky}$) would be against objects
with low $L_{IR}^{(cd)}$. To reiterate, our current work adopts
$T_{mbb}>3T_{mbb}^{err}$. While our sample does not have an uniform
$S_{250}$ threshold due to the varying survey limits in different
fields, our objects all have $S_{250}>\SI{50}{\milli\jansky}$. From
Figure~\ref{fig:lir_vs_tmbb_simu}, one can see that the objects that
have highest probability of being selected by our criteria would be
those with higher $L_{IR}^{(cd)}$ and lower $T_{mbb}$ than our data points,
however such objects are not presented in our sample, i.e., there is a
genuine lack of such objects.

Before discussing the lack of objects with (high $L_{IR}^{(cd)}$, low
$T_{mbb}$), let us first understand the increasing trend of $T_{mbb}$
with increasing $L_{IR}^{(cd)}$. Recall that for a perfect black body,
Stefan-Boltzmann law states that $L=4\pi R^2\sigma T^4$.  Motivated by
this, we integrated Equation~\eqref{eq:mbb} and found that the
equivalent for a general opacity MBB should follow $L_{IR}^{(cd)}=4\pi
R_{eff}^2\sigma T_{mbb}^{4.32}$, where $R_{eff}$ of a given galaxy
should be interpreted as the effective radius of the equivalent FIR
emitting region if we combine together \emph{all} its dust-enshrouded
star-forming regions. As shown in Figure~\ref{fig:lir_vs_tmbb}, our
data points can be explained by this relation with a family of
$R_{eff}$. At $L_{IR}^{(cd)}\lesssim\SI{d12}{\Lsun}$, the increasing in
$L_{IR}^{(cd)}$ is mostly dominated by the increasing in $R_{eff}$, which
range from $\sim 0.1$ to \SI{0.5}{\kilo\parsec}, and the increasing in
$T_{mbb}$ only plays a modest role. This is consistent with the
suggestion of \citet{Symeonidis2013} as summarized earlier. Naturally,
$R_{eff}$ cannot be increased indefinitely because the sizes of
galaxies are finite. Our data suggest that $R_{eff}$ reaches its
maximum of $\sim 0.5\text{--}\SI{0.7}{\kilo\parsec}$ at the ULIRG
luminosity. At $L_{IR}^{(cd)}\gtrsim\SI{d12}{\Lsun}$, the increasing in
$L_{IR}^{(cd)}$ is taken over by the increasing of $T_{mbb}$, which can be
due to more intense radiation field caused by more intense starburst
activity. The lack of (high $L_{IR}^{(cd)}$, low $T_{mbb}$) objects thus is
the result of the limit in $R_{eff}$.

This also suggests that the $T_{mbb}$-$L_{IR}^{(cd)}$ relation as seen in
Figure~\ref{fig:lir_vs_tmbb} for IR quasars is an envelope of the
general distribution of IR-luminous objects on the ($T_{mbb}$,
$L_{IR}^{(cd)}$)
plane. In other words, for a given $L_{IR}^{(cd)}$, the dust temperature
reached in IR quasars is the lowest among all possibilities
in IR-luminous objects. In fact, the data points of
\citet{Symeonidis2013} indeed are above ours in
Figure~\ref{fig:lir_vs_tmbb}, which is perfectly consistent with our
interpretation. However, we do not have an explanation on why this
envelope manifests itself in IR quasars.

\subsection{Dust mass and gas mass}

\begin{figure}[t]
\plotone{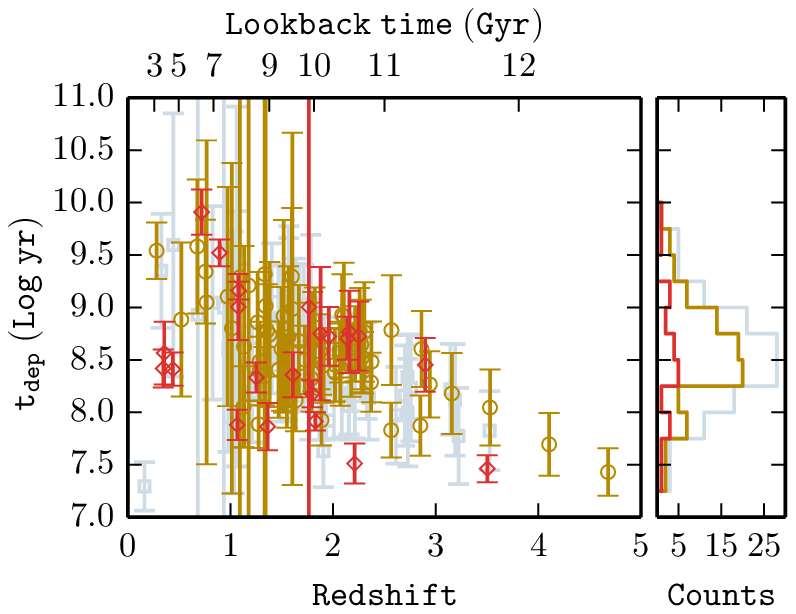}
\caption{Distribution of the gas depletion time $t_{dep}$ for the IR
    quasars. The left panel shows the distribution with respect to
    redshifts and look-back time. A number of objects have $t_{dep}$
    larger than the look-back time, which could mean that they would
    still have plenty of gas left when evolve to today. The middle
    panel shows the histogram of $t_{dep}$. The symbols are the same
    as in Figure~\ref{fig:mdust_vs_lir_tmbb}.}
\label{fig:tau_vs_z}
\end{figure}

The MBB fits also resulted in estimates of dust mass (hereafter
$M_{dust}$), whose distribution is shown in
Figure~\ref{fig:mdust_vs_lir_tmbb} with respect to $L_{IR}^{(cd)}$ and
$T_{mbb}$. As the calculation of $M_{dust}$ is strongly affected by
$T_{mbb}$ ($M_{dust}\propto L_{IR}^{(cd)}T_{mbb}^{-5}$; see
\citealt{Casey2012}), again only those with $T_{mbb}/T_{mbb}^{err}\geq
3$ are included in the plot. The distribution of $M_{dust}$ peaks at
$\sim \num{d8.75}\text{--}\SI{d9.5}{\Msun}$, which seems to be higher
than the dust contents of ULIRGs at these redshifts in general
\citep[see, e.g.,][]{Yan2014}. However, considering that a good
fraction of our objects have $L_{IR}^{(cd)}>\SI{d13}{\Lsun}$, such high dust
masses probably are not surprising. Interestingly, for the objects in
our sample, Figure~\ref{fig:mdust_vs_lir_tmbb} also suggests that
$M_{dust}$ is almost a flat distribution of $T_{mbb}$. This can be
explained by the observed $T_{mbb}$-$L_{IR}^{(cd)}$ relation of our IR
quasars, which can be approximated by $L_{IR}^{(cd)} \propto
    T_{mbb}^\alpha$,
where $\alpha \sim 4\text{--}5$. This means that for \emph{this group
    of objects} we should observe a flat distribution of $M_{dust}$
with respect to $T_{mbb}$, which is exactly what
Figure~\ref{fig:mdust_vs_lir_tmbb} shows. Therefore,
our results are self-consistent.

Adopting a nominal gas-to-dust ratio of 140, we obtain the gas masses
of these objects, which peak at $M_{gas}\sim (0.8\text{--}4.4)\times
10^{11}\si{\Msun}$. This indicates that the IR quasar host galaxies
are very rich in gas. If they could turn all their gas into stars
in their current ULIRG phase,
they would grow substantially in stellar masses. In fact, the added
stars alone would amount to the stellar masses of typical giant
elliptical galaxies in the local universe, which are to the order of
\SI{d11}{\Msun}. Figure~\ref{fig:tau_vs_z} shows the time scale,
$t_{dep}$, that their host galaxies would deplete the gas reservoir if
they would keep forming stars at the rates as seen in their current
ULIRG phase. Most of these objects have
$t_{dep}\lesssim\SI{560}{\mega\yr}$ ($\mathrm{Log}(t_{dep})\lesssim
8.75$), which are broadly consistent with the duration of ULIRGs and
therefore would suggest that they could indeed turn all their gas
reservoir in their current ULIRG phase. However, there are a few
objects (\num{22}, among which \num{5} are among those of the most
secure $T_{mbb}$ estimates) that have $t_{dep} > \SI{1}{\giga\yr}$
(among which one has $t_{dep} > \SI{5}{\giga\yr}$). It is unclear
whether such objects would be able to keep their extreme SFRs over such
a long period.

\subsection{Fraction of IR-luminous quasars}\label{sec:result_frac}

\begin{figure*}[t]
\plotone{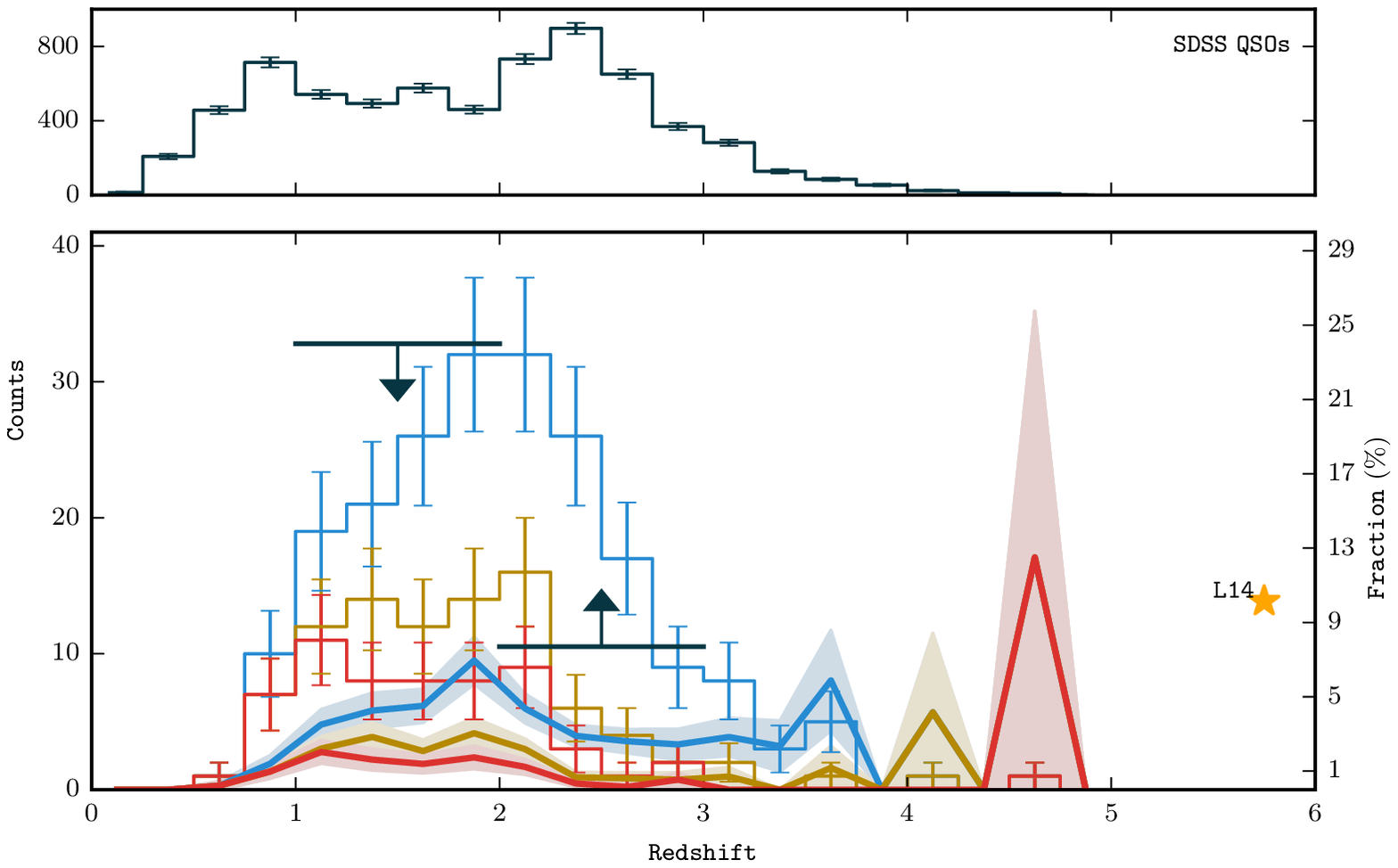}
\caption{Evolution of ``IR bright'' ($L_{IR}^{SB}>L_c$) quasars with
    respect to redshifts (lower panel). In this work, we adopt
    $L_c=\SI{d12.5}{\Lsun}$. For
    comparison, the redshift distribution of the parent SDSS quasar
    sample is reproduced in the upper panel.  The colored histograms
    in the lower panel show the distributions of IR bright quasars
    among the entire sample (blue), the SNR5 sample (yellow), and its
    bright subsample (red). The error bars denote the uncertainties
    based on the Poissonian statistics. In all cases, there is a peak
    at around $1.5\lesssim z\lesssim 2.5$, which does not follow the
    distribution of the parent SDSS quasar sample. To demonstrate it
    more clearly, the curves (with the same color coding as the
    histograms) show the \emph{fraction} of the IR bright quasars
    among the parent quasars. The shaded regions denotes the
    Poissonian uncertainties. The fraction has a prominent drop at
    around $z\approx 2$. The two arrows indicate the impact of the
    incompleteness correction as detailed in the text and
    Appendix~\ref{sec:app_stack}. For reference, such a fraction at
    $z\approx 6$ is also calculated using \citet{Leipski2014} and is
    shown as the yellow asterisk.}
\label{fig:ir_bright_vs_z}
\end{figure*}

A question of general interest is how many optically selected quasars
are FIR-bright. Table~\ref{tab:herschelcatalogues} has provided a
rough answer to this question: there are \num{6710} SDSS quasars in
our \Herschel{} fields, and 354 (5.3\%) of them have
\SI{250}{\micron} detections, among which \num{134} (2.0\%) are
detected at SNR $\geq 5$
\footnote{Our \SI{250}{\micron} detection rate of 5.3\% is lower than
    that of \citet{Dai2012}, who obtain a $\sim 10\%$
    \SI{250}{\micron} detection rate among their 326 quasars in the
    HerMES Lockman Hole field. Such a difference is due to the
    difference in the parent quasar samples. Most importantly, the
    sample of Dai et al. is pre-selected using the MIPS
    \SI{24}{\micron} detections, which presumably is biased to favor
    FIR detections.}.
However, little can be further inferred from
such statistics, because this is set by their \emph{detections} in
the \SI{250}{\micron} images, which are entangled by the varying
survey limits over the fields, the cosmological dimming effects and
the $k$-corrections.
To address this question in a better defined manner, one should
    discuss it in terms of luminosity. To take advantage of all the
    \SI{250}{\micron} detections for this analysis, here we use $L_{IR}^{SB}$
    because its derivation does not require all three SPIRE bands (as
    oppose to $L_{IR}^{mbb}$). This also means that we are discussing
    in terms of $L_{IR}$ rather than $L_{IR}^{(cd)}$ as we do in the
    previous sections. To minimize the systematic effects introduced
    by a particular set of templates, we adopt the average of the
    three derived values based on the three sets (SK07, CE01 and
    DH02). For the sake of simplicity, we still denote this quantity
    as $L_{IR}^{SB}$.

We introduce a ``critical'' luminosity, $L_c$, such that quasars with
$L_{IR}^{SB}>L_c$ are deemed as \emph{``IR-luminous''}. The exact choice of
$L_c$ is somewhat arbitrary, and here we adopt $L_c =
\SI{d12.5}{\Lsun}$. The redshift distribution of these IR-luminous
quasars and their \emph{fraction} among the SDSS quasars is shown in
Figure~\ref{fig:ir_bright_vs_z}. The yellow symbols are for the case
of the SNR5 sample (i.e., the IR-luminous quasars among the SNR5
sample), while the blue and red symbols are for the cases of the whole
IR quasar sample and the bright SNR5 sample, respectively, which
represent the most aggressive and the most conservative inclusions of
objects in the calculation. Obviously, the fraction of IR-luminous
quasars is \emph{not} flat over all redshifts, which is suggestive
of evolution with time. To be specific, this fraction peaks at around
$z\sim 1\text{--}2$, and drops sharply from $z\approx 2$ to higher
redshifts. While the ``spiky'' features at $z\gtrsim 3.5$ could be
attributed to the small numbers in both the IR quasar sample and its
parent SDSS quasar sample, it would be difficult to explain the
notable drop from $z\approx 2$ to $z\approx 3$ using this factor
alone.

Two factors could impact the above picture, however. First of all, the
exact fraction of IR-luminous quasars of course depends on the adopted
$L_c$. From Figure~\ref{fig:lir_vs_z}, it is clear that increasing
$L_c$ would lower the current peak and shift it slightly to higher
redshifts, because there would be fewer objects qualified as
IR-luminous at lower redshifts. However, $L_c$ should not be increased
arbitrarily. For example, if we were to adopt $L_c =\SI{d13}{\Lsun}$,
only a few objects would have survived and this analysis would not be
meaningful. On the other hand, lowering $L_c$ would not change the
fractions at $z>1$ because the number of IR-luminous quasars would not
change \emph{in our sample}. However, this is largely due to the
second factor, namely, the survey incompleteness. In fact, again from
Figure~\ref{fig:lir_vs_z} one can see that the incompleteness at
\SI{d12.5}{\Lsun} could become significant at $z\gtrsim 2.5$. We
investigated the effect of this incompleteness by stacking the SDSS
quasars that did not find matches in the \Herschel{} catalogs
(hereafter ``\Herschel{}-undetected quasars''). The details of this
stacking analysis is presented in Appendix~\ref{sec:app_stack}.

Basically, we need to take into account the objects that have
$L_{IR}^{SB}>L_c=\SI{d12.5}{\Lsun}$ and yet are missing from our sample
due to the survey limits. To focus on the drop at $z\approx 2$, we
only discuss the redshift ranges $1\leq z < 2$ and $2\leq z< 3$. While
the correction factors cannot be determined precisely, we obtained the
\emph{maximum factor possible} at $1\leq z <2$, $f_{cor}<5.4$, and the
\emph{minimum factor necessary} at $2\leq z <3$, $f_{cor}>3.1$. This
could potentially increase the fraction of IR-luminous quasars from
$\sim 4.5\%$ to $\lesssim 24\%$ at $1\leq z<2$, and will definitely
increase it from $\sim 2.5\%$ to $\lesssim 7.7\%$ at $2\leq z <3$.
Therefore, it is possible that the drop at $z\sim 2$ still persists
after the incompleteness correction, although our current data do not
allow us determine whether the drop is as steep as what inferred from
using only the \Herschel{}-detected objects.

\subsection{Contribution to luminosity density}\label{sec:result_lird}

\begin{figure*}[!t]
\plotone{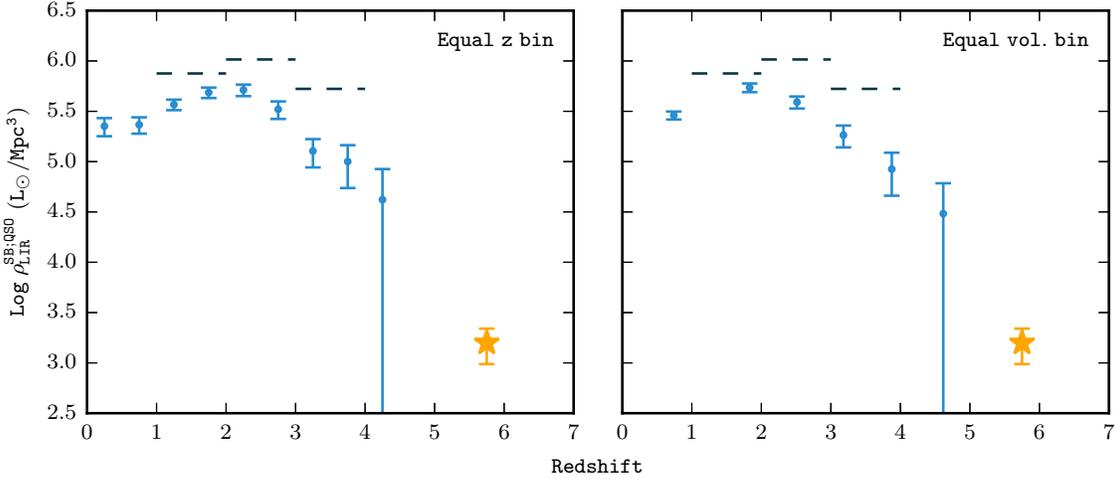}
\caption{Contribution of IR quasars to the IR luminosity density as a
    function of redshift, for the cases of equal redshift binning
    (left) and equal volume binning (right). The IR luminosities
        used here are based on the starburst models (i.e.,
        $L_{IR}^{SB}$; see \S\ref{sec:result_frac} for details).
    The error bars take into account
    both the errors in the derived $L_{IR}^{SB}$ values and the Poissonian uncertainties
    in the number counts. The yellow asterisk denotes the result
    derived using the high-$z$ IR quasar sample of \citet{Leipski2014}.
    The horizontal dashed lines are the contribution of the
    \Herschel{}-undetected quasars based on the stacking analysis as
    detailed in Appendix~\ref{sec:app_stack}.}
\label{fig:lird_vs_z_avg}
\end{figure*}

Here we investigate the contribution of IR quasars to the IR
luminosity density ($\rho_{LIR}^{SB;QSO}$). This is done by adding
$L_{IR}^{SB}$ of the IR quasars in a given redshift bin and then dividing
by the total survey volume in this bin. We do not intend to
correct for the incompleteness imposed by the \Herschel{} survey
limits, and thus what we can obtain would only be a strict lower limit
of the contribution from the optical quasar population.

While our sample is the largest one possible at this stage, its size
is still rather limited when being divided into subsamples by
redshifts, and thus the exact step-sizes in the redshift domain will
slightly affect the detailed results.  For this reason, we adopted two
types of binning in redshifts, one being a uniform division at a
step-size of $\Delta z=0.5$ and the other being equal volume
($\SI{1.46d9}{\mega\parsec\cubed}$) in the successive bins. The
results are shown in Figure~\ref{fig:lird_vs_z_avg}. While the
detailed features are somewhat different in these two schemes, the
overall characteristics are the same.

First of all, IR quasars among optical quasar population only
contribute a very small fraction to the total IR luminosity density.
One recent example to compare to is the study of \citet{Viero2013},
where the authors use a K-band selected galaxy sample to perform a
similar analysis in the FIR and find that the IR luminosity density
produced by galaxies peaks around $z=1\text{--}2$ at
$\sim\SI{d8.9}{\Lsun\per\mega\parsec\cubed}$. As
Figure~\ref{fig:lird_vs_z_avg} shows, dust emission due to star forming
activity of IR quasars
only contribute $\sim
0.06\%$ of this amount.

Second, $\rho_{LIR}^{SB;QSO}$ declines at $z\gtrsim 2$. This echoes the
decline of IR luminous quasar fraction at the same redshift as
discussed above (see Figure~\ref{fig:ir_bright_vs_z}). However, this
picture is also affected by the survey limits. To take the
\Herschel{}-undetected quasars into account, we used the stacking
results as described in Appendix~\ref{sec:app_stack}. While we were
only able to make corrections over four redshift bins with much larger
step-size ($\Delta z=1$), it is clear that
$\rho_{LIR}^{SB;QSO}$ after the correction peaks at $z\approx 2.5$
instead of $z\approx 1.5$ when uncorrected.

It is well known that optical quasar number density peaks at around
$z\approx 2\text{--}3$ (see e.g., \citealt{Osmer2004} for review).
Using the SDSS DR3 quasars, \citet{Richards2006} find the peak at
$z\approx 2.5$ when integrating the $i$-band luminosity function to
$M_i[z=2]=\SI{-27.6}{\mag}$. \citet{Jiang2006} use a faint quasar
sample in an area of \SI{3.9}{\deg\squared} within the SDSS Stripe 82
and find that the peak would shift to $z\approx 2$ if integrating to
$M_g=\SI{-22.5}{\mag}$ \citep[see also][]{Ross2013}. The UV luminosity density (and hence the
global SFR density) of normal galaxies rises sharply from $z=0$ to
higher redshifts, also peaks at $z\approx 2$, but keeps flat out to
$z\approx 4\text{--}5$ \citep[see e.g.,][]{Hopkins2006}. It is thus
intriguing that $\rho_{LIR}^{SB;QSO}$ peaks at around the same redshift,
which suggests that this could be the result of the co-evolution of
extreme star formation and quasars.

\subsection{Two quasars with increasing FIR SEDs}

In the course of the FIR SED analysis, we found two unique quasars that have
monolithically increasing SEDs from 250 to
\SI{500}{\micron}\footnote{These two objects have been excluded from
    our analysis elsewhere in the paper.}. They are
J090910.08+012135.6 (``HATLAS-SDP-001'' for short) at $z=1.02$ and
J012528.83-000555.9 (``HerS-080'') at $z=1.08$, respectively.
By searching the
NASA/IPAC Extragalactic Database (NED), we found that both of them are
blazars cataloged by \citet{Massaro2009}, with the object names of
BZQJ0909+0121 and BZQJ0125-0005, respectively. The FIR properties of
HATLAS-SDP-001 has also been discussed by \citet{Gonzalez-Nuevo2010}.

The monolithically increasing SEDs of these objects mimic the SEDs of
the so-called ``\SI{500}{\micron} peakers'', which are candidates of
FIR galaxies at $z>4$ \citep{Pope2010}. Apparently, such objects can
be non-negligible contaminators to such FIR high-z candidates, as they
pass the usual criterion of $S_{500}/S_{350}\geq 1.3$ \citep[see
e.g.,][]{Riechers2013}.

\section{Conclusion and Summary}\label{sec:summary}

In this work, we combined the SDSS quasars from DR7 and DR10 and
searched for their FIR counterparts in the \SI{172}{\deg\squared}
\Herschel{} wide survey fields where the high-level \Herschel{} maps
have been made public by the relevant survey teams. From the total of
\num{6170} SDSS quasars within the \Herschel{} field coverage, we
assembled a sample of \num{354} quasars that are detected in the
\Herschel{} SPIRE data, \num{134} of which are highly secure,
$\text{SNR}\geq 5$ sources in \SI{250}{\micron}. As we used a
stringent matching criterion, the contamination due to source blending
is minimal. This IR quasar sample spans a wide redshift range of
$0.14\leq z\leq 4.7$, and is the largest sample of optical quasars
that have FIR detections.

To investigate their properties, we analyzed their FIR SEDs using a modified
black body model (MBB) as well as three different sets of starburst (SB)
templates (SK07, CE01 and DH02). By focusing on the FIR emission, we confine
our discussion mainly to the cold-dust component of IR quasars, as the FIR
emission is predominantly due to this component. Our conclusions are
summarized below.

\begin{itemize}

    \item The results based on the MBB model (independent of the heating
        source) are consistent with those based on the SB models (legitimate only when
    the heating source is from star formation), which strongly suggests that
    the IR luminosity of the cold-dust component in these IR quasars
    ($L_{IR}^{(cd)}$) is mainly due to the heating of star formation rather than
        AGN\@. This is further strengthened by the additional supporting evidence that
    there is no positive correlation between $L_{IR}^{(cd)}$ and the absolute
    magnitudes or the black hole masses of the IR quasars, both of the latter
    being indicators of the strength of AGN\@. $L_{IR}^{(cd)}$, derived in this
    work as $L_{IR}^{mbb}$ based on the SPIRE photometry, underestimates the
    total IR luminosity because it does not include the contribution from other
    warmer dust components. However, as it is very unlikely being significantly
    contaminated by the AGN heating, this quantity is more preferable in
    inferring the SFR of the host galaxies.

    \item The derived $L_{IR}^{(cd)}$ values, adopting the more conservative ones
    based on the MBB fitting results,
    range from \num{d10.8} to
    \SI{d13.8}{\Lsun} (after discarding two objects whose SEDs are
    barely constrained), with $\sim 80\%$ being ULIRG
    ($>\SI{d12}{\Lsun}$) and $\sim15\%$ being HyLIRG
    ($>\SI{d13}{\Lsun}$). There is a general trend that $L_{IR}^{(cd)}$
    increases at larger redshifts, which is mostly due to the
    selection effect in our flux-limited sample. However, there is
    a lack of high $L_{IR}^{(cd)}$ objects at the lowest redshifts ($z<1$),
    which is broadly consistent with the picture that ULIRGs are
    scarce in the low-redshift universe.

    \item
    Due to the wavelength coverage of the SPIRE passbands, the dust
    temperature can only be well constrained for a fraction of the IR
    quasars whose SPIRE detections sample both the blue and the red
    sides to the peak of their FIR emissions. For these objects, the
    derived temperatures show an increasing trend with redshift, which
    is caused by this selection effect. We show that these objects
    distribute along a ``preferred'' region on the dust temperate
    versus redshift plane, and this temperature bias is common to any
    \Herschel{} sources that are limited by the SPIRE data.
    Nevertheless, we find that most of the IR quasars with well
    constrained temperatures have $T_{mbb} \sim$ $25$ to
    \SI{50}{\kelvin}, which is consistent with the picture that their
    FIR emission is mostly due to young stars heating the cold ISM.

    \item
    In spite of the dust temperature bias, the IR quasars with well
    constrained dust temperatures allow us to investigate the
    $T$-$L_{IR}^{(cd)}$ relation over a wide dynamic range in $L_{IR}^{(cd)}$. We
    find that there is a dramatic increase of dust temperature at
    $L_{IR}^{(cd)}>\SI{d12}{\Lsun}$. Through simulations, we show that this
    trend cannot be due to the selection effect of our sample.
    Instead, this trend, which holds for IR quasars, seems to be the
    envelope of the general distribution of IR objects on the ($T$,
    $L_{IR}^{(cd)}$) plane. At the low luminosity end along this envelope,
    the increasing of $L_{IR}^{(cd)}$ is largely due to the increasing in the
    effective radius of the heated region (or equivalently, the
    enclosed dust mass). The behavior of the trend shows that the size of
    the heated region cannot be arbitrarily increased, and any further
    increasing of $L_{IR}^{(cd)}$ must be largely driven by the increased
    heating (i.e., more intense star formation rate per unit volume).

    \item
        The SFR values inferred from $L_{IR}^{(cd)}$ range from
    \num{6.3} to
    \SI{6310}{\Msun\per\yr} (for a Chabrier IMF; the values would be
    $1.7\times$ higher if using a Salpeter IMF). From the dust mass
    derived via MBB fitting, and using a nominal gas-to-dust ratio of
    140, we have inferred the gas mass for the IR quasars, most of
    which being within $\sim
    (0.8\text{--}4.4)\times10^{11}\si{\Msun}$. Given the SFR and the
    gas mass, for most of the objects in our sample, the time scale
    that their host galaxies would deplete the gas reservoir is
    $t_{dep} \lesssim\SI{560}{\mega\yr}$, while a few of them could
    have $t_{dep} > \SI{1}{\giga\yr}$.

    \item
    The fraction of IR-luminous optical quasars evolves with time. For
    a fiducial threshold of $L_c=\SI{d12.5}{\Lsun}$, the fraction of
    quasars with $L_{IR}>L_c$ peaks at around $z\approx 2$, an epoch in
    general agreement with the peak of the global star formation rate
    density evolution.

    \item
    IR quasars, even counting those that are not detected in the
    current {\it Herschel}\, surveys, only contribute a very small
    fraction to the total IR
    luminosity density. This contribution also peaks at around
    $z\approx 2\text{--}3$.

\end{itemize}

The full catalog of our sample, including all the derived physical
properties, is available as the online data of this paper.

\acknowledgments{
    We thank the referees for the helpful comments. We acknowledge the support
of the University of Missouri Research Board Grant RB 15-22 and NASA's
Astrophysics Data Analysis Program under grant number NNX15AM92G. We thank
the useful discussions with Drs. Yu Gao, Aigen Li, and Rui Xue. This
research has used the data taken by the \Herschel{} observatory.
{Herschel} is an ESA space observatory with science instruments
provided by European-led Principal Investigator consortia and with
important participation from NASA\@. The quasar catalogs are from the
SDSS, SDSS-II and SDSS-III\@. We have also made use of the NASA/IPAC
Extragalactic Database (NED), which is operated by the Jet Propulsion
Laboratory, California Institute of Technology, under contract with
NASA\@.}

{\it Facilities:\/} \facility{Herschel}, \facility{Sloan}.

\appendix

\section{Impact of $\lambda_0$ in MBB models}\label{sec:app_lambda0}

\begin{figure*}[t]
\plotone{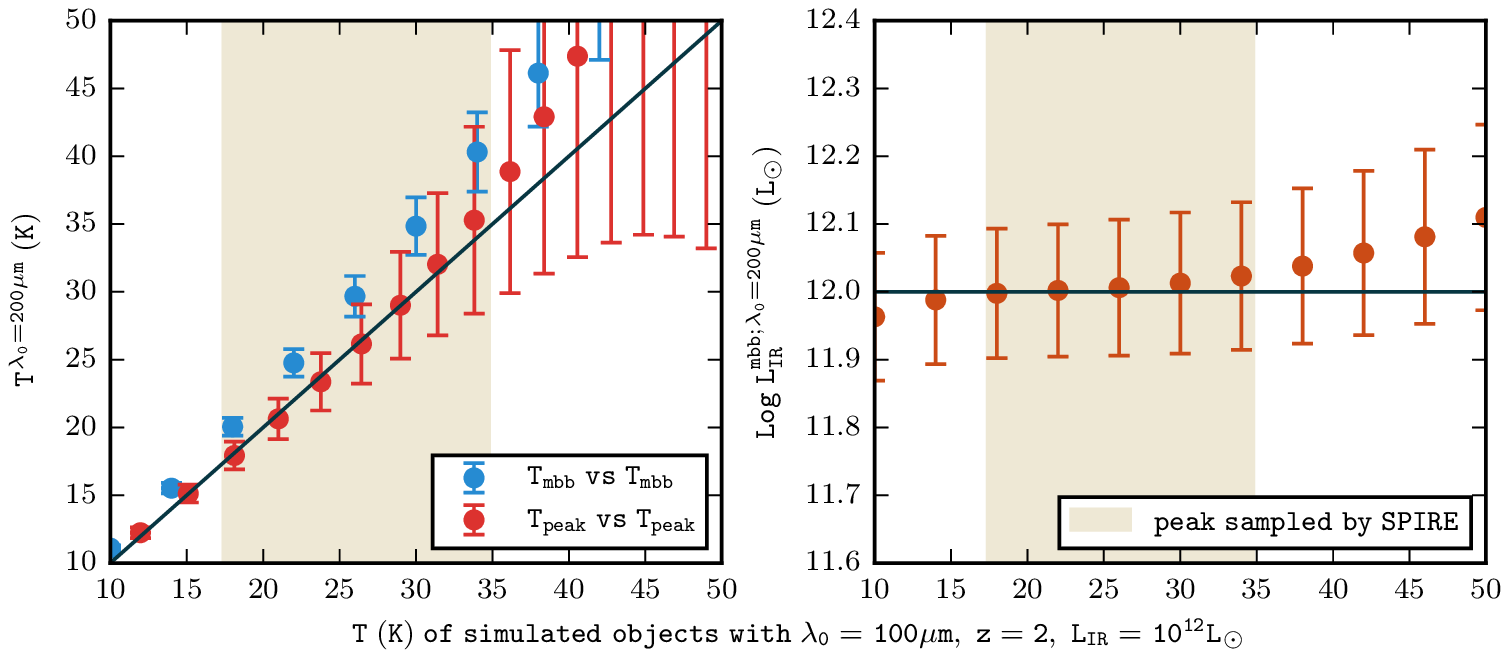}
\caption{Impact of the different choices of $\lambda_0$ to the dust
    temperature and the IR luminosity estimates, using a
    representative case where the input MBB models using
    $\lambda_0=\SI{100}{\micron}$ are at $z=2$ and have
    $L_{IR}^{mbb}=\SI{d12}{\Lsun}$. The left panel shows the comparison of
    the output $T_{mbb}$ (blue symbols) and $T_{peak}$ (red symbols)
    estimates, derived using the MBB fit with
    $\lambda_0=\SI{200}{\micron}$, to the input temperature.
    Similarly, the right panel shows the comparison of
    $L_{IR}^{mbb}$
    estimates to the input value. In both panels, the shaded region
    indicates the temperature range within which the peak region of
    the MBB spectra are well sampled by the SPIRE bands.}
\label{fig:simu_lambda0}
\end{figure*}

As detailed in \S\ref{sec:cmcirsed}, we adopted the MBB model of the
\texttt{cmcirsed} code \citep{Casey2012}, which takes the following
form:
\begin{equation*}
    \displaystyle S(\lambda) = N_{mbb}\frac{(1-\mathrm{e}^
        {-(\frac{\lambda_0}{\lambda})^{\beta}})(\frac{c}{\lambda})^3}
    {\mathrm{e}^{hc/(\lambda kT_{mbb})}-1}
    \,.
\end{equation*}
The term $1-\mathrm{e}^{-(\frac{\lambda_0}{\lambda})^{\beta}}$
modifies the black body term in that it includes a
wavelength-dependent optical depth $\tau(\lambda)$, which is assumed
to follow a power-law of
$\tau(\lambda)=(\frac{\lambda_0}{\lambda})^{\beta}$. The parameter
$\lambda_0$ is the wavelength where the optical depth is unity. By
default, \texttt{cmcirsed} sets $\lambda_0=\SI{200}{\micron}$. In this
work, we followed \citet{Draine2006} and adopted
$\lambda_0=\SI{100}{\micron}$.

The difference in the choice of $\lambda_0$ impacts the derived
$T_{mbb}$ significantly. This is because the ``modifying'' term varies
significantly when different $\lambda_0$ is adopted. As an example,
let us consider a FIR SED that peaks at restframe
$\lambda=\SI{100}{\micron}$. The ``modifying'' term is 0.63 at this
peak wavelength when $\lambda_0=\SI{100}{\micron}$, but becomes 0.94
when $\lambda_0=\SI{200}{\micron}$. Therefore, in order to fit the
peak flux density, the black body term will need to be smaller in the
case of $\lambda_0=\SI{100}{\micron}$, which means that the derived
$T_{mbb}$ must be smaller. For further demonstration, we generated a
series of MBB models of varying $T_{mbb}$ using
$\lambda_0=\SI{100}{\micron}$, convolved them with the SPIRE band
response curves, and then fitted the simulated photometry using the
MBB models with $\lambda_0=\SI{200}{\micron}$. A representative case
is given in Figure~\ref{fig:simu_lambda0}, where the simulated objects
are at $z=2$ and all have $L_{IR}^{mbb}=\SI{d12}{\Lsun}$. The left panel
shows the relation of the derived $T_{mbb}$ values and the input
values. The peak temperature $T_{peak}$, on the other hand, will not
be affected significantly because the fitting procedure, regardless of the
choice of $\lambda_0$, will always find the model that best matches
the given SED, and therefore the peak wavelength and the peak
temperature based on the Wien's displacement law, will not change
much.

Similarly, the choice of $\lambda_0$ has little impact to the derived
$L_{IR}^{mbb}$ values because the shape of the best-fit model is governed by
the observed FIR SED\@. When its peak region is well sampled by the
three SPIRE bands, the small differences in the fitted SED beyond the
peak will only have a small contribution to $L_{IR}^{mbb}$ and thus have
little influence. The right panel of Figure~\ref{fig:simu_lambda0}
shows the comparison of $L_{IR}^{mbb}$ based on the same simulation.

\section{Summary of online data table and examples of SED fitting}\label{sec:snr5_sed}

\begin{table*}[t]
\centering
\caption{Summary of the online data table}
    \label{tab:onlinedata}
\begin{tabular}{lll}
\tophline
Column          &  Name     & Description                 \\
\midhline
1      & SDSS\_IAU\_name     &   SDSS quasar IAU name  \\
2      & Herschel\_IAU\_name &   IAU name of the Herschel counterpart as in the original catalogs  \\
3      & str\_id        &   Object ID assigned in this work  \\
4      & RA\_250        &   Herschel RA (J2000)  \\
5      & Dec\_250       &   Herschel Dec (J2000)  \\
6      & F250           &   \SI{250}{\micron} flux (mJy)  \\
7      & E250           &   \SI{250}{\micron} flux error (mJy)  \\
8      & F350           &   \SI{350}{\micron} flux (mJy)  \\
9      & E350           &   \SI{350}{\micron} flux error (mJy)  \\
10     & F500           &   \SI{500}{\micron} flux (mJy)  \\
11     & E500           &   \SI{500}{\micron} flux error (mJy)  \\
12     & is\_SNR5       &   1: SNR in \SI{250}{\micron} $>= 5$; 0: SNR in \SI{250}{\micron} $<5$  \\
13     & RA\_100        &   \SI{100}{\micron} RA (J2000)  \\
14     & Dec\_100       &   \SI{100}{\micron} Dec (J2000)  \\
15     & F100           &   \SI{100}{\micron} flux (mJy)  \\
16     & E100           &   \SI{100}{\micron} flux error (mJy)  \\
17     & RA\_160        &   \SI{160}{\micron} RA (J2000)  \\
18     & Dec\_160       &   \SI{160}{\micron} Dec (J2000)  \\
19     & F160           &   \SI{160}{\micron} flux (mJy)  \\
20     & E160           &   \SI{160}{\micron} flux error (mJy)  \\
21     & RA\_DR7        &   SDSS DR7 RA (J2000)  \\
22     & Dec\_DR7       &   SDSS DR7 Dec (J2000)  \\
23     & z\_DR7         &   SDSS DR7 redshift  \\
24     & Mi\_DR7        &   SDSS DR7 $i'$-band absolute magnitude (at z=0)  \\
25     & RA\_DR10       &   SDSS DR10 RA (J2000)  \\
26     & Dec\_DR10      &   SDSS DR10 Dec (J2000)  \\
27     & z\_DR10        &   SDSS DR10 redshift  \\
28     & Mi\_DR10       &   SDSS DR10 $i'$-band absolute magnitude (at z=2)  \\
29     & LOG\_MBH       &   Black hole mass measured by \citet{Shen2011} (\si{Log \Msun})  \\
30     & LOG\_MBHERR    &   Error of the black hole mass  \\
31     & LOG\_LX        &  Hard-band X-ray luminosity in the restframe $2\text{--}\SI{10}{\keV}$ (\si{erg\per\second}) \\
32     & GAMMA          &  Photon index of the X-ray SED used to derive X-ray luminosity \\
33     & XRAY\_REF       &  References of the X-ray data. CSC: \citet{Evans2010}; 3XMM: \citet{Rosen2015} \\
34    & LOG\_LIR\_MBB     &   $L_{IR}^{(cd)}$ measured using MBB fitting (\si{Log \Lsun}), integrated over $8\text{--}\SI{1000}{\micron}$  \\
35    & LOG\_LIRERR\_MBB  &   Error of the measured $L_{IR}^{(cd)}$  \\
36    & CHISQ\_MBB        &   $\chi^2$ of the MBB fitting  \\
37    & TMBB              &   Derived black body temperature of the best fit model (K)  \\
38    & TMBBERR           &   Error of $T_{MBB}$ (K)  \\
39    & TPEAK             &   Derived peak temperature of the best fit model (K)  \\
40    & LOG\_MDUST         &   Dust mass (\si{Log \Msun})  \\
41    & LOG\_MDUSTERR      &   Error of dust mass  \\
42    & LOG\_SFR           &   SFR using \citet{Kennicutt1998}, modified for Chabrier IMF (\si{Log \Msun\per\yr})  \\
43    & LOG\_MGAS          &   Gas mass converted from dust mass (assuming gas-to-mass-ratio of 140)  \\
44    & LOG\_TDEP          &   Gas depletion time scale (\si{Log yr})  \\
45    & LOG\_LIR\_SK07     &   $L_{IR}$ measured using SK07 templates (\si{Log \Lsun})  \\
46    & LOG\_LIRERR\_SK07  &   Error of the SK07 $L_{IR}$  \\
47    & CHISQ\_SK07        &   $\chi^2$ of the SK07 fitting  \\
48    & LOG\_LIR\_CE01     &   $L_{IR}$ measured using CE01 templates (\si{Log \Lsun})  \\
49    & LOG\_LIRERR\_CE01  &   Error of the CE01 $L_{IR}$  \\
50    & CHISQ\_CE01        &   $\chi^2$ of the CE01 fitting  \\
51   & LOG\_LIR\_DH02     &   $L_{IR}$ measured using DH02 templates (\si{Log \Lsun})  \\
52   & LOG\_LIRERR\_DH02  &   Error of the DH02 $L_{IR}$  \\
53   & CHISQ\_DH02        &   $\chi^2$ of the DH02 fitting  \\
\bottomhline
\end{tabular}
\end{table*}

\begin{figure*}[t]
    \epsscale{1.1}
    \plotone{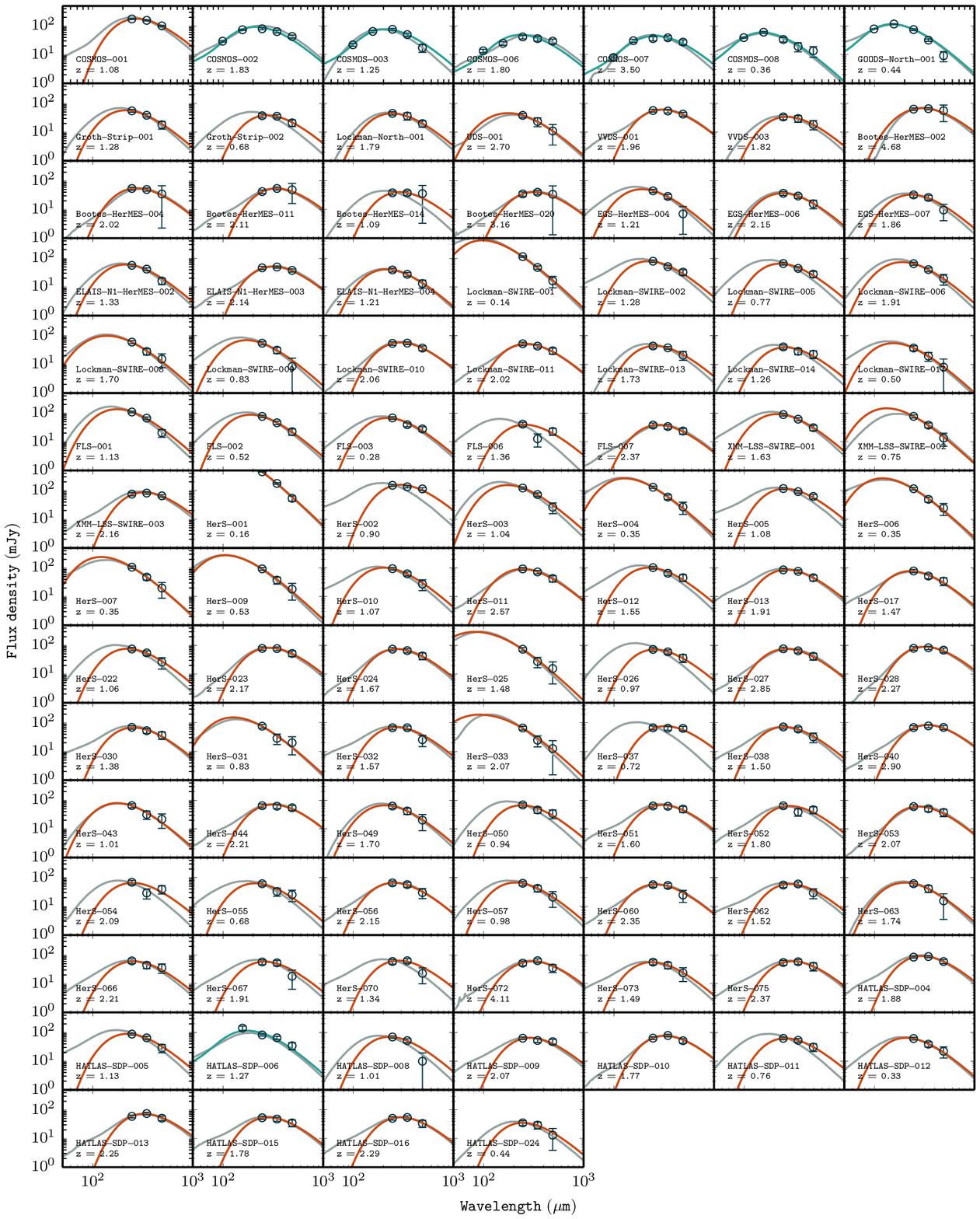}
    \caption {FIR SEDs and their best-fit models for the IR quasars in
        the SNR5 subsample. The dark blue points are the observed SEDs,
        while the gray and the red curves are for the best-fit SK07
        and MBB models, respectively. For objects with PACS data, the
        best-fit MBB+PL models are shown in blue.}
    \label{fig:snr5_sed}
\end{figure*}

The header information of the online data table is given in
Table~\ref{tab:onlinedata}. We note that this table includes all the
\num{354} sources as summarized in \S~\ref{sec:samplesummary} and
Table~\ref{tab:herschelcatalogues}.

The SEDs and the best-fit models for the IR quasars in the SNR5 subsample are
shown in Figure~\ref{fig:snr5_sed} as examples. For clarity, we only plot the
MBB (red curves)
and the SK07 (grey curves) model fits. While there are 134 objects in this
subsample, only 102 of them have photometry in all the three SPIRE bands to
allow for the MBB fit, and hence we only show these 102 objects in this
figure.
The other 32 objects in the SNR5 subsample still have $L_{IR}$ values as
determined using the starburst models (see \S\ref{sec:modeling}),
which are included in the
online data table. We note that seven objects in this figure also have PACS
data, for which we also have MBB+PL fits (shown as the blue curves).

\section{Test of AGN/Starburst
    decomposition}\label{sec:app_decomp}

\begin{figure*}[t!]
    \epsscale{0.9}
    \plotone{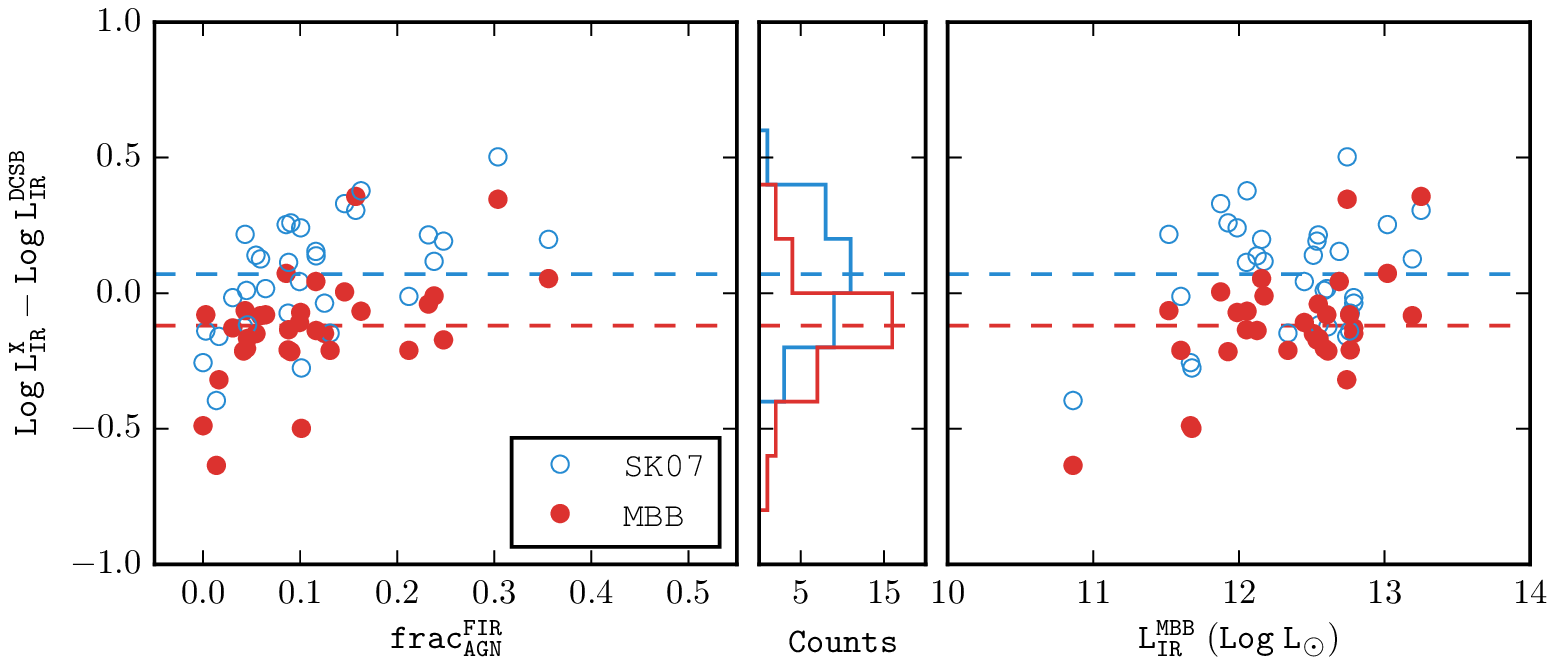}
    \caption {Comparisons of derived IR luminosities using only the FIR
        SEDs to those obtained through the AGN/SF decomposition method of
        \citet{Mullaney2011} incorporating additional mid-IR data. For
        simplicity, only $L_{IR}^{mbb}$ and $L_{IR}^{SK07}$ are compared to
        $L_{IR}^{DCSB}$ here, plotted in red and blue, respectively. The
        left panel shows the differences with respect to the AGN fraction
        to $L_{IR}$ (over $8\text{--}\SI{1000}{\micron}$) as determined by
        the decomposition, while the right panel shows these differences
        with respect to $L_{IR}^{mbb}$ (effectively $L_{IR}^{(cd)}$). The
        red and the blue dashed lines show the mean values of the
        differences. The histogram of the differences are shown in the
        middle panel.}
    \label{fig:decompir_comp}
\end{figure*}

The availability of mid-to-far IR data in the recent years have
allowed better characterization of AGN SEDs in this regime
\citep[e.g.][]{Shi2007, Netzer2007, Maiolino2007, Shang2011,
    Mullaney2011, Shi2013, Dale2014}. In light of these improvements,
it has been advocated that the mid-to-far IR contributions from the
AGN and the host galaxy star formation can be separated through the
decomposition of the SEDs \citep[e.g.][]{Mullaney2011, Dale2014}. Such
decomposition approaches, however, still have caveats.
An important one is that
the AGN templates in use can only be constructed by
subtracting the star formation contributions from the observed SEDs,
where such contributions are determined through the calibration of
some indicators of star formation activities (such as the PAH
features). It is unclear whether such calibration is universally
applicable given the complexity of the star forming regions.
Furthermore, the far-IR part of the templates is just the
extrapolation from the mid-IR part, and a cutoff has to be applied
beyond a certain wavelength in FIR that is somewhat arbitrarily
chosen. This means that such decomposition schemes are not appropriate to
address the question of AGN contribution in FIR, because it is already
assumed to be minimal by design.

With these caveats in mind, we tested the AGN and the star
    formation (AGN/SF) decomposition of our objects in order to
    further check the consistency of our conclusion that
    $L_{IR}^{(cd)}$ is mainly due to the heating from star formation.
    We only carried out this test for the objects that have the PACS
    data and/or those that have the \Spitzer{} MIPS
24/\SI{70}{\micron} data readily available from the releases of
    the relevant \Spitzer{} Legacy Survey programs residing in the
    IPAC Infrared Science Archinve (IRSA). In brief, there are in
    total 87 (COSMOS 33, Lockman 26, FLS 14, XMM-LSS-SWIRE 14) objects
    in our IR quasar sample that have at least MIPS \SI{24}{\micron}
    data, among which 37 objects have detections in all the three
    SPIRE bands, and thus are suitable for the test and the
    comparison.

We used the software suite of \citet{Mullaney2011} in this test. This
particular flavor of decomposition involves one AGN template and five
starburst templates (denoted from ``SB1'' to ``SB5''), and in each
fitting run the routine combines the AGN template and one of the five
starburst templates and determines the normalizations through the
least-$\chi^2$ fitting. The AGN template is forced to cut off at a
wavelength between 20 and \SI{100}{\micron} in the form of a black
body, and the exact cut-off wavelength is determined during the
fitting.
The exact AGN+SB combination that gave the smallest $\chi^2$ was
    deemed to be the best fit, from which the total IR luminosities
    (integrating over $8\text{--}\SI{1000}{\micron}$) due to the
    AGN and the SB heating were then obtained. Here we denote these as
    $L_{IR}^{DCAGN}$ and $L_{IR}^{DCSB}$, respectively, and denote the
    fraction of the AGN contribution as $frac_{IR}^{AGN}\equiv
    L_{IR}^{DCAGN}/(L_{IR}^{DCAGN}+L_{IR}^{DCSB})$.
Figure~\ref{fig:decompir_comp} compares $L_{IR}^{DCSB}$ thus obtained
to $L_{IR}^{mbb}$ (filled red circles) and $L_{IR}^{SK07}$ (open blue
circles) that we derived based on the fit to the SPIRE data as
described in \S\ref{sec:modeling} and \S\ref{sec:result}. The mean
difference between $L_{IR}^{SK07}$ ($L_{IR}^{mbb}$) and
$L_{IR}^{DCSB}$ is \SI{0.07}{\dex} (\SI{-0.12}{\dex}) with a sample
standard deviation \SI{0.2}{\dex} (\SI{0.2}{\dex}). The results
    are fully consistent with what we derived using solely the SPIRE
    bands, i.e., $L_{IR}^{SK07}$ and $L_{IR}^{DCSB}$, which both
    measure the total IR luminosity due to star formation, agree to
    within $\sim 17\%$, while $L_{IR}^{mbb}$, which measures only the
    total IR luminosity from the cold-dust component, is less than
    $L_{IR}^{DCSB}$ by $\sim 32\%$. Considering the differences in the
    methods and the template sets employed, we conclude that the
    decomposition test also provides results that are consistent with
    our interpretation that $L_{IR}^{(cd)}$ is mainly due to the
    heating of star formation.

\section{Stacking analysis of {\it Herschel}-undetected SDSS quasars}
\label{sec:app_stack}

To investigate how the survey limits impact our results, we performed
a stacking analysis of the SDSS quasars that are not matched in the
adopted HerMES, H-ATLAS and HerS catalogs. While a fraction of these
unmatched SDSS quasars could be blended sources rejected by our
stringent matching criterion, the majority should be those that are
undetected in the current SPIRE \SI{250}{\micron} data. It is
difficult to stack \emph{all} such objects from all fields, because
different fields have different survey sensitivities. Therefore, we
only stacked on a field-by-field basis. We show here the results in
the HerMES L6-XMM-LSS and L5-Bootes fields and the HerS field, which
are representative of the entire data sets in terms of survey limit.

\subsection{Stacking procedure and photometry}

\begin{figure*}[t]
\plotone{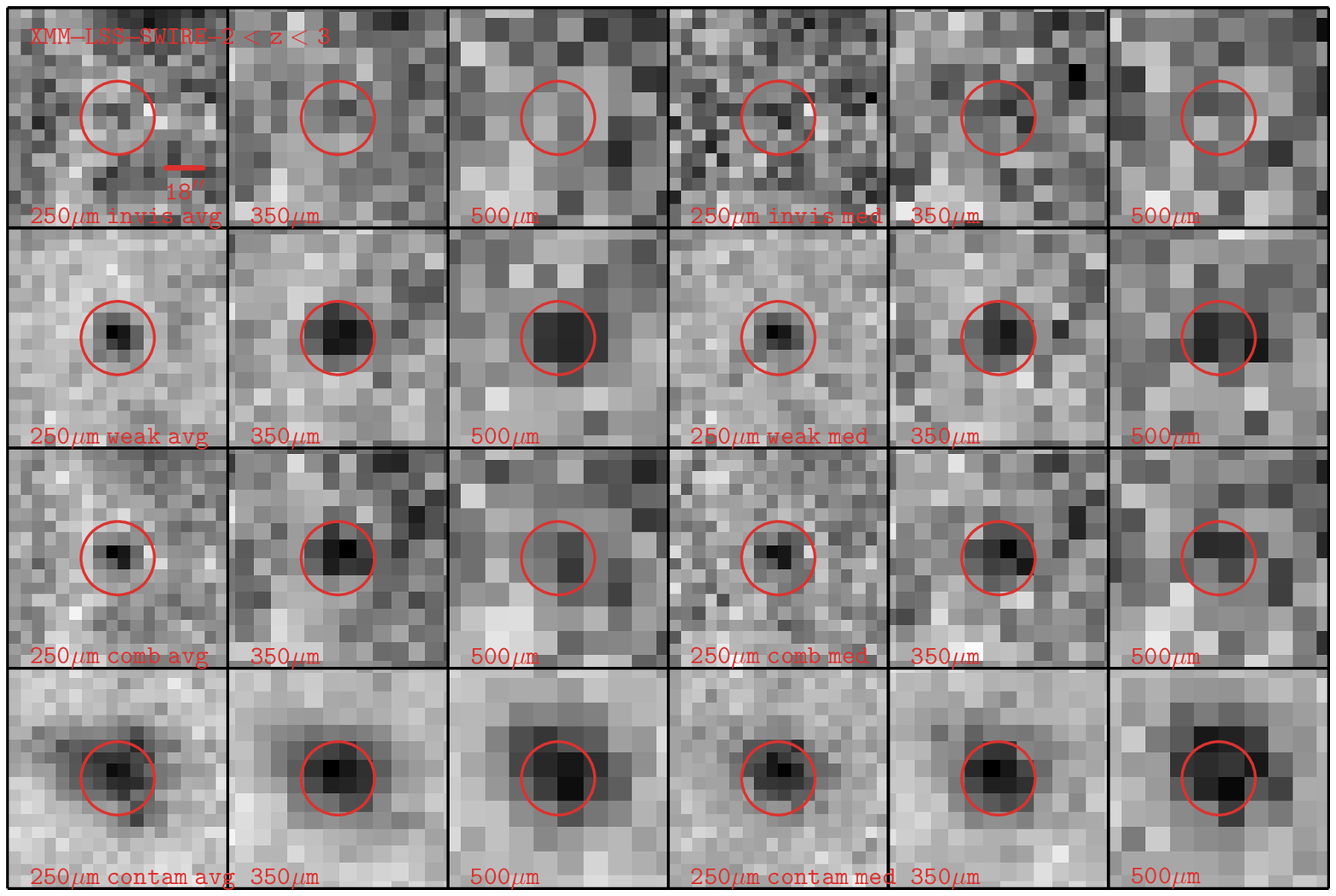}
\caption{Stacked SPIRE image stamps of the \Herschel{}-undetected
    quasars in the HerMES L6-XMM-LSS-SWIRE field at redshift
    $2<z\leq3$. The left panel shows the stamps generated using
    average in stacking, while the right panel shows those derived
    using median. In each panel, the stacks of the ``invisible'',
    ``weak'', ``combined'' and ``contaminated'' sets are shown from
    the top to the bottom.} \label{fig:stack_image}
\end{figure*}

The unmatched SDSS quasars in each field were first split into
different bins by redshifts. For simplicity, we only discuss the
``equal redshift'' case but not the ``equal volume'' one. We adopted a
large step-size, $\Delta z=1$, to ensure enough number of objects for
stacking at the most critical redshifts. The objects were then
visually inspected in \SI{250}{\micron} and were subsequently divided
into three categories: \emph{``weak''} objects seem to have some weak
enhancement at the source locations, \emph{``invisible''} objects do
not have any sign of detection, and \emph{``contaminated''} objects
are likely contaminated by nearby sources. The results from this
classification, while done in \SI{250}{\micron}, were assumed to be
valid in 350 and \SI{500}{\micron} as well. The detailed numbers are
listed in Table~\ref{tab:stacking_photometry} for these three fields. The
stacking was done for these three categories separately in each
redshift bin, using both average and median as the combining methods.
We also combined the ``invisible'' and the ``weak'' objects into the
``combined'' set and stacked them together. In most cases, the
``invisible'' stacks still do not result in significant detections,
while the ``weak'' stacks usually result in positive detections in at
least 250 and \SI{350}{\micron}. The ``combined'' stacks, on the other
hand, are weaker than the ``weak'' stacks, but are still positive
detections in at least \SI{250}{\micron}.  The ``contaminated''
stacks, as expected, often show contamination from the residuals of
the blended neighbors, and hence were excluded from further
discussion. For demonstration, Figure~\ref{fig:stack_image} shows the
average and the median \SI{250}{\micron} stacks in these four cases
for the quasars at $2<z\leq 3$ in the L6-XMM-LSS-SWIRE field.

We used the Herschel Interactive Processing Environment (HIPE)
software package to do photometry on the stacks. The \Herschel{}
science and uncertainty frames of the image cutouts at the undetected
SDSS quasar coordinates were stacked separately and inputted to HIPE
task \texttt{sourceExtractorSussextractor}, which then performed PSF
fitting to measure the fluxes and associated errors. A \SI{3}{\sig}
detection limit was adopted during the photometry. The measured
results are also listed in Table~\ref{tab:stacking_photometry}.

As the ``invisible" set usually does not result in detections, it will
not be discussed further as an independent set. We will focus on the
``weak'' and the ``comb'' sets.

\begin{table*}[t]
\centering
\caption{Summary of photometry of stacked image}
    \label{tab:stacking_photometry}
\setlength{\tabcolsep}{2pt}
\begin{tabular}{lrrrr|rrrr|rrrr}
\tophline
                          & \multicolumn{4}{c|}{invisible set}                         & \multicolumn{4}{c|}{weak set}                          & \multicolumn{4}{c}{combined set}                     \\
  z                       & $N$ & $S_{250}$ & $S_{350}$ & $S_{500}$         &   $N$ & $S_{250}$ & $S_{350}$ & $S_{500}$            & $N$ & $S_{250}$ & $S_{350}$ & $S_{500}$ \\
\midhline
\multicolumn{12}{c}{Bootes-HerMES}                                                                                                         \\
\midhline
(0, 1] & 24 &      --      &      --      &      --      & 10 & $10.00\pm0.78$ & $2.53\pm0.78$ &      --      & 34 & $4.04\pm0.54$ &      --      & $2.48\pm0.64$ \\
       &     &      --      & $2.48\pm0.68$ &      --      &     & $9.32\pm0.79$ &      --      &      --      &     & $4.73\pm0.54$ &      --      &      --      \\
\grouphline
(1, 2] & 21 &      --      &      --      &      --      & 15 & $11.19\pm0.56$ & $7.17\pm0.56$ & $9.14\pm0.65$ & 36 & $4.40\pm0.47$ & $1.70\pm0.46$ & $1.68\pm0.51$ \\
       &     &      --      &      --      & $3.43\pm0.75$ &     & $11.80\pm0.56$ & $8.80\pm0.56$ & $7.48\pm0.66$ &     & $4.76\pm0.47$ & $2.53\pm0.46$ & $3.23\pm0.55$ \\
\grouphline
(2, 3] & 66 &      --      &      --      &      --      & 22 & $6.71\pm0.48$ & $5.59\pm0.48$ & $3.11\pm0.57$ & 88 & $1.17\pm0.30$ &      --      &      --      \\
       &     &      --      &      --      &      --      &     & $7.76\pm0.48$ & $7.05\pm0.48$ & $2.92\pm0.57$ &     & $1.67\pm0.30$ & $1.27\pm0.29$ &      --      \\
\grouphline
(3, 4] & 21 &      --      &      --      &      --      & 4 & $10.07\pm1.18$ & $14.20\pm1.19$ & $17.73\pm1.47$ & 25 &      --      &      --      & $3.38\pm0.58$ \\
       &     &      --      &      --      &      --      &     & $11.43\pm1.19$ & $8.99\pm1.16$ & $13.32\pm1.52$ &     &      --      &      --      & $2.91\pm0.57$ \\
\grouphline
(4, 5] & 2 &      --      &      --      & $18.48\pm2.00$ & 0 &      --      &      --      &      --      & 2 &      --      &      --      & $18.48\pm2.00$ \\
       &     &      --      &      --      & $18.48\pm2.00$ &     &      --      &      --      &      --      &     &      --      &      --      & $18.48\pm2.00$ \\
\midhline
\multicolumn{12}{c}{XMM-LSS-SWIRE}                                                                                                         \\
\midhline
(0, 1] & 31 &      --      &      --      &      --      & 12 & $9.44\pm0.78$ & $6.37\pm0.79$ & $4.39\pm0.89$ & 43 & $1.45\pm0.44$ &      --      &      --      \\
       &     &      --      &      --      &      --      &     & $9.26\pm0.78$ & $6.10\pm0.79$ & $3.85\pm0.89$ &     &      --      &      --      & $2.71\pm0.51$ \\
\grouphline
(1, 2] & 10 &      --      &      --      &      --      & 8 & $12.47\pm1.01$ & $6.85\pm0.99$ &      --      & 18 & $5.93\pm0.74$ &      --      &      --      \\
       &     &      --      &      --      &      --      &     & $12.45\pm1.02$ & $9.00\pm0.99$ &      --      &     & $6.47\pm0.74$ & $3.87\pm0.70$ &      --      \\
\grouphline
(2, 3] & 113 &      --      &      --      &      --      & 41 & $9.35\pm0.45$ & $5.92\pm0.44$ & $4.36\pm0.53$ & 154 & $2.31\pm0.23$ & $1.70\pm0.22$ &      --      \\
       &     &      --      & $1.14\pm0.26$ &      --      &     & $9.27\pm0.45$ & $5.45\pm0.44$ & $4.76\pm0.53$ &     & $2.71\pm0.23$ & $2.42\pm0.22$ & $1.34\pm0.27$ \\
\grouphline
(3, 4] & 18 & $3.28\pm0.83$ &      --      &      --      & 7 & $9.16\pm1.12$ & $9.39\pm1.05$ & $7.28\pm1.24$ & 25 & $2.35\pm0.67$ &      --      &      --      \\
       &     &      --      &      --      &      --      &     & $9.72\pm1.12$ & $13.25\pm1.05$ & $6.72\pm1.24$ &     & $3.63\pm0.67$ &      --      & $2.98\pm0.72$ \\
\grouphline
(4, 5] & 1 &      --      &      --      & $24.32\pm2.93$ & 0 &      --      &      --      &      --      & 1 &      --      &      --      &      --      \\
       &     &      --      &      --      & $24.32\pm2.93$ &     &      --      &      --      &      --      &     &      --      &      --      &      --      \\
\midhline
\multicolumn{12}{c}{HerS}                                                                                                         \\
\midhline
(0, 1] & 904 & $7.12\pm0.63$ & $4.14\pm0.84$ & $2.22\pm0.73$ & 0 &      --      &      --      &      --      & 904 & $7.12\pm0.63$ & $4.14\pm0.84$ & $2.22\pm0.73$ \\
       &     & $6.88\pm0.63$ & $4.26\pm0.84$ & $2.53\pm0.73$ &     &      --      &      --      &      --      &     & $6.88\pm0.63$ & $4.26\pm0.84$ & $2.53\pm0.73$ \\
\grouphline
(1, 2] & 699 &      --      &      --      &      --      & 452 & $17.17\pm0.82$ & $11.22\pm1.10$ & $6.67\pm0.99$ & 1151 & $7.44\pm0.54$ & $5.13\pm0.71$ & $2.82\pm0.63$ \\
       &     &      --      &      --      &      --      &     & $17.35\pm0.82$ & $11.31\pm1.10$ & $6.62\pm0.98$ &     & $7.66\pm0.54$ & $5.50\pm0.71$ & $3.07\pm0.63$ \\
\grouphline
(2, 3] & 1046 &      --      &      --      &      --      & 350 & $14.97\pm1.00$ & $11.54\pm1.37$ & $8.46\pm1.15$ & 1396 & $4.24\pm0.53$ & $3.88\pm0.70$ & $3.39\pm0.61$ \\
       &     &      --      &      --      &      --      &     & $15.37\pm1.00$ & $11.75\pm1.37$ & $8.10\pm1.15$ &     & $4.54\pm0.53$ & $3.84\pm0.70$ & $3.30\pm0.61$ \\
\grouphline
(3, 4] & 220 &      --      &      --      &      --      & 77 & $13.11\pm1.83$ & $10.46\pm2.52$ & $10.55\pm2.13$ & 297 & $3.96\pm1.04$ & $4.79\pm1.41$ & $4.53\pm1.27$ \\
       &     &      --      &      --      &      --      &     & $13.31\pm1.83$ & $10.91\pm2.52$ & $12.15\pm2.13$ &     & $3.99\pm1.04$ & $4.78\pm1.41$ &      --      \\
\grouphline
(4, 5] & 27 &      --      &      --      &      --      & 0 &      --      &      --      &      --      & 27 &      --      &      --      &      --      \\
       &     &      --      &      --      &      --      &     &      --      &      --      &      --      &     &      --      &      --      &      --      \\
\bottomhline
\end{tabular}
\tablecomments{Two algorithms, namely, average and median, were
    employed while stacking. The first row for each redshift lists the
    photometry using the average stacks, while the second row lists
    those using the median stacks. The fluxes are given in
    \si{\milli\jansky}; None detections are denoted using ``--''.}
\end{table*}

\subsection{IR luminosity}

The FIR SEDs of the stacks could be fit in the usual way. There is a
caveat, however. The implicit assumption of any stacking analysis is
always that the sources being stacked can be scaled linearly. This is
not necessary true in our case, because these \Herschel{}-undetected
SDSS quasars could be of different dust temperatures and hence their
intrinsic FIR SEDs could vary significantly in shape. Therefore, the
physical properties inferred from the SED fitting of the stacks should
be interpreted with caution. Here we only focus on the total IR
luminosities of the stacks, and assume that these are the
representative values of the stacked population.

In most cases, only the \SI{250}{\micron} stacks have detections of
sufficient S/N, and this prevented us from doing the full suite of FIR SED
fitting as in \S\ref{sec:modeling}. First of all, we had to abandon the MBB
fit because it would require at least three bands to obtain reasonable
constraints. This left the three starburst models to use, which all allow
the estimation of $L_{IR}^{SB}$ by searching for the template whose
associated $L_{IR}$ is the closest match to the one or two-band flux
densities at the given redshift. However, we also opted to abandon the SK07
models because in some cases the results had large discrepancies with
respect to those based on CE01 and DH02. This is largely due to the wide
dynamic range and the fine grid of the SK07 templates, which could make
their results highly degenerated in case when only one or two bands are
available. As we only have a very limited number of bins here, including
these deviant SK07 results would significantly skew the picture. Therefore,
we used the results from the CE01 and the DH02 templates for this analysis.
These results are summarized in Table~\ref{tab:stacking_analysis}.

\begin{table*}[t]
\centering
\caption{Summary of stacking analysis}
    \label{tab:stacking_analysis}
\begin{tabular}{lrrrr|rrrr}
\tophline
                          & \multicolumn{4}{c|}{weak set}                          & \multicolumn{4}{c}{combined set}                     \\
                          & $0<z<1$     & $1<z<2$     & $2<z<3$     & $3<z<4$     & $0<z<1$     & $1<z<2$     & $2<z<3$     & $3<z<4$     \\
\midhline
\multicolumn{9}{c}{Bootes-HerMES}                                                                                                         \\
\midhline
$N_{obj}$                 & 10          & 15          & 22          & 4           & 34          & 36          & 88          & 25          \\
$\mathrm{L_{IR}^{CE01}}$  & 11.5 (11.5*)& 12.0 (12.0) & 12.2 (12.3) & 12.8 (12.7) & 11.2 (11.2*)& 11.5 (11.6) & 11.5*(11.5) & 12.2*(12.2*)\\
$\mathrm{L_{IR}^{DH02}}$  & 11.4 (11.4*)& 12.0 (12.2) & 12.2 (12.2) & 12.8 (12.6) & 11.2 (11.2*)& 11.4 (11.7) & 11.4*(11.7) & 12.2*(12.2*)\\
\midhline
\multicolumn{9}{c}{XMM-LSS-SWIRE}                                                                                                         \\
\midhline
$N_{obj}$                 & 12          &  8          & 41          & 7           & 43          & 18          & 154         & 25          \\
$\mathrm{L_{IR}^{CE01}}$  & 11.6 (11.5) & 12.1 (12.1) & 12.3 (12.3) & 12.6 (12.7) & 10.5*(11.6*)& 11.8*(11.8) & 11.7 (11.8) & 12.1*(12.2) \\
$\mathrm{L_{IR}^{DH02}}$  & 11.4 (11.4) & 12.2 (12.2) & 12.4 (12.2) & 12.6 (12.6) & 10.6*(11.2*)& 11.7*(11.7) & 11.7 (11.7) & 12.0*(12.2) \\
\midhline
\multicolumn{9}{c}{HerS}                                                                                                                  \\
\midhline
$N_{obj}$                 &             & 452         & 350         & 77          &             & 1151        & 1396        & 297         \\
$\mathrm{L_{IR}^{CE01}}$  &             & 12.2 (12.2) & 12.6 (12.6) & 12.7 (12.8) &             & 11.8 (11.9) & 12.0 (12.0) & 12.3 (12.3) \\
$\mathrm{L_{IR}^{DH02}}$  &             & 12.2 (12.2) & 12.6 (12.6) & 12.8 (12.8) &             & 12.0 (12.0) & 12.0 (12.0) & 12.4 (12.2) \\
\bottomhline
\end{tabular}
\tablecomments{The listed IR luminosities ($\mathrm{Log
        L_{IR}^{SB}/L_{\odot}}$) with and without parenthesis are derived
    from the median and the average stacks, respectively. The
    asterisks denote that only one photometry point is available when
    doing the SED fitting.}
\end{table*}

\subsection{Correction to fraction of IR-luminous quasars}

One key objective of this stacking analysis was to study how the
distribution of the fraction of IR-luminous optical quasars (see
Figure~\ref{fig:ir_bright_vs_z}) will change if the undetected
population is taken into account. In particular, we are interested in
understanding whether the sharp drop from $z\approx 2$ to higher
redshifts is caused by the potential bias when using only the
SPIRE-detected objects for the statistics.

This IR-luminous fraction depends on the exact choice of $L_c$, which
could be below the survey limit of the data, and therefore ideally we
should compensate for the number of objects that are below the survey
limit and yet still have $L_{IR}^{SB}>L_c$. However, as
Table~\ref{tab:stacking_analysis} shows, this is not always possible
because we cannot control the input undetected objects so that they
will stack to the exact $L_c$ required. Therefore, we need to make
corrections in a different way.

As we have chosen $L_c=\SI{d12.5}{\Lsun}$, it is only necessary to
analyze the ``weak'' sets, because from
Table~\ref{tab:stacking_analysis} it is obvious that the individually
undetected objects that still have $L_{IR}^{SB}>L_c$ can only be found in these
sets. We first start from the $2\leq z<3$ bin. In the L5-Bo\"{o}tes
field, there are 18 SPIRE-detected quasars in this bin that have
$L_{IR}^{SB}\geq\SI{d12.5}{\Lsun}$. The stack of 22 undetected objects in
this bin, on the other hand, has the median of
$L_{IR}^{SB;u}=10^{12.2\text{--}12.3}\si{\Lsun}$ from the fits using CE01
or DH02 models. We use the superscript ``$u$'' to denoted that this
quantity is for the ``undetected'' objects. By the definition of
median, half (i.e., 11) of these
undetected objects should have $L_{IR}^{SB}>\SI{d12.2}{\Lsun}$. As
$L_c>\SI{d12.2}{\Lsun}$, we conclude that \emph{at most} there are 11
objects that can be added to the existing 18 objects. In other words,
the number of objects above the threshold $L_c$ after this correction,
$N_{cor}$, should be $18\leq N_{cor}<29$, and the correction factor,
$f_{cor}$, should be $f_{cor}<1.6$. Similarly, for the bin of $1\leq
z<2$, there are 5 objects that have $L_{IR}^{SB}>\SI{d12.5}{\Lsun}$, and
the stack of 15 undetected objects has the median of
$L_{IR}^{SB;u}=\SI{d12.0}{\Lsun}$. Therefore we get $5\leq
N_{cor}<12$,
and $f_{cor}<2.4$.

We can repeat the similar analysis for all the fields, and obtain the
constrained corrections for each redshift bin. If $L_{IR}^{SB;u} < L_c$,
such a correction is ``maximum possible'', and if $L_{IR}^{SB;u} > L_c$,
such a correction is ``minimum necessary''. At $1\leq z<2$, the
corrections are all ``maximum possible". At $2\leq z<3$, the
correction is ``minimum necessary'' for the HerS field, and are
``maximum possible'' for all other fields. However, the HerS field
dominates the statistics here at $2\leq z<3$: the stack of 350
\Herschel{}-undetected quasars has median
$L_{IR}^{SB;u}=\SI{d12.6}{\Lsun}$, which means that we need to add 175
objects for the correction by this field alone. The total number of
\Herschel{}-detected quasars that have $L_{IR}^{SB}>\SI{d12.5}{\Lsun}$ in
this redshift bin is 84, and if we ignore the ``maximum possible''
corrections in other fields, the ``minimum necessary'' corrections
will be at least $(175+84)/84\approx 3.1$ for the entire sample.

\subsection{Correction to IR luminosity density}

As expected, the \Herschel{}-undetected SDSS quasars are significantly
less luminous than the \Herschel{}-detected ones at the same
redshifts. However, their collective contributions to
$\rho_{IR}^{SB;QSO}$ are comparable to those from the
\Herschel{}-detected population, simply because of their larger
number. This is particularly true at $z=2\text{--}3$, where their
contribution is even higher by a factor of $\sim 2\times$.

The correction in each redshift bin was done by adding the product of
the average stacked $L_{IR}^{SB}$ values of the ``combined'' sets multiplied by the
number of input objects in these sets. After such corrections, the evolution of
$\rho_{IR}^{SB;QSO}$ will peak at $z\approx 2.5$ instead of $z\approx
1.5$.

\bibliographystyle{apj.bst}
\bibliography{reflib.bib}

\begin{thebibliography}{}
\expandafter\ifx\csname natexlab\endcsname\relax\def\natexlab#1{#1}\fi

\bibitem[{{Aaronson} \& {Olszewski}(1984)}]{Aaronson1984}
{Aaronson}, M., \& {Olszewski}, E.~W. 1984, \nat, 309, 414

\bibitem[{{Abazajian} {et~al.}(2009){Abazajian}, {Adelman-McCarthy},
  {Ag{\"u}eros}, {Allam}, {Allende Prieto}, {An}, {Anderson}, {Anderson},
  {Annis}, {Bahcall}, \& et~al.}]{Abazajian2009}
{Abazajian}, K.~N., {Adelman-McCarthy}, J.~K., {Ag{\"u}eros}, M.~A., {et~al.}
  2009, \apjs, 182, 543

\bibitem[{{Annis} {et~al.}(2014){Annis}, {Soares-Santos}, {Strauss}, {Becker},
  {Dodelson}, {Fan}, {Gunn}, {Hao}, {Ivezi{\'c}}, {Jester}, {Jiang},
  {Johnston}, {Kubo}, {Lampeitl}, {Lin}, {Lupton}, {Miknaitis}, {Seo}, {Simet},
  \& {Yanny}}]{Annis2014}
{Annis}, J., {Soares-Santos}, M., {Strauss}, M.~A., {et~al.} 2014, \apj, 794,
  120

\bibitem[{{Azadi} {et~al.}(2015){Azadi}, {Aird}, {Coil}, {Moustakas}, {Mendez},
  {Blanton}, {Cool}, {Eisenstein}, {Wong}, \& {Zhu}}]{Azadi2015}
{Azadi}, M., {Aird}, J., {Coil}, A.~L., {et~al.} 2015, \apj, 806, 187

\bibitem[{{Blain} {et~al.}(2002){Blain}, {Smail}, {Ivison}, {Kneib}, \&
  {Frayer}}]{Blain2002}
{Blain}, A.~W., {Smail}, I., {Ivison}, R.~J., {Kneib}, J.-P., \& {Frayer},
  D.~T. 2002, \physrep, 369, 111

\bibitem[{{Bovy} {et~al.}(2011){Bovy}, {Hennawi}, {Hogg}, {Myers},
  {Kirkpatrick}, {Schlegel}, {Ross}, {Sheldon}, {McGreer}, {Schneider}, \&
  {Weaver}}]{Bovy2011}
{Bovy}, J., {Hennawi}, J.~F., {Hogg}, D.~W., {et~al.} 2011, \apj, 729, 141

\bibitem[{{Cao} {et~al.}(2008){Cao}, {Xia}, {Wu}, {Mao}, {Hao}, \&
  {Deng}}]{Cao2008}
{Cao}, C., {Xia}, X.~Y., {Wu}, H., {et~al.} 2008, \mnras, 390, 336

\bibitem[{{Carter}(1984)}]{Carter1984}
{Carter}, D. 1984, Astronomy Express, 1, 61

\bibitem[{{Casey}(2012)}]{Casey2012}
{Casey}, C.~M. 2012, \mnras, 425, 3094

\bibitem[{{Chary} \& {Elbaz}(2001)}]{Chary2001}
{Chary}, R., \& {Elbaz}, D. 2001, \apj, 556, 562

\bibitem[{{Daddi} {et~al.}(2005){Daddi}, {Dickinson}, {Chary}, {Pope},
  {Morrison}, {Alexander}, {Bauer}, {Brandt}, {Giavalisco}, {Ferguson}, {Lee},
  {Lehmer}, {Papovich}, \& {Renzini}}]{Daddi2005}
{Daddi}, E., {Dickinson}, M., {Chary}, R., {et~al.} 2005, \apjl, 631, L13

\bibitem[{{Dai} {et~al.}(2012){Dai}, {Bergeron}, {Elvis}, {Omont}, {Huang},
  {Bock}, {Cooray}, {Fazio}, {Hatziminaoglou}, {Ibar}, {Magdis}, {Oliver},
  {Page}, {Perez-Fournon}, {Rigopoulou}, {Roseboom}, {Scott}, {Symeonidis},
  {Trichas}, {Vieira}, {Willmer}, \& {Zemcov}}]{Dai2012}
{Dai}, Y.~S., {Bergeron}, J., {Elvis}, M., {et~al.} 2012, \apj, 753, 33

\bibitem[{{Dale} \& {Helou}(2002)}]{Dale2002}
{Dale}, D.~A., \& {Helou}, G. 2002, \apj, 576, 159

\bibitem[{{Dale} {et~al.}(2014){Dale}, {Helou}, {Magdis}, {Armus},
  {D{\'{\i}}az-Santos}, \& {Shi}}]{Dale2014}
{Dale}, D.~A., {Helou}, G., {Magdis}, G.~E., {et~al.} 2014, \apj, 784, 83

\bibitem[{{de Grijp} {et~al.}(1985){de Grijp}, {Miley}, {Lub}, \& {de
  Jong}}]{deGrijp1985}
{de Grijp}, M.~H.~K., {Miley}, G.~K., {Lub}, J., \& {de Jong}, T. 1985, \nat,
  314, 240

\bibitem[{{Draine}(2006)}]{Draine2006}
{Draine}, B.~T. 2006, \apj, 636, 1114

\bibitem[{{Draine} {et~al.}(2007){Draine}, {Dale}, {Bendo}, {Gordon}, {Smith},
  {Armus}, {Engelbracht}, {Helou}, {Kennicutt}, {Li}, {Roussel}, {Walter},
  {Calzetti}, {Moustakas}, {Murphy}, {Rieke}, {Bot}, {Hollenbach}, {Sheth}, \&
  {Teplitz}}]{Draine2007}
{Draine}, B.~T., {Dale}, D.~A., {Bendo}, G., {et~al.} 2007, \apj, 663, 866

\bibitem[{{Eales} {et~al.}(2010){Eales}, {Dunne}, {Clements}, {Cooray}, {de
  Zotti}, {Dye}, {Ivison}, {Jarvis}, {Lagache}, {Maddox}, {Negrello},
  {Serjeant}, {Thompson}, {van Kampen}, {Amblard}, {Andreani}, {Baes},
  {Beelen}, {Bendo}, {Benford}, {Bertoldi}, {Bock}, {Bonfield}, {Boselli},
  {Bridge}, {Buat}, {Burgarella}, {Carlberg}, {Cava}, {Chanial}, {Charlot},
  {Christopher}, {Coles}, {Cortese}, {Dariush}, {da Cunha}, {Dalton}, {Danese},
  {Dannerbauer}, {Driver}, {Dunlop}, {Fan}, {Farrah}, {Frayer}, {Frenk},
  {Geach}, {Gardner}, {Gomez}, {Gonz{\'a}lez-Nuevo}, {Gonz{\'a}lez-Solares},
  {Griffin}, {Hardcastle}, {Hatziminaoglou}, {Herranz}, {Hughes}, {Ibar},
  {Jeong}, {Lacey}, {Lapi}, {Lawrence}, {Lee}, {Leeuw}, {Liske},
  {L{\'o}pez-Caniego}, {M{\"u}ller}, {Nandra}, {Panuzzo}, {Papageorgiou},
  {Patanchon}, {Peacock}, {Pearson}, {Phillipps}, {Pohlen}, {Popescu},
  {Rawlings}, {Rigby}, {Rigopoulou}, {Robotham}, {Rodighiero}, {Sansom},
  {Schulz}, {Scott}, {Smith}, {Sibthorpe}, {Smail}, {Stevens}, {Sutherland},
  {Takeuchi}, {Tedds}, {Temi}, {Tuffs}, {Trichas}, {Vaccari}, {Valtchanov},
  {van der Werf}, {Verma}, {Vieria}, {Vlahakis}, \& {White}}]{Eales2010}
{Eales}, S., {Dunne}, L., {Clements}, D., {et~al.} 2010, \pasp, 122, 499

\bibitem[{{Elston} {et~al.}(1985){Elston}, {Cornell}, \&
  {Lebofsky}}]{Elston1985}
{Elston}, R., {Cornell}, M.~E., \& {Lebofsky}, M.~J. 1985, \apj, 296, 106

\bibitem[{{Evans} {et~al.}(2001){Evans}, {Frayer}, {Surace}, \&
  {Sanders}}]{Evans2001}
{Evans}, A.~S., {Frayer}, D.~T., {Surace}, J.~A., \& {Sanders}, D.~B. 2001,
  \aj, 121, 1893

\bibitem[{{Evans} {et~al.}(2010){Evans}, {Primini}, {Glotfelty}, {Anderson},
  {Bonaventura}, {Chen}, {Davis}, {Doe}, {Evans}, {Fabbiano}, {Galle}, {Gibbs},
  {Grier}, {Hain}, {Hall}, {Harbo}, {(Helen He}, {Houck}, {Karovska},
  {Kashyap}, {Lauer}, {McCollough}, {McDowell}, {Miller}, {Mitschang},
  {Morgan}, {Mossman}, {Nichols}, {Nowak}, {Plummer}, {Refsdal}, {Rots},
  {Siemiginowska}, {Sundheim}, {Tibbetts}, {Van Stone}, {Winkelman}, \&
  {Zografou}}]{Evans2010}
{Evans}, I.~N., {Primini}, F.~A., {Glotfelty}, K.~J., {et~al.} 2010, \apjs,
  189, 37

\bibitem[{{Gonz{\'a}lez-Nuevo} {et~al.}(2010){Gonz{\'a}lez-Nuevo}, {de Zotti},
  {Andreani}, {Barton}, {Bertoldi}, {Birkinshaw}, {Bonavera}, {Buttiglione},
  {Cooke}, {Cooray}, {Danese}, {Dunne}, {Eales}, {Fan}, {Jarvis},
  {Kl{\"o}ckner}, {Hatziminaoglou}, {Herranz}, {Hughes}, {Lapi}, {Lawrence},
  {Leeuw}, {Lopez-Caniego}, {Massardi}, {Mauch}, {Micha{\l}owski}, {Negrello},
  {Rawlings}, {Rodighiero}, {Samui}, {Serjeant}, {Vieira}, {White}, {Amblard},
  {Auld}, {Baes}, {Bonfield}, {Burgarella}, {Cava}, {Clements}, {Dariush},
  {Dye}, {Frayer}, {Fritz}, {Ibar}, {Ivison}, {Lagache}, {Maddox}, {Pascale},
  {Pohlen}, {Rigby}, {Sibthorpe}, {Smith}, {Temi}, {Thompson}, {Valtchanov}, \&
  {Verma}}]{Gonzalez-Nuevo2010}
{Gonz{\'a}lez-Nuevo}, J., {de Zotti}, G., {Andreani}, P., {et~al.} 2010, \aap,
  518, L38

\bibitem[{{Griffin} {et~al.}(2010){Griffin}, {Abergel}, {Abreu}, {Ade},
  {Andr{\'e}}, {Augueres}, {Babbedge}, {Bae}, {Baillie}, {Baluteau}, {Barlow},
  {Bendo}, {Benielli}, {Bock}, {Bonhomme}, {Brisbin}, {Brockley-Blatt},
  {Caldwell}, {Cara}, {Castro-Rodriguez}, {Cerulli}, {Chanial}, {Chen},
  {Clark}, {Clements}, {Clerc}, {Coker}, {Communal}, {Conversi}, {Cox},
  {Crumb}, {Cunningham}, {Daly}, {Davis}, {de Antoni}, {Delderfield}, {Devin},
  {di Giorgio}, {Didschuns}, {Dohlen}, {Donati}, {Dowell}, {Dowell}, {Duband},
  {Dumaye}, {Emery}, {Ferlet}, {Ferrand}, {Fontignie}, {Fox}, {Franceschini},
  {Frerking}, {Fulton}, {Garcia}, {Gastaud}, {Gear}, {Glenn}, {Goizel},
  {Griffin}, {Grundy}, {Guest}, {Guillemet}, {Hargrave}, {Harwit}, {Hastings},
  {Hatziminaoglou}, {Herman}, {Hinde}, {Hristov}, {Huang}, {Imhof}, {Isaak},
  {Israelsson}, {Ivison}, {Jennings}, {Kiernan}, {King}, {Lange}, {Latter},
  {Laurent}, {Laurent}, {Leeks}, {Lellouch}, {Levenson}, {Li}, {Li},
  {Lilienthal}, {Lim}, {Liu}, {Lu}, {Madden}, {Mainetti}, {Marliani}, {McKay},
  {Mercier}, {Molinari}, {Morris}, {Moseley}, {Mulder}, {Mur}, {Naylor},
  {Nguyen}, {O'Halloran}, {Oliver}, {Olofsson}, {Olofsson}, {Orfei}, {Page},
  {Pain}, {Panuzzo}, {Papageorgiou}, {Parks}, {Parr-Burman}, {Pearce},
  {Pearson}, {P{\'e}rez-Fournon}, {Pinsard}, {Pisano}, {Podosek}, {Pohlen},
  {Polehampton}, {Pouliquen}, {Rigopoulou}, {Rizzo}, {Roseboom}, {Roussel},
  {Rowan-Robinson}, {Rownd}, {Saraceno}, {Sauvage}, {Savage}, {Savini},
  {Sawyer}, {Scharmberg}, {Schmitt}, {Schneider}, {Schulz}, {Schwartz},
  {Shafer}, {Shupe}, {Sibthorpe}, {Sidher}, {Smith}, {Smith}, {Smith},
  {Spencer}, {Stobie}, {Sudiwala}, {Sukhatme}, {Surace}, {Stevens}, {Swinyard},
  {Trichas}, {Tourette}, {Triou}, {Tseng}, {Tucker}, {Turner}, {Vaccari},
  {Valtchanov}, {Vigroux}, {Virique}, {Voellmer}, {Walker}, {Ward}, {Waskett},
  {Weilert}, {Wesson}, {White}, {Whitehouse}, {Wilson}, {Winter}, {Woodcraft},
  {Wright}, {Xu}, {Zavagno}, {Zemcov}, {Zhang}, \& {Zonca}}]{Griffin2010}
{Griffin}, M.~J., {Abergel}, A., {Abreu}, A., {et~al.} 2010, \aap, 518, L3

\bibitem[{{Haas} {et~al.}(2003){Haas}, {Klaas}, {M{\"u}ller}, {Bertoldi},
  {Camenzind}, {Chini}, {Krause}, {Lemke}, {Meisenheimer}, {Richards}, \&
  {Wilkes}}]{Haas2003}
{Haas}, M., {Klaas}, U., {M{\"u}ller}, S.~A.~H., {et~al.} 2003, \aap, 402, 87

\bibitem[{{Hao} {et~al.}(2007){Hao}, {Weedman}, {Spoon}, {Marshall},
  {Levenson}, {Elitzur}, \& {Houck}}]{Hao2007}
{Hao}, L., {Weedman}, D.~W., {Spoon}, H.~W.~W., {et~al.} 2007, \apjl, 655, L77

\bibitem[{{Heckman} {et~al.}(1987){Heckman}, {Armus}, \& {Miley}}]{Heckman1987}
{Heckman}, T.~M., {Armus}, L., \& {Miley}, G.~K. 1987, \aj, 93, 276

\bibitem[{{Hopkins} \& {Beacom}(2006)}]{Hopkins2006}
{Hopkins}, A.~M., \& {Beacom}, J.~F. 2006, \apj, 651, 142

\bibitem[{{Houck} {et~al.}(1985){Houck}, {Schneider}, {Danielson},
  {Neugebauer}, {Soifer}, {Beichman}, \& {Lonsdale}}]{Houck1985}
{Houck}, J.~R., {Schneider}, D.~P., {Danielson}, G.~E., {et~al.} 1985, \apjl,
  290, L5

\bibitem[{{Houck} {et~al.}(1984){Houck}, {Soifer}, {Neugebauer}, {Beichman},
  {Aumann}, {Clegg}, {Gillett}, {Habing}, {Hauser}, {Low}, {Miley},
  {Rowan-Robinson}, \& {Walker}}]{Houck1984a}
{Houck}, J.~R., {Soifer}, B.~T., {Neugebauer}, G., {et~al.} 1984, \apjl, 278,
  L63

\bibitem[{{Ibar} {et~al.}(2010){Ibar}, {Ivison}, {Cava}, {Rodighiero},
  {Buttiglione}, {Temi}, {Frayer}, {Fritz}, {Leeuw}, {Baes}, {Rigby}, {Verma},
  {Serjeant}, {M{\"u}ller}, {Auld}, {Dariush}, {Dunne}, {Eales}, {Maddox},
  {Panuzzo}, {Pascale}, {Pohlen}, {Smith}, {de Zotti}, {Vaccari}, {Hopwood},
  {Cooray}, {Burgarella}, \& {Jarvis}}]{Ibar2010}
{Ibar}, E., {Ivison}, R.~J., {Cava}, A., {et~al.} 2010, \mnras, 409, 38

\bibitem[{{Ivison} {et~al.}(2007){Ivison}, {Greve}, {Dunlop}, {Peacock},
  {Egami}, {Smail}, {Ibar}, {van Kampen}, {Aretxaga}, {Babbedge}, {Biggs},
  {Blain}, {Chapman}, {Clements}, {Coppin}, {Farrah}, {Halpern}, {Hughes},
  {Jarvis}, {Jenness}, {Jones}, {Mortier}, {Oliver}, {Papovich},
  {P{\'e}rez-Gonz{\'a}lez}, {Pope}, {Rawlings}, {Rieke}, {Rowan-Robinson},
  {Savage}, {Scott}, {Seigar}, {Serjeant}, {Simpson}, {Stevens}, {Vaccari},
  {Wagg}, \& {Willott}}]{Ivison2007}
{Ivison}, R.~J., {Greve}, T.~R., {Dunlop}, J.~S., {et~al.} 2007, \mnras, 380,
  199

\bibitem[{{Jiang} {et~al.}(2006){Jiang}, {Fan}, {Cool}, {Eisenstein}, {Zehavi},
  {Richards}, {Scranton}, {Johnston}, {Strauss}, {Schneider}, \&
  {Brinkmann}}]{Jiang2006}
{Jiang}, L., {Fan}, X., {Cool}, R.~J., {et~al.} 2006, \aj, 131, 2788

\bibitem[{{Kennicutt}(1998)}]{Kennicutt1998}
{Kennicutt}, Jr., R.~C. 1998, \araa, 36, 189

\bibitem[{{Kim} \& {Sanders}(1998)}]{Kim1998}
{Kim}, D.-C., \& {Sanders}, D.~B. 1998, \apjs, 119, 41

\bibitem[{{Lawrence} {et~al.}(1999){Lawrence}, {Rowan-Robinson}, {Ellis},
  {Frenk}, {Efstathiou}, {Kaiser}, {Saunders}, {Parry}, {Xiaoyang}, \&
  {Crawford}}]{Lawrence1999}
{Lawrence}, A., {Rowan-Robinson}, M., {Ellis}, R.~S., {et~al.} 1999, \mnras,
  308, 897

\bibitem[{{Leipski} {et~al.}(2010){Leipski}, {Meisenheimer}, {Klaas}, {Walter},
  {Nielbock}, {Krause}, {Dannerbauer}, {Bertoldi}, {Besel}, {de Rosa}, {Fan},
  {Haas}, {Hutsemekers}, {Jean}, {Lemke}, {Rix}, \& {Stickel}}]{Leipski2010a}
{Leipski}, C., {Meisenheimer}, K., {Klaas}, U., {et~al.} 2010, \aap, 518, L34

\bibitem[{{Leipski} {et~al.}(2013){Leipski}, {Meisenheimer}, {Walter}, {Besel},
  {Dannerbauer}, {Fan}, {Haas}, {Klaas}, {Krause}, \& {Rix}}]{Leipski2013}
{Leipski}, C., {Meisenheimer}, K., {Walter}, F., {et~al.} 2013, \apj, 772, 103

\bibitem[{{Leipski} {et~al.}(2014){Leipski}, {Meisenheimer}, {Walter}, {Klaas},
  {Dannerbauer}, {De Rosa}, {Fan}, {Haas}, {Krause}, \& {Rix}}]{Leipski2014}
---. 2014, \apj, 785, 154

\bibitem[{{Lonsdale} {et~al.}(2006){Lonsdale}, {Farrah}, \&
  {Smith}}]{Lonsdale2006}
{Lonsdale}, C.~J., {Farrah}, D., \& {Smith}, H.~E. 2006, {Ultraluminous
  Infrared Galaxies}, ed. J.~W. {Mason}, 285

\bibitem[{{Lutz} {et~al.}(2011){Lutz}, {Poglitsch}, {Altieri}, {Andreani},
  {Aussel}, {Berta}, {Bongiovanni}, {Brisbin}, {Cava}, {Cepa}, {Cimatti},
  {Daddi}, {Dominguez-Sanchez}, {Elbaz}, {F{\"o}rster Schreiber}, {Genzel},
  {Grazian}, {Gruppioni}, {Harwit}, {Le Floc'h}, {Magdis}, {Magnelli},
  {Maiolino}, {Nordon}, {P{\'e}rez Garc{\'{\i}}a}, {Popesso}, {Pozzi},
  {Riguccini}, {Rodighiero}, {Saintonge}, {Sanchez Portal}, {Santini}, {Shao},
  {Sturm}, {Tacconi}, {Valtchanov}, {Wetzstein}, \& {Wieprecht}}]{Lutz2011}
{Lutz}, D., {Poglitsch}, A., {Altieri}, B., {et~al.} 2011, \aap, 532, A90

\bibitem[{{Magnelli} {et~al.}(2009){Magnelli}, {Elbaz}, {Chary}, {Dickinson},
  {Le Borgne}, {Frayer}, \& {Willmer}}]{Magnelli2009}
{Magnelli}, B., {Elbaz}, D., {Chary}, R.~R., {et~al.} 2009, \aap, 496, 57

\bibitem[{{Maiolino} {et~al.}(2007){Maiolino}, {Shemmer}, {Imanishi}, {Netzer},
  {Oliva}, {Lutz}, \& {Sturm}}]{Maiolino2007}
{Maiolino}, R., {Shemmer}, O., {Imanishi}, M., {et~al.} 2007, \aap, 468, 979

\bibitem[{{Massaro} {et~al.}(2009){Massaro}, {Giommi}, {Leto}, {Marchegiani},
  {Maselli}, {Perri}, {Piranomonte}, \& {Sclavi}}]{Massaro2009}
{Massaro}, E., {Giommi}, P., {Leto}, C., {et~al.} 2009, \aap, 495, 691

\bibitem[{{Mullaney} {et~al.}(2011){Mullaney}, {Alexander}, {Goulding}, \&
  {Hickox}}]{Mullaney2011}
{Mullaney}, J.~R., {Alexander}, D.~M., {Goulding}, A.~D., \& {Hickox}, R.~C.
  2011, \mnras, 414, 1082

\bibitem[{{Netzer} {et~al.}(2014){Netzer}, {Mor}, {Trakhtenbrot}, {Shemmer}, \&
  {Lira}}]{Netzer2014}
{Netzer}, H., {Mor}, R., {Trakhtenbrot}, B., {Shemmer}, O., \& {Lira}, P. 2014,
  \apj, 791, 34

\bibitem[{{Netzer} {et~al.}(2007){Netzer}, {Lutz}, {Schweitzer}, {Contursi},
  {Sturm}, {Tacconi}, {Veilleux}, {Kim}, {Rupke}, {Baker}, {Dasyra},
  {Mazzarella}, \& {Lord}}]{Netzer2007}
{Netzer}, H., {Lutz}, D., {Schweitzer}, M., {et~al.} 2007, \apj, 666, 806

\bibitem[{{Oliver} {et~al.}(2012){Oliver}, {Bock}, {Altieri}, {Amblard},
  {Arumugam}, {Aussel}, {Babbedge}, {Beelen}, {B{\'e}thermin}, {Blain},
  {Boselli}, {Bridge}, {Brisbin}, {Buat}, {Burgarella},
  {Castro-Rodr{\'{\i}}guez}, {Cava}, {Chanial}, {Cirasuolo}, {Clements},
  {Conley}, {Conversi}, {Cooray}, {Dowell}, {Dubois}, {Dwek}, {Dye}, {Eales},
  {Elbaz}, {Farrah}, {Feltre}, {Ferrero}, {Fiolet}, {Fox}, {Franceschini},
  {Gear}, {Giovannoli}, {Glenn}, {Gong}, {Gonz{\'a}lez Solares}, {Griffin},
  {Halpern}, {Harwit}, {Hatziminaoglou}, {Heinis}, {Hurley}, {Hwang}, {Hyde},
  {Ibar}, {Ilbert}, {Isaak}, {Ivison}, {Lagache}, {Le Floc'h}, {Levenson},
  {Faro}, {Lu}, {Madden}, {Maffei}, {Magdis}, {Mainetti}, {Marchetti},
  {Marsden}, {Marshall}, {Mortier}, {Nguyen}, {O'Halloran}, {Omont}, {Page},
  {Panuzzo}, {Papageorgiou}, {Patel}, {Pearson}, {P{\'e}rez-Fournon}, {Pohlen},
  {Rawlings}, {Raymond}, {Rigopoulou}, {Riguccini}, {Rizzo}, {Rodighiero},
  {Roseboom}, {Rowan-Robinson}, {S{\'a}nchez Portal}, {Schulz}, {Scott},
  {Seymour}, {Shupe}, {Smith}, {Stevens}, {Symeonidis}, {Trichas}, {Tugwell},
  {Vaccari}, {Valtchanov}, {Vieira}, {Viero}, {Vigroux}, {Wang}, {Ward},
  {Wardlow}, {Wright}, {Xu}, \& {Zemcov}}]{Oliver2012}
{Oliver}, S.~J., {Bock}, J., {Altieri}, B., {et~al.} 2012, \mnras, 424, 1614

\bibitem[{{Osmer}(2004)}]{Osmer2004}
{Osmer}, P.~S. 2004, Coevolution of Black Holes and Galaxies, 324

\bibitem[{{Osterbrock} \& {De Robertis}(1985)}]{Osterbrock1985}
{Osterbrock}, D.~E., \& {De Robertis}, M.~M. 1985, \pasp, 97, 1129

\bibitem[{{P{\^a}ris} {et~al.}(2014){P{\^a}ris}, {Petitjean}, {Aubourg},
  {Ross}, {Myers}, {Streblyanska}, {Bailey}, {Hall}, {Strauss}, {Anderson},
  {Bizyaev}, {Borde}, {Brinkmann}, {Bovy}, {Brandt}, {Brewington},
  {Brownstein}, {Cook}, {Ebelke}, {Fan}, {Filiz Ak}, {Finley}, {Font-Ribera},
  {Ge}, {Hamann}, {Ho}, {Jiang}, {Kinemuchi}, {Malanushenko}, {Malanushenko},
  {Marchante}, {McGreer}, {McMahon}, {Miralda-Escud{\'e}}, {Muna},
  {Noterdaeme}, {Oravetz}, {Palanque-Delabrouille}, {Pan}, {Perez-Fournon},
  {Pieri}, {Riffel}, {Schlegel}, {Schneider}, {Simmons}, {Viel}, {Weaver},
  {Wood-Vasey}, {Y{\`e}che}, \& {York}}]{Paris2014}
{P{\^a}ris}, I., {Petitjean}, P., {Aubourg}, {\'E}., {et~al.} 2014, \aap, 563,
  A54

\bibitem[{{Pascale} {et~al.}(2011){Pascale}, {Auld}, {Dariush}, {Dunne},
  {Eales}, {Maddox}, {Panuzzo}, {Pohlen}, {Smith}, {Buttiglione}, {Cava},
  {Clements}, {Cooray}, {Dye}, {de Zotti}, {Fritz}, {Hopwood}, {Ibar},
  {Ivison}, {Jarvis}, {Leeuw}, {L{\'o}pez-Caniego}, {Rigby}, {Rodighiero},
  {Scott}, {Smith}, {Temi}, {Vaccari}, \& {Valtchanov}}]{Pascale2011}
{Pascale}, E., {Auld}, R., {Dariush}, A., {et~al.} 2011, \mnras, 415, 911

\bibitem[{{Pilbratt} {et~al.}(2010){Pilbratt}, {Riedinger}, {Passvogel},
  {Crone}, {Doyle}, {Gageur}, {Heras}, {Jewell}, {Metcalfe}, {Ott}, \&
  {Schmidt}}]{Pilbratt2010}
{Pilbratt}, G.~L., {Riedinger}, J.~R., {Passvogel}, T., {et~al.} 2010, \aap,
  518, L1

\bibitem[{{Poglitsch} {et~al.}(2010){Poglitsch}, {Waelkens}, {Geis},
  {Feuchtgruber}, {Vandenbussche}, {Rodriguez}, {Krause}, {Renotte}, {van
  Hoof}, {Saraceno}, {Cepa}, {Kerschbaum}, {Agn{\`e}se}, {Ali}, {Altieri},
  {Andreani}, {Augueres}, {Balog}, {Barl}, {Bauer}, {Belbachir}, {Benedettini},
  {Billot}, {Boulade}, {Bischof}, {Blommaert}, {Callut}, {Cara}, {Cerulli},
  {Cesarsky}, {Contursi}, {Creten}, {De Meester}, {Doublier}, {Doumayrou},
  {Duband}, {Exter}, {Genzel}, {Gillis}, {Gr{\"o}zinger}, {Henning},
  {Herreros}, {Huygen}, {Inguscio}, {Jakob}, {Jamar}, {Jean}, {de Jong},
  {Katterloher}, {Kiss}, {Klaas}, {Lemke}, {Lutz}, {Madden}, {Marquet},
  {Martignac}, {Mazy}, {Merken}, {Montfort}, {Morbidelli}, {M{\"u}ller},
  {Nielbock}, {Okumura}, {Orfei}, {Ottensamer}, {Pezzuto}, {Popesso},
  {Putzeys}, {Regibo}, {Reveret}, {Royer}, {Sauvage}, {Schreiber}, {Stegmaier},
  {Schmitt}, {Schubert}, {Sturm}, {Thiel}, {Tofani}, {Vavrek}, {Wetzstein},
  {Wieprecht}, \& {Wiezorrek}}]{Poglitsch2010}
{Poglitsch}, A., {Waelkens}, C., {Geis}, N., {et~al.} 2010, \aap, 518, L2

\bibitem[{{Pope} \& {Chary}(2010)}]{Pope2010}
{Pope}, A., \& {Chary}, R.-R. 2010, \apjl, 715, L171

\bibitem[{{Richards} {et~al.}(2002){Richards}, {Fan}, {Newberg}, {Strauss},
  {Vanden Berk}, {Schneider}, {Yanny}, {Boucher}, {Burles}, {Frieman}, {Gunn},
  {Hall}, {Ivezi{\'c}}, {Kent}, {Loveday}, {Lupton}, {Rockosi}, {Schlegel},
  {Stoughton}, {SubbaRao}, \& {York}}]{Richards2002}
{Richards}, G.~T., {Fan}, X., {Newberg}, H.~J., {et~al.} 2002, \aj, 123, 2945

\bibitem[{{Richards} {et~al.}(2006){Richards}, {Strauss}, {Fan}, {Hall},
  {Jester}, {Schneider}, {Vanden Berk}, {Stoughton}, {Anderson}, {Brunner},
  {Gray}, {Gunn}, {Ivezi{\'c}}, {Kirkland}, {Knapp}, {Loveday}, {Meiksin},
  {Pope}, {Szalay}, {Thakar}, {Yanny}, {York}, {Barentine}, {Brewington},
  {Brinkmann}, {Fukugita}, {Harvanek}, {Kent}, {Kleinman}, {Krzesi{\'n}ski},
  {Long}, {Lupton}, {Nash}, {Neilsen}, {Nitta}, {Schlegel}, \&
  {Snedden}}]{Richards2006}
{Richards}, G.~T., {Strauss}, M.~A., {Fan}, X., {et~al.} 2006, \aj, 131, 2766

\bibitem[{{Riechers} {et~al.}(2013){Riechers}, {Bradford}, {Clements},
  {Dowell}, {P{\'e}rez-Fournon}, {Ivison}, {Bridge}, {Conley}, {Fu}, {Vieira},
  {Wardlow}, {Calanog}, {Cooray}, {Hurley}, {Neri}, {Kamenetzky}, {Aguirre},
  {Altieri}, {Arumugam}, {Benford}, {B{\'e}thermin}, {Bock}, {Burgarella},
  {Cabrera-Lavers}, {Chapman}, {Cox}, {Dunlop}, {Earle}, {Farrah}, {Ferrero},
  {Franceschini}, {Gavazzi}, {Glenn}, {Solares}, {Gurwell}, {Halpern},
  {Hatziminaoglou}, {Hyde}, {Ibar}, {Kov{\'a}cs}, {Krips}, {Lupu}, {Maloney},
  {Martinez-Navajas}, {Matsuhara}, {Murphy}, {Naylor}, {Nguyen}, {Oliver},
  {Omont}, {Page}, {Petitpas}, {Rangwala}, {Roseboom}, {Scott}, {Smith},
  {Staguhn}, {Streblyanska}, {Thomson}, {Valtchanov}, {Viero}, {Wang},
  {Zemcov}, \& {Zmuidzinas}}]{Riechers2013}
{Riechers}, D.~A., {Bradford}, C.~M., {Clements}, D.~L., {et~al.} 2013, \nat,
  496, 329

\bibitem[{{Rigby} {et~al.}(2011){Rigby}, {Maddox}, {Dunne}, {Negrello},
  {Smith}, {Gonz{\'a}lez-Nuevo}, {Herranz}, {L{\'o}pez-Caniego}, {Auld},
  {Buttiglione}, {Baes}, {Cava}, {Cooray}, {Clements}, {Dariush}, {de Zotti},
  {Dye}, {Eales}, {Frayer}, {Fritz}, {Hopwood}, {Ibar}, {Ivison}, {Jarvis},
  {Panuzzo}, {Pascale}, {Pohlen}, {Rodighiero}, {Serjeant}, {Temi}, \&
  {Thompson}}]{Rigby2011}
{Rigby}, E.~E., {Maddox}, S.~J., {Dunne}, L., {et~al.} 2011, \mnras, 415, 2336

\bibitem[{{Roseboom} {et~al.}(2010){Roseboom}, {Oliver}, {Kunz}, {Altieri},
  {Amblard}, {Arumugam}, {Auld}, {Aussel}, {Babbedge}, {B{\'e}thermin},
  {Blain}, {Bock}, {Boselli}, {Brisbin}, {Buat}, {Burgarella},
  {Castro-Rodr{\'{\i}}guez}, {Cava}, {Chanial}, {Chapin}, {Clements}, {Conley},
  {Conversi}, {Cooray}, {Dowell}, {Dwek}, {Dye}, {Eales}, {Elbaz}, {Farrah},
  {Fox}, {Franceschini}, {Gear}, {Glenn}, {Solares}, {Griffin}, {Halpern},
  {Harwit}, {Hatziminaoglou}, {Huang}, {Ibar}, {Isaak}, {Ivison}, {Lagache},
  {Levenson}, {Lu}, {Madden}, {Maffei}, {Mainetti}, {Marchetti}, {Marsden},
  {Mortier}, {Nguyen}, {O'Halloran}, {Omont}, {Page}, {Panuzzo},
  {Papageorgiou}, {Patel}, {Pearson}, {P{\'e}rez-Fournon}, {Pohlen},
  {Rawlings}, {Raymond}, {Rigopoulou}, {Rizzo}, {Rowan-Robinson}, {Portal},
  {Schulz}, {Scott}, {Seymour}, {Shupe}, {Smith}, {Stevens}, {Symeonidis},
  {Trichas}, {Tugwell}, {Vaccari}, {Valtchanov}, {Vieira}, {Vigroux}, {Wang},
  {Ward}, {Wright}, {Xu}, \& {Zemcov}}]{Roseboom2010}
{Roseboom}, I.~G., {Oliver}, S.~J., {Kunz}, M., {et~al.} 2010, \mnras, 409, 48

\bibitem[{{Rosen} {et~al.}(2015){Rosen}, {Webb}, {Watson}, {Ballet}, {Barret},
  {Braito}, {Carrera}, {Ceballos}, {Coriat}, {Della Ceca}, {Denkinson},
  {Esquej}, {Farrell}, {Freyberg}, {Gris{\'e}}, {Guillout}, {Heil},
  {Law-Green}, {Lamer}, {Lin}, {Martino}, {Michel}, {Motch}, {Nebot
  Gomez-Moran}, {Page}, {Page}, {Page}, {Pakull}, {Pye}, {Read}, {Rodriguez},
  {Sakano}, {Saxton}, {Schwope}, {Scott}, {Sturm}, {Traulsen}, {Yershov}, \&
  {Zolotukhin}}]{Rosen2015}
{Rosen}, S.~R., {Webb}, N.~A., {Watson}, M.~G., {et~al.} 2015, ArXiv e-prints,
  arXiv:1504.07051

\bibitem[{{Ross} {et~al.}(2012){Ross}, {Myers}, {Sheldon}, {Y{\`e}che},
  {Strauss}, {Bovy}, {Kirkpatrick}, {Richards}, {Aubourg}, {Blanton}, {Brandt},
  {Carithers}, {Croft}, {da Silva}, {Dawson}, {Eisenstein}, {Hennawi}, {Ho},
  {Hogg}, {Lee}, {Lundgren}, {McMahon}, {Miralda-Escud{\'e}},
  {Palanque-Delabrouille}, {P{\^a}ris}, {Petitjean}, {Pieri}, {Rich}, {Roe},
  {Schiminovich}, {Schlegel}, {Schneider}, {Slosar}, {Suzuki}, {Tinker},
  {Weinberg}, {Weyant}, {White}, \& {Wood-Vasey}}]{Ross2012}
{Ross}, N.~P., {Myers}, A.~D., {Sheldon}, E.~S., {et~al.} 2012, \apjs, 199, 3

\bibitem[{{Ross} {et~al.}(2013){Ross}, {McGreer}, {White}, {Richards}, {Myers},
  {Palanque-Delabrouille}, {Strauss}, {Anderson}, {Shen}, {Brandt},
  {Y{\`e}che}, {Swanson}, {Aubourg}, {Bailey}, {Bizyaev}, {Bovy}, {Brewington},
  {Brinkmann}, {DeGraf}, {Di Matteo}, {Ebelke}, {Fan}, {Ge}, {Malanushenko},
  {Malanushenko}, {Mandelbaum}, {Maraston}, {Muna}, {Oravetz}, {Pan},
  {P{\^a}ris}, {Petitjean}, {Schawinski}, {Schlegel}, {Schneider}, {Silverman},
  {Simmons}, {Snedden}, {Streblyanska}, {Suzuki}, {Weinberg}, \&
  {York}}]{Ross2013}
{Ross}, N.~P., {McGreer}, I.~D., {White}, M., {et~al.} 2013, \apj, 773, 14

\bibitem[{{Rowan-Robinson}(1995)}]{Rowan-Robinson1995}
{Rowan-Robinson}, M. 1995, \mnras, 272, 737

\bibitem[{{Rowan-Robinson} {et~al.}(2004){Rowan-Robinson}, {Lari},
  {Perez-Fournon}, {Gonzalez-Solares}, {La Franca}, {Vaccari}, {Oliver},
  {Gruppioni}, {Ciliegi}, {H{\'e}raudeau}, {Serjeant}, {Efstathiou},
  {Babbedge}, {Matute}, {Pozzi}, {Franceschini}, {Vaisanen}, {Afonso-Luis},
  {Alexander}, {Almaini}, {Baker}, {Basilakos}, {Barden}, {del Burgo},
  {Bellas-Velidis}, {Cabrera-Guerra}, {Carballo}, {Cesarsky}, {Clements},
  {Crockett}, {Danese}, {Dapergolas}, {Drolias}, {Eaton}, {Egami}, {Elbaz},
  {Fadda}, {Fox}, {Genzel}, {Goldschmidt}, {Gonzalez-Serrano}, {Graham},
  {Granato}, {Hatziminaoglou}, {Herbstmeier}, {Joshi}, {Kontizas}, {Kontizas},
  {Kotilainen}, {Kunze}, {Lawrence}, {Lemke}, {Linden-V{\o}rnle}, {Mann},
  {M{\'a}rquez}, {Masegosa}, {McMahon}, {Miley}, {Missoulis}, {Mobasher},
  {Morel}, {N{\o}rgaard-Nielsen}, {Omont}, {Papadopoulos}, {Puget},
  {Rigopoulou}, {Rocca-Volmerange}, {Sedgwick}, {Silva}, {Sumner}, {Surace},
  {Vila-Vilaro}, {van der Werf}, {Verma}, {Vigroux}, {Villar-Martin},
  {Willott}, {Carrami{\~n}ana}, \& {Mujica}}]{Rowan-Robinson2004}
{Rowan-Robinson}, M., {Lari}, C., {Perez-Fournon}, I., {et~al.} 2004, \mnras,
  351, 1290

\bibitem[{{Sanders} {et~al.}(1989){Sanders}, {Phinney}, {Neugebauer}, {Soifer},
  \& {Matthews}}]{Sanders1989}
{Sanders}, D.~B., {Phinney}, E.~S., {Neugebauer}, G., {Soifer}, B.~T., \&
  {Matthews}, K. 1989, \apj, 347, 29

\bibitem[{{Sanders} {et~al.}(1988){Sanders}, {Soifer}, {Elias}, {Madore},
  {Matthews}, {Neugebauer}, \& {Scoville}}]{Sanders1988}
{Sanders}, D.~B., {Soifer}, B.~T., {Elias}, J.~H., {et~al.} 1988, \apj, 325, 74

\bibitem[{{Schneider} {et~al.}(2010){Schneider}, {Richards}, {Hall}, {Strauss},
  {Anderson}, {Boroson}, {Ross}, {Shen}, {Brandt}, {Fan}, {Inada}, {Jester},
  {Knapp}, {Krawczyk}, {Thakar}, {Vanden Berk}, {Voges}, {Yanny}, {York},
  {Bahcall}, {Bizyaev}, {Blanton}, {Brewington}, {Brinkmann}, {Eisenstein},
  {Frieman}, {Fukugita}, {Gray}, {Gunn}, {Hibon}, {Ivezi{\'c}}, {Kent}, {Kron},
  {Lee}, {Lupton}, {Malanushenko}, {Malanushenko}, {Oravetz}, {Pan}, {Pier},
  {Price}, {Saxe}, {Schlegel}, {Simmons}, {Snedden}, {SubbaRao}, {Szalay}, \&
  {Weinberg}}]{Schneider2010}
{Schneider}, D.~P., {Richards}, G.~T., {Hall}, P.~B., {et~al.} 2010, \aj, 139,
  2360

\bibitem[{{Schweitzer} {et~al.}(2006){Schweitzer}, {Lutz}, {Sturm}, {Contursi},
  {Tacconi}, {Lehnert}, {Dasyra}, {Genzel}, {Veilleux}, {Rupke}, {Kim},
  {Baker}, {Netzer}, {Sternberg}, {Mazzarella}, \& {Lord}}]{Schweitzer2006}
{Schweitzer}, M., {Lutz}, D., {Sturm}, E., {et~al.} 2006, \apj, 649, 79

\bibitem[{{Scoville} {et~al.}(2003){Scoville}, {Frayer}, {Schinnerer}, \&
  {Christopher}}]{Scoville2003}
{Scoville}, N.~Z., {Frayer}, D.~T., {Schinnerer}, E., \& {Christopher}, M.
  2003, \apjl, 585, L105

\bibitem[{{Serjeant} {et~al.}(2010){Serjeant}, {Bertoldi}, {Blain}, {Clements},
  {Cooray}, {Danese}, {Dunlop}, {Dunne}, {Eales}, {Falder}, {Hatziminaoglou},
  {Hughes}, {Ibar}, {Jarvis}, {Lawrence}, {Lee}, {Micha{\l}owski}, {Negrello},
  {Omont}, {Page}, {Pearson}, {van der Werf}, {White}, {Amblard}, {Auld},
  {Baes}, {Bonfield}, {Burgarella}, {Buttiglione}, {Cava}, {Dariush}, {de
  Zotti}, {Dye}, {Frayer}, {Fritz}, {Gonzalez-Nuevo}, {Herranz}, {Ivison},
  {Lagache}, {Leeuw}, {Lopez-Caniego}, {Maddox}, {Pascale}, {Pohlen}, {Rigby},
  {Rodighiero}, {Samui}, {Sibthorpe}, {Smith}, {Temi}, {Thompson},
  {Valtchanov}, \& {Verma}}]{Serjeant2010a}
{Serjeant}, S., {Bertoldi}, F., {Blain}, A.~W., {et~al.} 2010, \aap, 518, L7

\bibitem[{{Shang} {et~al.}(2011){Shang}, {Brotherton}, {Wills}, {Wills},
  {Cales}, {Dale}, {Green}, {Runnoe}, {Nemmen}, {Gallagher}, {Ganguly},
  {Hines}, {Kelly}, {Kriss}, {Li}, {Tang}, \& {Xie}}]{Shang2011}
{Shang}, Z., {Brotherton}, M.~S., {Wills}, B.~J., {et~al.} 2011, \apjs, 196, 2

\bibitem[{{Shen} {et~al.}(2011){Shen}, {Richards}, {Strauss}, {Hall},
  {Schneider}, {Snedden}, {Bizyaev}, {Brewington}, {Malanushenko},
  {Malanushenko}, {Oravetz}, {Pan}, \& {Simmons}}]{Shen2011}
{Shen}, Y., {Richards}, G.~T., {Strauss}, M.~A., {et~al.} 2011, \apjs, 194, 45

\bibitem[{{Shi} {et~al.}(2013){Shi}, {Helou}, {Armus}, {Stierwalt}, \&
  {Dale}}]{Shi2013}
{Shi}, Y., {Helou}, G., {Armus}, L., {Stierwalt}, S., \& {Dale}, D. 2013, \apj,
  764, 28

\bibitem[{{Shi} {et~al.}(2007){Shi}, {Ogle}, {Rieke}, {Antonucci}, {Hines},
  {Smith}, {Low}, {Bouwman}, \& {Willmer}}]{Shi2007}
{Shi}, Y., {Ogle}, P., {Rieke}, G.~H., {et~al.} 2007, \apj, 669, 841

\bibitem[{{Siebenmorgen} \& {Kr{\"u}gel}(2007)}]{Siebenmorgen2007}
{Siebenmorgen}, R., \& {Kr{\"u}gel}, E. 2007, \aap, 461, 445

\bibitem[{{Smith} {et~al.}(2011){Smith}, {Dunne}, {Maddox}, {Eales},
  {Bonfield}, {Jarvis}, {Sutherland}, {Fleuren}, {Rigby}, {Thompson}, {Baldry},
  {Bamford}, {Buttiglione}, {Cava}, {Clements}, {Cooray}, {Croom}, {Dariush},
  {de Zotti}, {Driver}, {Dunlop}, {Fritz}, {Hill}, {Hopkins}, {Hopwood},
  {Ibar}, {Ivison}, {Jones}, {Kelvin}, {Leeuw}, {Liske}, {Loveday}, {Madore},
  {Norberg}, {Panuzzo}, {Pascale}, {Pohlen}, {Popescu}, {Prescott}, {Robotham},
  {Rodighiero}, {Scott}, {Seibert}, {Sharp}, {Temi}, {Tuffs}, {van der Werf},
  \& {van Kampen}}]{Smith2011}
{Smith}, D.~J.~B., {Dunne}, L., {Maddox}, S.~J., {et~al.} 2011, \mnras, 416,
  857

\bibitem[{{Solomon} \& {Vanden Bout}(2005)}]{Solomon2005}
{Solomon}, P.~M., \& {Vanden Bout}, P.~A. 2005, \araa, 43, 677

\bibitem[{{Symeonidis} {et~al.}(2013){Symeonidis}, {Vaccari}, {Berta}, {Page},
  {Lutz}, {Arumugam}, {Aussel}, {Bock}, {Boselli}, {Buat}, {Capak}, {Clements},
  {Conley}, {Conversi}, {Cooray}, {Dowell}, {Farrah}, {Franceschini},
  {Giovannoli}, {Glenn}, {Griffin}, {Hatziminaoglou}, {Hwang}, {Ibar},
  {Ilbert}, {Ivison}, {Floc'h}, {Lilly}, {Kartaltepe}, {Magnelli}, {Magdis},
  {Marchetti}, {Nguyen}, {Nordon}, {O'Halloran}, {Oliver}, {Omont},
  {Papageorgiou}, {Patel}, {Pearson}, {P{\'e}rez-Fournon}, {Pohlen}, {Popesso},
  {Pozzi}, {Rigopoulou}, {Riguccini}, {Rosario}, {Roseboom}, {Rowan-Robinson},
  {Salvato}, {Schulz}, {Scott}, {Seymour}, {Shupe}, {Smith}, {Valtchanov},
  {Wang}, {Xu}, {Zemcov}, \& {Wuyts}}]{Symeonidis2013}
{Symeonidis}, M., {Vaccari}, M., {Berta}, S., {et~al.} 2013, \mnras, 431, 2317

\bibitem[{{Symeonidis} {et~al.}(2014){Symeonidis}, {Georgakakis}, {Page},
  {Bock}, {Bonzini}, {Buat}, {Farrah}, {Franceschini}, {Ibar}, {Lutz},
  {Magnelli}, {Magdis}, {Oliver}, {Pannella}, {Paolillo}, {Rosario},
  {Roseboom}, {Vaccari}, \& {Villforth}}]{Symeonidis2014}
{Symeonidis}, M., {Georgakakis}, A., {Page}, M.~J., {et~al.} 2014, \mnras, 443,
  3728

\bibitem[{{Viero} {et~al.}(2013){Viero}, {Moncelsi}, {Quadri}, {Arumugam},
  {Assef}, {B{\'e}thermin}, {Bock}, {Bridge}, {Casey}, {Conley}, {Cooray},
  {Farrah}, {Glenn}, {Heinis}, {Ibar}, {Ikarashi}, {Ivison}, {Kohno},
  {Marsden}, {Oliver}, {Roseboom}, {Schulz}, {Scott}, {Serra}, {Vaccari},
  {Vieira}, {Wang}, {Wardlow}, {Wilson}, {Yun}, \& {Zemcov}}]{Viero2013}
{Viero}, M.~P., {Moncelsi}, L., {Quadri}, R.~F., {et~al.} 2013, \apj, 779, 32

\bibitem[{{Viero} {et~al.}(2014){Viero}, {Asboth}, {Roseboom}, {Moncelsi},
  {Marsden}, {Mentuch Cooper}, {Zemcov}, {Addison}, {Baker}, {Beelen}, {Bock},
  {Bridge}, {Conley}, {Devlin}, {Dor{\'e}}, {Farrah}, {Finkelstein},
  {Font-Ribera}, {Geach}, {Gebhardt}, {Gill}, {Glenn}, {Hajian}, {Halpern},
  {Jogee}, {Kurczynski}, {Lapi}, {Negrello}, {Oliver}, {Papovich}, {Quadri},
  {Ross}, {Scott}, {Schulz}, {Somerville}, {Spergel}, {Vieira}, {Wang}, \&
  {Wechsler}}]{Viero2014}
{Viero}, M.~P., {Asboth}, V., {Roseboom}, I.~G., {et~al.} 2014, \apjs, 210, 22

\bibitem[{{Wang} {et~al.}(2013){Wang}, {Viero}, {Clarke}, {Bock}, {Buat},
  {Conley}, {Farrah}, {Heinis}, {Magdis}, {Marchetti}, {Marsden}, {Norberg},
  {Oliver}, {Roehlly}, {Roseboom}, {Schulz}, {Smith}, {Vaccari}, \&
  {Zemcov}}]{Wang2013o}
{Wang}, L., {Viero}, M., {Clarke}, C., {et~al.} 2013, ArXiv e-prints,
  arXiv:1312.0552

\bibitem[{{Wang} {et~al.}(2010){Wang}, {Carilli}, {Neri}, {Riechers}, {Wagg},
  {Walter}, {Bertoldi}, {Menten}, {Omont}, {Cox}, \& {Fan}}]{Wang2010}
{Wang}, R., {Carilli}, C.~L., {Neri}, R., {et~al.} 2010, \apj, 714, 699

\bibitem[{{Wang} {et~al.}(2011{\natexlab{a}}){Wang}, {Wagg}, {Carilli},
  {Walter}, {Riechers}, {Willott}, {Bertoldi}, {Omont}, {Beelen}, {Cox},
  {Strauss}, {Bergeron}, {Forveille}, {Menten}, \& {Fan}}]{Wang2011a}
{Wang}, R., {Wagg}, J., {Carilli}, C.~L., {et~al.} 2011{\natexlab{a}}, \apjl,
  739, L34

\bibitem[{{Wang} {et~al.}(2011{\natexlab{b}}){Wang}, {Wagg}, {Carilli}, {Neri},
  {Walter}, {Omont}, {Riechers}, {Bertoldi}, {Menten}, {Cox}, {Strauss}, {Fan},
  \& {Jiang}}]{Wang2011}
---. 2011{\natexlab{b}}, \aj, 142, 101

\bibitem[{{Xia} {et~al.}(2012){Xia}, {Gao}, {Hao}, {Tan}, {Mao}, {Omont},
  {Flaquer}, {Leon}, \& {Cox}}]{Xia2012}
{Xia}, X.~Y., {Gao}, Y., {Hao}, C.-N., {et~al.} 2012, \apj, 750, 92

\bibitem[{{Yan} {et~al.}(2014){Yan}, {Stefanon}, {Ma}, {Willner}, {Somerville},
  {Ashby}, {Dav{\'e}}, {P{\'e}rez-Gonz{\'a}lez}, {Cava}, {Wiklind}, {Kocevski},
  {Rafelski}, {Kartaltepe}, {Cooray}, {Koekemoer}, \& {Grogin}}]{Yan2014}
{Yan}, H., {Stefanon}, M., {Ma}, Z., {et~al.} 2014, \apjs, 213, 2

\bibitem[{{York} {et~al.}(2000){York}, {Adelman}, {Anderson}, {Anderson},
  {Annis}, {Bahcall}, {Bakken}, {Barkhouser}, {Bastian}, {Berman}, {Boroski},
  {Bracker}, {Briegel}, {Briggs}, {Brinkmann}, {Brunner}, {Burles}, {Carey},
  {Carr}, {Castander}, {Chen}, {Colestock}, {Connolly}, {Crocker}, {Csabai},
  {Czarapata}, {Davis}, {Doi}, {Dombeck}, {Eisenstein}, {Ellman}, {Elms},
  {Evans}, {Fan}, {Federwitz}, {Fiscelli}, {Friedman}, {Frieman}, {Fukugita},
  {Gillespie}, {Gunn}, {Gurbani}, {de Haas}, {Haldeman}, {Harris}, {Hayes},
  {Heckman}, {Hennessy}, {Hindsley}, {Holm}, {Holmgren}, {Huang}, {Hull},
  {Husby}, {Ichikawa}, {Ichikawa}, {Ivezi{\'c}}, {Kent}, {Kim}, {Kinney},
  {Klaene}, {Kleinman}, {Kleinman}, {Knapp}, {Korienek}, {Kron}, {Kunszt},
  {Lamb}, {Lee}, {Leger}, {Limmongkol}, {Lindenmeyer}, {Long}, {Loomis},
  {Loveday}, {Lucinio}, {Lupton}, {MacKinnon}, {Mannery}, {Mantsch}, {Margon},
  {McGehee}, {McKay}, {Meiksin}, {Merelli}, {Monet}, {Munn}, {Narayanan},
  {Nash}, {Neilsen}, {Neswold}, {Newberg}, {Nichol}, {Nicinski}, {Nonino},
  {Okada}, {Okamura}, {Ostriker}, {Owen}, {Pauls}, {Peoples}, {Peterson},
  {Petravick}, {Pier}, {Pope}, {Pordes}, {Prosapio}, {Rechenmacher}, {Quinn},
  {Richards}, {Richmond}, {Rivetta}, {Rockosi}, {Ruthmansdorfer}, {Sandford},
  {Schlegel}, {Schneider}, {Sekiguchi}, {Sergey}, {Shimasaku}, {Siegmund},
  {Smee}, {Smith}, {Snedden}, {Stone}, {Stoughton}, {Strauss}, {Stubbs},
  {SubbaRao}, {Szalay}, {Szapudi}, {Szokoly}, {Thakar}, {Tremonti}, {Tucker},
  {Uomoto}, {Vanden Berk}, {Vogeley}, {Waddell}, {Wang}, {Watanabe},
  {Weinberg}, {Yanny}, {Yasuda}, \& {SDSS Collaboration}}]{York2000}
{York}, D.~G., {Adelman}, J., {Anderson}, Jr., J.~E., {et~al.} 2000, \aj, 120,
  1579

\end{thebibliography}

\end{document}